\documentclass{aa}
\usepackage{amsmath}
\usepackage{amssymb}
\usepackage{graphicx}
\graphicspath{{figures/}}
\usepackage{natbib}
\usepackage[colorlinks=True,
           linkcolor=blue,
            citecolor= blue]{hyperref}

\newcommand{\Pra}{\mbox{Pr}}

\newcommand{\Ro}{\mbox{Ro}}

\newcommand{\LophiL}{{\mbox{\underbar{Lo}}}_\varphi} 
\newcommand{\mi}{\mathrm{i}}
\newcommand{\pdv}[2]{\frac{\partial {#1}}{\partial {#2}}}
\newcommand{\md}{\eta} 
\newcommand{\mpv}{\mu_0} 
\newcommand{\vA}{u_{{\mbox{\tiny A}}}} 
\newcommand{\Rmd}{\mbox{R}_m}
\newcommand{\ks}{k_s}
\newcommand{\kz}{k_z}
\newcommand{\kr}{k_r}
\newcommand{\kt}{k_\theta}
\newcommand{\wls}{\lambda_s}
\newcommand{\wlz}{\lambda_z}

\newcommand{\gr}{\gamma}
\newcommand{\iu}{{i\mkern1mu}}
\newcommand{\Bphi}{B_\varphi}
\newcommand{\Alfven}{Alfv\'en}
\newcommand{\beq}{\begin{equation}}
\newcommand{\omegaND}{\widetilde{\omega}}
\newcommand{\refp}[1]{(\ref{#1})}
\newcommand{\beql}[1]{\begin{equation}\label{#1}}
\newcommand{\eeq}{\end{equation}}
\newcommand{\eeqp}{\, .\end{equation}}
\newcommand{\eeqc}{\, ,\end{equation}}
\newcommand{\beqa}{\begin{eqnarray}}
\newcommand{\eeqa}{\end{eqnarray}}

\newcommand{\AfphiL}{\underline{\omega}_{\mbox{\tiny A}\varphi}} 

\title{Interplay between magnetic fields and differential rotation in a stably stratified stellar radiative zone}
\author {L. Jouve\inst{\ref{irap1}}
\and F. Ligni\`eres\inst{\ref{irap1}}
\and M. Gaurat \inst{\ref{irap1}}
}

\institute{Universit\'e de Toulouse, CNRS, Institut de Recherche en Astrophysique et Plan\'etologie, 14 Avenue Edouard Belin, 31400 Toulouse, France\\
email:ljouve@irap.omp.eu \label{irap1}}

\date{Received /
Accepted}

\abstract
{The interactions between magnetic fields and differential rotation in stellar radiative interiors could play a major role at explaining the magnetism of intermediate-mass and massive stars, as well as understanding the differential rotation profile observed in red-giant stars.}
{The present study aims at studying the flow and field produced by a stellar radiative zone which is initially made to rotate differentially in the presence of a large-scale poloidal magnetic field threading the whole domain. We focus both on the axisymmetric configurations produced by the initial winding-up of the magnetic field lines and on the possible instabilities of those configurations. The effects of the stable stratification and thermal diffusion are investigated in detail, we aim in particular at assessing the role of the stratification at stabilising the system.}
{We perform 2D and 3D global Boussinesq numerical simulations started from an initial radial or cylindrical differential rotation and a large-scale poloidal magnetic field. Under the conditions of a large rotation frequency compared to the Alfv\'en frequency, a magnetic configuration strongly dominated by its toroidal component is built.
We then perturb this configuration to observe the development of non-axisymmetric instabilities.}
{The parameters of the simulations are chosen to respect the ordering of time scales of a typical stellar radiative zone. In this framework, the axisymmetric evolution is studied by varying the relative effects of the thermal diffusion, the Br\"unt-V\"ais\"al\"a frequency, the rotation and the initial poloidal field strength. After a transient time and using a suitable adimensionalisation, we find that the axisymmetric state only depends on  $t_{es}/t_{Ap}$ the ratio between the Eddington-Sweet circulation time scale and the Alfv\'en time scale. A scale analysis of the Boussinesq magnetohydrodynamical equations allows us to recover this result. In the cylindrical case, a magneto-rotational instability develops when the thermal diffusivity is sufficiently high to enable the favored wavenumbers to be insensitive to the effects of the stable stratification. In the radial case, the magneto-rotational instability is driven by the latitudinal shear created by the back-reaction of the Lorentz force on the flow. Increasing the level of stratification then leaves the growth rate of the instability mainly unaffected while its horizontal length scale grows.}
{Non-axisymmetric instabilities are likely to exist in stellar radiative zones despite the stable stratification. They could be at the origin of the magnetic dichotomy observed in intermediate-mass and massive stars and are unavoidable candidates for the transport of angular momentum in red giant stars.}


\begin{document}

\authorrunning{Jouve Ligni\`eres Gaurat}
\titlerunning{Magnetic fields in stably stratified radiative zones}

\maketitle


\section{Introduction}


Considerable progress has been made recently about the knowledge of magnetic fields at the surface of stars, mostly thanks to the ground-based instruments NARVAL at the Pic du Midi observatory in France and ESPaDOnS at the Mauna Kea Observatory in Hawa\"i. It has been known for more than a century now that the Sun harbors a strong magnetic field which manifests itself as spots popping-up at the solar surface \citep{hale08}. There is also now a general consensus on the fact that this magnetic field is produced by dynamo action inside the convective envelope of the Sun and that such a process should be quite general for all solar-like stars \citep{parker55, moffatt78}. The magnetism of intermediate-mass and massive stars has also been thoroughly investigated. It is however expected to differ strongly from that of cool stars because of the presence of the outer radiative zone. Indeed, if a convective dynamo is at play in the core of hot stars, it might be more difficult for the magnetic field created in the convective core to travel all the way to the surface so that observers from Earth could see it. In intermediate-mass and massive stars, the magnetism is indeed quite different from what is observed on cool stars: 5 to 10\% of these stars do exhibit a strong surface magnetic field above $300\rm G$ and these stars are also the ones which show chemical peculiarities in their spectra (Ap/Bp stars). Thanks to recent spectropolarimetric observations, detections of a much smaller amplitude field (at the sub-Gauss level) have been obtained on stars like Vega, Sirius A, Alhena, $\beta$-Uma or $\theta$-Leo \citep{lignieres2009,blazere2016a,blazere2016b}, leading to the idea that 2 classes of magnetism could exist in intermediate-mass and massive stars: the strong dipolar field of Ap/Bp stars and the ultra-weak Vega-like magnetic field. A sound explanation for the existence of these 2 types of magnetism and the absence of stars possessing fields with amplitudes between approximately $1\rm G$ and $300 \rm G$ is still lacking. A possible scenario was proposed by \citet{auriere2007}, relying on the existence of a magnetic instability which could develop only for weak enough dipolar fields and which would lead to the disruption of the axisymmetric magnetic configuration. In that scenario, a crucial role is given to the differential rotation which acts on the dipolar magnetic field to produce a configuration dominated by the toroidal component, very likely to be unstable to a magnetohydrodynamical (MHD) instability. If such an instability existed, not only would it possibly explain the minimum field of Ap/Bp stars, but it could also be at the origin of dynamo action in the radiative zones of Vega-like stars. Various studies have indeed recently focused on the appealing idea that dynamo action would not require convective motions but only non-axisymmetric hydro or MHD instabilities which, in conjonction with the differential rotation, would produce the electromotive force needed to close the dynamo loop \citep{spruit2002, braithwaite2006b, zahn07, guervilly2010, marcotte2016}. It is for now still debated if such a radiative zone dynamo could exist in stars.

The interplay between differential rotation and magnetic fields which is at the heart of the \citet{auriere2007} explanation of the magnetism of hot stars is also invoked to interpret the recent asteroseismic observations of more than 300 red giants provided by the Kepler satellite in the last decade. Indeed, in those stars, the radiative zone contracts below the H-burning shell and expands above, naturally leading to a spin-up of the innermost regions and a braking of the layers above. This is indeed what is observed, a differential rotation is established between the inner and outer shells in these stars because of the contraction/expansion phenomena \citep{deheuvels2012,deheuvels2014}. However, simple models assuming conservation of angular momentum considerably overestimate the level of differential rotation produced. More puzzling is the fact that even sophisticated stellar evolution models including the rotationally-induced transport of angular momentum fail at reproducing the observations \citep{eggenberger2012a, eggenberger2012b, ceillier2013, marques2013}. A more efficient transport of angular momentum seems then to be at play in those stellar radiative zones and magnetic fields are seriously considered as interesting candidates to play this role. In particular, the transport by travelling Alfv\'en waves could strongly modify the level of differential rotation, through, for example, the phase-mixing mechanism \citep{ionson78,spruit99}. Moreover, the development of magnetohydrodynamical (MHD) instabilities could lead to a turbulent transport which would efficiently redistribute the angular momentum. This possibility has been studied recently \citep{cantiello2014, fuller2019, eggenberger2019b} with unclear conclusions so far.


Instabilities of a differentially rotating stellar radiative zone with or without the presence of a magnetic field have also been widely investigated theoretically, experimentally and numerically. In hydrodynamical situations, differential rotation can be unstable to various types of instabilities, such as centrifugal or shear instabilities. Centrifugal (or inertial) instabilities require strong enough gradient while weak shear instabilities tend to be stabilized by the Coriolis force \citep[e.g.][]{knobloch82}. In the MHD case, a shear flow which is hydrodynamically stable can become unstable because of the presence of a large-scale magnetic field. This has been studied in various configurations and in particular when the differential rotation is forced through the boundaries. This is the case of the Taylor Couette flow in cylindrical geometry (or the equivalent spherical Couette flow in spherical geometry). A detailed review of the various MHD instabilities which can arise in Taylor-Couette flows for different rotation rates of the inner and outer cylinders has been published recently by \citet{rudiger2018}. The main instabilities described in that review are the current-driven Tayler instability \citep{tayler73, markey73} which is purely magnetic and the magnetorotational instability \citep{velikhov59, chandra60, Acheson78, BH92} which necessitates a gradient of rotation and is thus shear-driven. As described in \citet{rudiger2018}, the MRI exists for various large-scale magnetic field geometries: the standard MRI is found for purely axial fields, the so-called azimuthal-MRI for purely azimuthal fields and the so-called helical-MRI for a mixed axial/azimuthal configuration. It could be argued that the Tayler instability is the most relevant for stellar interiors since it only requires a magnetic configuration sufficiently dominated by its toroidal or its poloidal component and a rather weak rotation or differential rotation \citep{spruit99}. Detailed studies have been conducted using linear stability analysis for purely toroidal fields with various latitudinal dependences in rotating or differentially rotating radiative zones \citep{Kitchatinov08, Rudiger10, Rudiger16}. These analysis were local in radius but global in the horizontal directions and took into account the effects of stratification, focusing in particular on a realistic stellar regime where the heat conductivity is high. The $m=1$ Tayler instability was found to develop even for a large rotation rate compared to the toroidal Alfv\'en frequency but with very weak growth rates.

In this work, we do not focus on the instability of a purely toroidal field but wish to study the global 3D evolution of an initally poloidal field wound-up into a toroidal field by an initial differential rotation. The system containing all the physical ingredients of a stellar radiative zone (i.e. stratification, axisymmetric meridional flow, shear, global rotation, a mixed poloidal/toroidal magnetic field configuration, heat conductivity, viscosity and magnetic diffusivity) is then let free to evolve into potentially unstable equilibria. Our recent numerical studies \citep{jouve2015, meduri2019} show that it is in fact the MRI which is favored in these specific conditions. In these calculations, the initial poloidal field is wound-up by the differential rotation imposed initially for \citet{jouve2015} and forced trough the boundaries in \citet{meduri2019} until the Maxwell stresses feed back on the flow. In these situations, the toroidal Alfv\'en frequency always remains small compared to the rotation frequency and the dynamics associated with the rotation and the shear dominate. The growth rate of the Tayler instability is thus probably strongly reduced by the rotation, as shown by \cite{PT85} or \cite{Kitchatinov08} but the conditions for the development of the MRI are gathered so that the instability grows on a rotation time-scale. However, the important effects of stable stratification are omitted in the 3D numerical calculations cited above. Only a few recent 3D global numerical studies have focused on the effect of stable stratification on MHD instabilities in specific cases, like for example \citet{Philidet2019} for spherical Couette flows, \citet{Guerrero2019} for the Tayler instability in a non-rotating spherical shell or \cite{Szklarski13} for the Tayler instability of a toroidal field produced by the winding-up of an initial poloidal field. In this last study, very similar to what is presented in this paper, the wound-up magnetic field is found to be unstable only if the feedback on the differential rotation is inhibited until the ratio of toroidal Alfv\'en frequency to rotation frequency becomes sufficiently large so that the Tayler instability sets in. The MRI has thus probably been stabilized by the stable stratification in these particular calculations. In this paper, we investigate the possibility that high heat conductivities could let the MRI develop again in the same type of numerical setup.


In fact, in most studies dedicated to instabilities of MHD flows with differential rotation, the effect of the stable stratification is often neglected. For the application to stellar interiors, this is yet a crucial ingredient which may suppress a large number of instabilities, in particular the MRI \citep{spruit99}. Indeed, in order to avoid doing work against the stable stratification, the unstable displacements must be nearly horizontal and thus the vertical wavenumber must be high, at which point the diffusive effects will act to make the perturbations decay away. 
 However, in the hydrodynamical case, it has been shown that the largest growth rates of the instability of a horizonthal shear flow would be mostly unaffected by the presence of a large Br\"unt-V\"ais\"al\"a frequency $N$ (\citet{deloncle2007} for the inflectional instability, \citet{kloosterziel2008} for the inertial instability).  
Moreover, non-adiabatic effects should also be considered: if the thermal diffusivity is large, which is the case for stellar radiative zones, the effect of the stable stratification can be strongly reduced and some instabilities may survive for higher values of $N$ (see \citet{townsend58, zahn92} for the case of a vertical shear in a stably stratified atmosphere). The possible effects of a high thermal diffusion on MHD instabilities have been discussed theoretically for example by \citet{Acheson78} or \citet{spruit99} but very few global numerical simulations exist where MHD states containing mixed poloidal/toroidal fields and meridional flows and differential rotation subject to the Lorentz force feedback in a stably stratified environment with a varying thermal diffusivity have been analysed in detail. This is what we present in this article. This work is a follow-up on \citet{jouve2015} where the following intial value problem was considered: an initially imposed large-scale poloidal field is wound-up by an initially imposed differential rotation to produce an axisymmetric toroidal field. After approximately an Alfv\'en time-scale, the magnetic field back-reacts on the differential rotation and the dynamics is dominated by Alfv\'en waves which progressively damp the differential rotation. We focused in this last work on the possible development of non-axisymmetric instabilities during this whole process. We now study the effects of the stable stratification with various values of the Br\"unt-V\"ais\"al\"a frequency, when the thermal diffusivity is also allowed to vary. In particular, we wish to determine the characteristics of the new axisymmetric MHD states and whether the MRI found in \citet{jouve2015} can survive in a stably stratified environment. 

The paper is organized as follows: in Sect. \ref{sec_model} we present the model and the numerical code used to solve the MHD equations. Sect.\ref{sec_axi} then discusses the axisymmetric joint evolution of the magnetic field and the flow. We then investigate the stability of this axisymmetric configuration in Sect.\ref{sec_stab} and finally conclude in Sect.\ref{sec_conclu}.


\section{Numerical model}
\label{sec_model}

We wish to explore the interplay between magnetic fields and differential rotation in a 3D spherical shell with stable stratification, to mimic the physical processes at play in a differentially rotating stellar radiative zone. To do so, we choose to focus on an initial value problem where a magnetic field and differential rotation will be initially prescribed and then let free to evolve with time, according to the MHD equations in the Boussinesq approximation. Indeed, for now, we neglect the effects of a varying density. This will be considered in future works. The details of the equations are given in Sect. \ref{sub_eq}, the initial and boundary conditions are then discussed in Sections \ref{sub_init1} and \ref{sub_init2} and the numerical method is finally briefly described in Sect. \ref{sub_code}.

\subsection{Governing equations}
\label{sub_eq}

Assuming uniform dynamic viscosity $\mu$, magnetic diffusivity $\eta$, thermal conductivity $\chi$ and neglecting the local sources of heat and the centrifugal force, the governing equations under the Boussinesq approximation of a magnetized flow are
    \begin{gather}
     \label{incomp_bouss}
     \vphantom{\left(\frac{B^2}{1}\right)}\vec{\nabla}\cdot\vec{v}=0\ \ \ ,\\
     \begin{split}
     \label{ns_bouss}
     \vphantom{\left(\frac{B^2}{1}\right)}\frac{D\vec{v}}{Dt}=&-2\vec{\Omega_0}\times\vec{v}-\alpha\,T_1\,\vec{g}-\frac{1}{\rho}\vec{\nabla}\left(p_1+\frac{B^2}{8\pi}\right)\\
     &+\frac{1}{4\pi\rho}(\vec{B}\cdot\vec{\nabla})\,\vec{B}+\nu\Delta\vec{v}\ \ \ ,
     \end{split}\\
     \label{energie_bous}
     \vphantom{\left(\frac{B^2}{1}\right)}\frac{DT_1}{Dt}+\vec{v}\cdot\vec{\nabla}\overline{T}=\kappa\,\Delta T_1\ \ \ ,\\
     \label{induction_bouss}
     \vphantom{\left(\frac{B^2}{1}\right)}\frac{\partial \vec{B}}{\partial t}=\vec{\nabla}\times(\vec{v}\times\vec{B})+\eta\,\Delta\vec{B}\ \ \ ,
    \end{gather}
where $\vec{v}$ is the velocity field, $\vec{B}$ is the magnetic field, $\vec{\Omega_0}$ is the rotation rate at the rotation axis, $T(r,\theta,t)=\overline{T}(r)+T_1(r,\theta,t)$ is the temperature field with $\overline{T}(r)$ the temperature of the reference state and $T_1$ its fluctuation, $\rho$ is the uniform density of the reference state, $p_1$ is the pressure fluctuation, gravity is proportional to $1/r^2$, $\alpha$ is the coefficient of thermal expansion, $\nu=\mu/\rho$ is the kinematic viscosity and $\kappa=\chi/(\rho \, c_p)$ is the thermal diffusivity where $c_p$ is the heat capacity at constant pressure. 

These equations are then non-dimensionalised using $d=r_o-r_i$ (where $r_i$ and $r_o$ are respectively the inner and outer radii of the spherical shell) the thickness of the spherical domain, as the length unit, the poloidal Alfv\'en time $t_{Ap}=d\sqrt{4\pi\rho}/B_0$ as the time unit where the surface radial magnetic field at the poles $B_0$ is the poloidal magnetic field unit, $d\Omega_0\sqrt{4\pi\rho}$ as the toroidal magnetic field unit, $V_{Ap}=d/t_{Ap}$ as the meridional circulation unit, $d\Omega_0$ as the azimuthal velocity flow unit, $\Delta T=T_o-T_i$ as the temperature unit where $T_o$ and $T_i$ are respectively the temperature at the outer and at the inner radius of the spherical shell and $d^2\Omega^2_0\rho$ as the pressure unit. The full set of governing equations of the problem is given in appendix \ref{sec_eqnonaxi}, namely the equations for the 3 components of the velocity field, for the 3 components of the magnetic field and for the temperature field.

Five dimensionless numbers appear in the set of equations:
    \begin{gather}
     \label{tapto}
    Lo=\vphantom{\sqrt{\frac{\alpha g \Delta T}{R}}}\frac{t_{\Omega}}{t_{Ap}}=\frac{B_0}{d\Omega_0\sqrt{4\pi\rho}}\ \ \ ,\\
     \label{nsuro}
     \vphantom{\sqrt{\frac{\alpha g \Delta T}{R}}}\frac{N}{\Omega_0}=\frac{1}{\Omega_0}\sqrt{\frac{\alpha g \varDelta T}{d}}\ \ \ ,\\
     \label{lundquist}
     \vphantom{\sqrt{\frac{\alpha g \Delta T}{R}}}Lu=\frac{t_\eta}{t_{Ap}}=\frac{dB_0}{\eta\sqrt{4\pi\rho}}\ \ \ ,\\
     \label{prandtl}
     \vphantom{\sqrt{\frac{\alpha g \Delta T}{R}}}Pr=\frac{t_{\kappa}}{t_{\nu}}=\frac{\nu}{\kappa}\ \ \ ,\\
     \label{prandtl_mag}
     \vphantom{\sqrt{\frac{\alpha g \Delta T}{R}}}Pm=\frac{t_\eta}{t_{\nu}}=\frac{\nu}{\eta}\ \ \ ,
    \end{gather}

The Lorentz number $Lo$ measures the ratio between the rotation time-scale $t_\Omega$ and the Alfv\'en time-scale based on the poloidal field $t_{Ap}$, $N/\Omega_0$ is the ratio between the Br\"unt-V{\"a}is{\"a}l{\"a} frequency and the rotation frequency, the Lundquist number $Lu$ measures the ratio between the poloidal Alfv\'en time-scale and the magnetic diffusion time $t_\eta=d^2/\eta$ and finally the Prandtl numbers quantify the ratio of diffusivities or the ratio of diffusive time-scales where $t_\nu=d^2/\nu$ is the viscous time-scale and $t_\kappa=d^2/\kappa$ is the thermal diffusive time-scale.

We can add to these numbers, the definition of the Ekman number, which will be mentioned in the text, and measures the ratio of rotation to viscous time-scales: $E_k = \nu/\Omega d^2 = Lo/(Pm Lu)$.
We choose in this study to fix the values of 2 dimensionless numbers, namely the Lundquist number $Lu$ and the magnetic Prandtl number $Pm$. We then focus on the effects of the 3 other parameters: the Lorentz number $Lo$, the ratio $N/\Omega_0$ and the Prandtl number $Pr$. We shall see that in the axisymmetric case, the number of relevant dimensionless parameters can in fact be, in some limit cases, reduced to only one.
    
\subsection{Initial and boundary conditions}
\label{sub_init1}

In this work, we focus on initial conditions which will produce a large-scale magnetic field, likely to be unstable to MHD instabilities under certain circumstances. To do so, we start from a poloidal field which will be acted upon by an initial differential rotation. The winding-up of the initial poloidal field by the differential rotation will naturally produce a toroidal magnetic field. We propose to focus on the conditions for stability of such a magnetic configuration embedded in a stably stratified atmosphere. 

Initially, we thus choose the magnetic field to be axisymmetric, purely poloidal with a constant current density. The detailed expression of the initial magnetic field then reads:
\begin{equation}
\begin{split}
 \label{ci_mf}
 \vec{B}(r,\theta,t=0)
 =&\vec{B}_p(r,\theta,t=0)\\
 =&\frac{3 r \cos\theta B_0 }{r_o}\,\frac{1-4r_o/3r+r_i^4/3r^4}{1-(r_i/r_o)^4}\vec{e_r}\\
 &-\frac{3 r \sin\theta B_0 }{2 r_o}\,\frac{3-8r_o/3r-r_i^4/3r^4}{1-(r_i/r_o)^4}\vec{e_\theta}\ \ \ . 
\end{split}
\end{equation}
With this choice of normalization, $B_0$ is the value of the radial field on the axis of rotation at the outer shell $r=r_o$. For the boundary conditions, we impose that the magnetic field matches continuously to a potential field at both inner and outer boundaries.


The velocity field is also initially axisymmetric but purely azimuthal 
\begin{equation}
 \vec{v}(r,\theta,t=0)=v_\varphi(r,\theta,t=0)\,\vec{e_\varphi}=r\sin\theta\,\Omega(r,\theta,t=0)\,\vec{e_\varphi}\ \ \
\end{equation}
and two different initial rotation profiles will be used. They are discussed in the following section. The boundary conditions for the velocity field are chosen to be impenetrable and stress-free at both inner and outer shells.

The initial temperature field is a purely radial solution satisfying the thermal equilibrium $\vec \nabla^2 \overline{T}=0$. Fixed values are imposed for the temperature at both boundaries :
    \begin{equation}
     \label{ci_t}
     T(r,t=0)=\overline{T}(r)=T_i+\Delta T \frac{1-r_i/r}{1-r_i/r_o}\ \ \
    \end{equation}
Finally, to ensure that the flow is stable with respect to convection, the Br\"unt-V{\"a}is{\"a}l{\"a} frequency $N$ must be real which means that $\Delta T>0$.

\subsection{Radial VS cylindrical differential rotation}
\label{sub_init2}

The evolution of the toroidal field originating from the winding-up of an initial poloidal field by the differential rotation is expected to be strongly dependent on the differential rotation profile and magnetic configuration. Indeed, the term producing the toroidal field, known as the $\Omega$-effect is proportional to ${ \bf B_p \cdot \nabla} \Omega$ and the angle between the poloial field lines and the isocontours of $\Omega$ will thus determine the amount of toroidal field created. The efficiency of this $\Omega$-effect is quite important for our study since the ratio between toroidal and poloidal fields is known to be crucial for the stability of the magnetic configurations. That is why we choose to study two different profiles for the initial differential rotation, namely one dependent on the cylindrical radius only and the other one dependent on the spherical radius only. The expressions of both rotation profiles are given below:

\begin{gather}
    \label{cylindrical}
     \Omega(r\sin\theta,t=0)=\Omega_0\sqrt{\frac{2}{1+(r\sin\theta/r_o)^4}}\ \ \ ,\\
    \label{radial}
     \Omega(r,t=0)=\Omega_0\frac{1-c_1 \, (r-r_i)^2 \, r_o/r^3-c_2 \, (r-r_i)^2/(r\,r_o)}{1-(c_1+c_2)(1-r_i/r_o)^2}\ \ \ .
    \end{gather}
\noindent where $\Omega_0$ is the rotation rate at the equator at $r=r_o$ and where $c_1=0.980$ and $c_2=0.214$ are chosen such that the contrast in the rotation rate between the inner and outer shells is approximately the same for both profiles, namely $(\Omega_i-\Omega_0)/\Omega_0 \approx 1$. We thus choose to initially impose a strong differential rotation which is then let free to evolve without any forcing. The transport of angular momentum resulting from the system dynamics will then naturally modify this initial profile.

\begin{figure}[h!]
\begin{center}
  \includegraphics[width=0.22\textwidth]{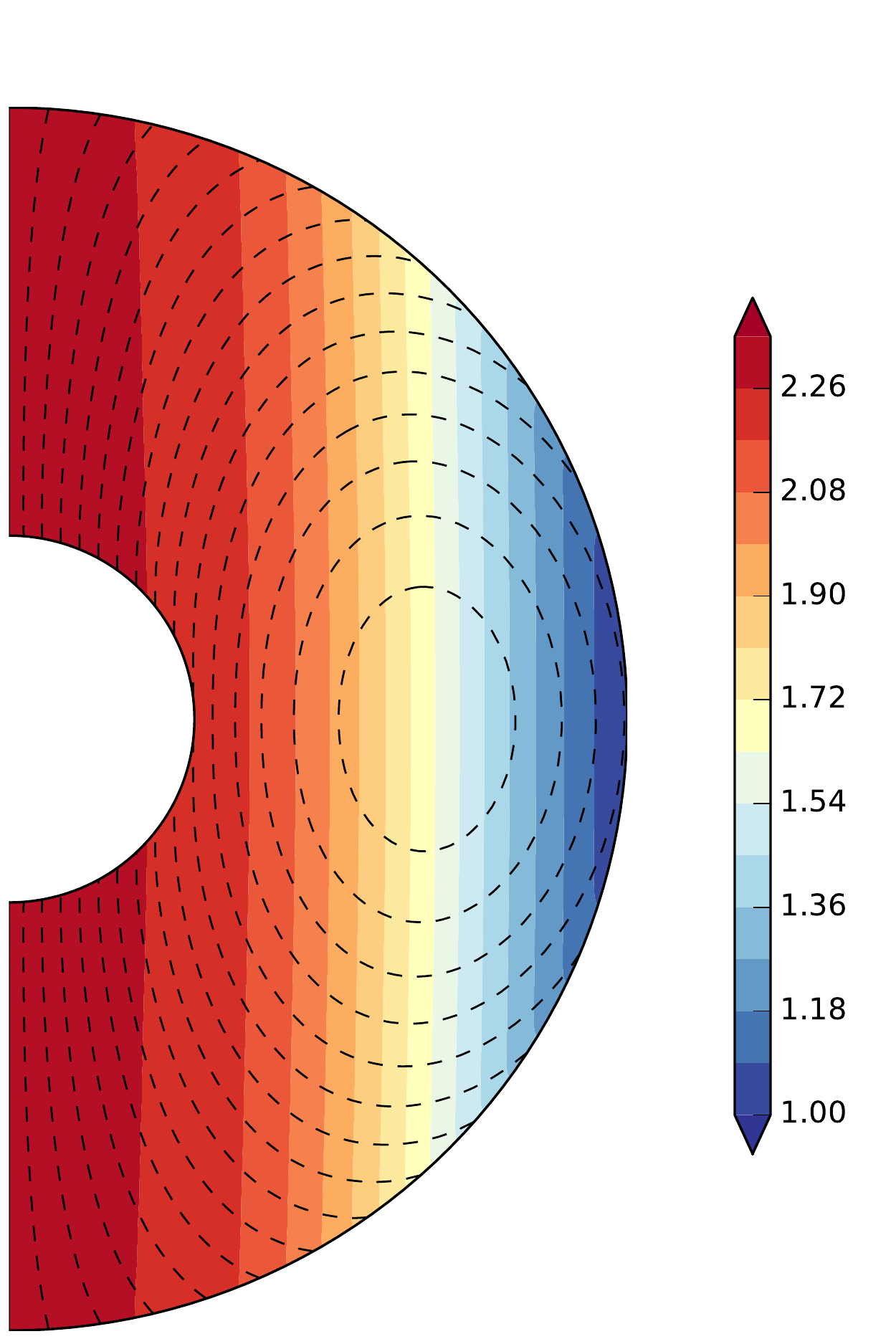}
  \includegraphics[width=0.22\textwidth]{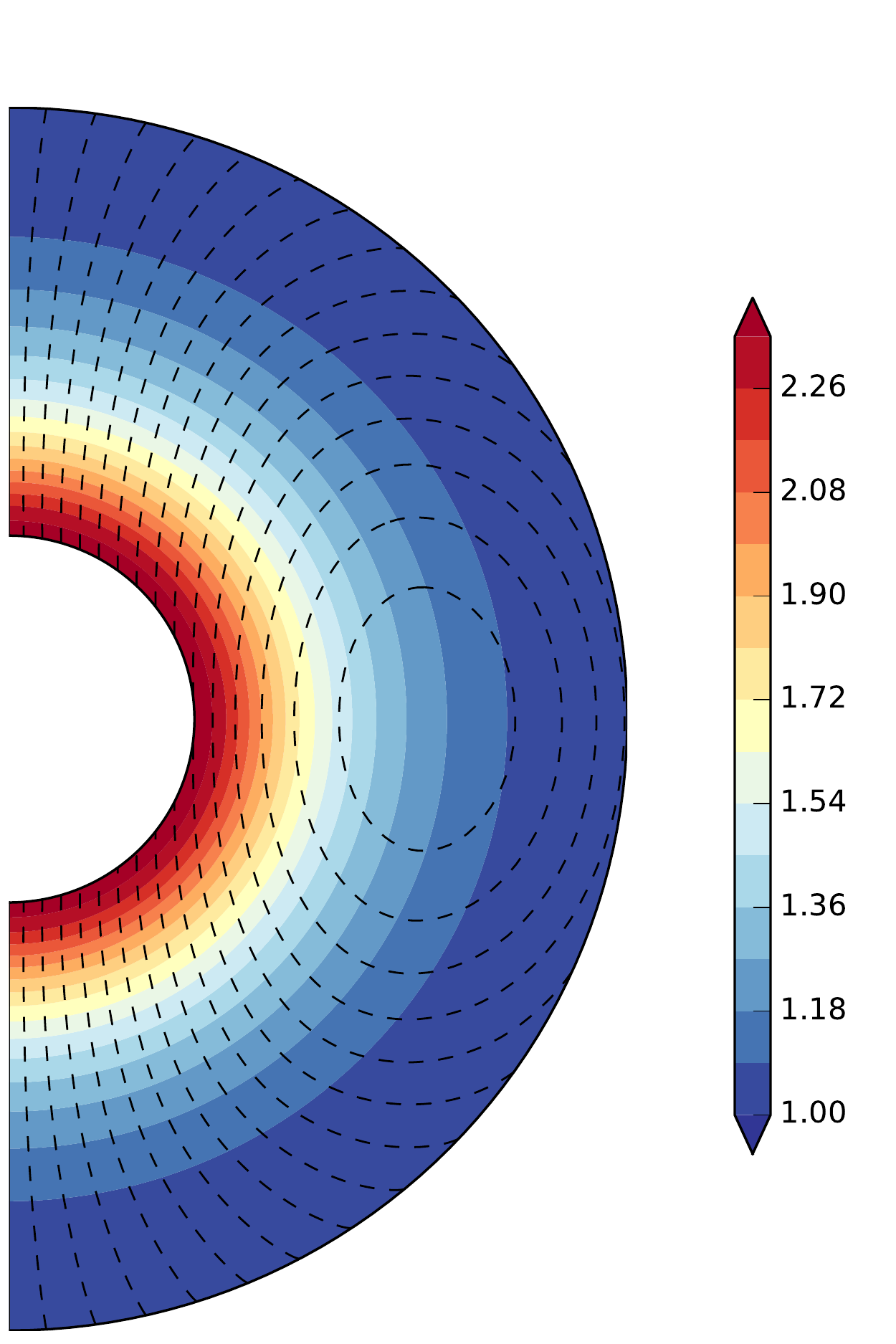}
  \caption{Initial configurations for the cylindrical rotation profile (left) and radial one (right). $\Omega$ is scaled with $\Omega_0$ and superimposed are the poloidal magnetic field lines.}
 \label{fig_init}
\end{center}
\end{figure}

Figure \ref{fig_init} illustrates these two different profiles of differential rotation and enables to envision the interaction with the initial poloidal field since it is overplotted in black dashed lines. For example, we can tell that the radial differential rotation profile is likely to produce a strong toroidal field in the bulk of our domain since this is where the poloidal magnetic field lines are almost perpendicular to the isocontours of $\Omega$. On the contrary, in the cylindrical case, the orthogonality is more confined to a region close to the top boundary at mid-latitudes and the toroidal field will thus be mostly produced in this region. Another consequence of our initial setup is that the resulting toroidal field will be antisymmetric with respect to the equator, positive in the Northern hemisphere, negative in the Southern hemisphere and vanishing at the equator.

\subsection{Numerical method}
\label{sub_code}

The numerical simulations were computed with the numerical code MagIC \citep{Wicht2002, Gastine12}. MagIC is a fully documented, publicly available code (\url{https://github.com/magic-sph/magic}) which solves the MHD equations in a spherical shell using a poloidal toroidal decomposition for the mass flux and the magnetic fields:

\begin{equation}
{\bf \rho u} = {\bf \nabla \times \nabla \times} \, (W \, {\bf e_r}) + {\bf \nabla \times} \, (Z \, {\bf e_r}),
\end{equation}

\begin{equation}
{\bf B} = {\bf \nabla \times \nabla \times} \, (C \, {\bf e_r}) + {\bf \nabla 
\times} \, (D \, {\bf e_r}),
\end{equation} where $W$ ($C$) and $Z$ ($D$) are the poloidal and toroidal 
potentials. The scalar potentials $W, Z, C, D$ and the pressure $p$ are 
further expanded in spherical harmonic functions up to degree $l_{max}$ in 
colatitude $\theta$ and longitude $\varphi$ and in Chebyshev polynomials up to 
degree $N_r$ in the radial direction. An exhaustive description of the 
complete numerical technique can be found in \citep{GG81}. We also make use of the spherical harmonic transforms contained in the SHTns library \citep{Schaeffer13} which greatly decreases the computational time for our calculations. Typical 
numerical resolutions employed in this study range from ($N_r=65$, 
$l_{max}=170$) for the more diffusive cases to  ($N_r=129$, $l_{max}=341$) for the less diffusive ones. The considered spherical shell extends 
from longitude $\varphi=0$ to $\varphi=2\pi$, from colatitude $\theta=0$ to 
$\theta=\pi$ and from radius $r=r_i=\eta/(1-\eta)$ to $r=r_o=1/(1-\eta)$ where 
$\eta=r_i/r_o$ is the aspect ratio of the shell, equal to $0.3$ in all our calculations.\\

\section{Axisymmetric evolution}
\label{sec_axi}

In this section, we consider the axisymmetric configurations resulting from the evolution of a poloidal field in a differentially rotating stably stratified spherical shell. We perform a parametric study to get an overview of the different configurations that can be reached. 
In analyzing the simulation results, we shall benefit from previous studies by \citet{gaurat2015} and \citet{jouve2015} where the same problem has been considered although with simplifying assumptions. In \citet{gaurat2015} meridional motions were neglected altogether, while they were taken into account in \citet{jouve2015} but without the effects of a stable stratification.

As shown on Fig. \ref{fig_init}, the initial differential rotation is either cylindrical or radial, but has the same maximum contrast $\Delta \Omega /\Omega$. We vary the non-dimensional numbers $Lo = t_\Omega/t_{Ap}$, $N/\Omega_0$ and $Pr = \nu/\kappa$, while the two others $Lu$ and $Pm=\nu/\eta$ are fixed to $Lu=50$ and $Pm=\nu/\eta=1$. We also restrict ourselves to initial poloidal fields that are weak enough (that is low $Lo$ value) so that the $\Omega$-effect produces magnetic configurations dominated by the toroidal component of the magnetic field. Indeed, the toroidal field will grow linearly with time until the magnetic field back-reacts on the differential rotation. The winding time-scale for the toroidal field to get to the same amplitude as the poloidal field is defined as $t_w=1/q\Omega$ where $q=r ||\nabla \Omega|| / \Omega$ is the shear parameter, of order 1 in our case, such that for our situation, $t_w \approx t_\Omega$. For the Maxwell stresses to back-react on the flow, a time-scale equal to $t_{Ap}$ is needed. If $t_\Omega << t_{Ap}$ and thus $Lo<<1$, the toroidal field will then have time to grow significantly above the poloidal field value before the differential rotation profile is affected by the magnetic field.

In the following, we first discuss the range of parameters that is relevant to stellar radiative zones and then specify the parameters of the present  simulations (Sect. \ref{subpara}). The typical magnetic configurations obtained from radial vs cylindrical initial differential rotation are described in Sect. \ref{submag}. The effects of the stable stratification on these configurations are analyzed in Sect. \ref{submeri}.

\subsection{Physical parameters in stellar radiative zones and the parameter range of our numerical simulations}
\label{subpara}

Parameters such as the Br\"unt V\"ais\"al\"a frequency and the Prandtl number come from stellar evolution models, whereas rotation rates are obtained from observations. For main sequence massive and intermediate-mass stars, the ratio $N/\Omega_0$ is typically much larger than one, except for stars rotating near break-up velocity for which $N/\Omega_0 \sim 1$,
meanwhile the Prandtl number $\Pra$ is always much smaller than $1$. For example, stellar structure models of a $3 M_{\sun}$ star indicate that, during the main sequence, $N \sim 1-2 \times 10^{-3} s^{-1}$ away from the convective core \citep{talon2008} while $\Pra \leq 4 \times 10^{-6}$ \citep{garaud2015}. For rotation periods between $1$ and $2.7$ days, 
$N/\Omega_0$ is then comprised between $\sim 10$ and $75$.
For comparison, this ratio is much higher, $N/\Omega_0 \sim 300$, in the radiative zone of the Sun. 
As we shall see below, the product $\Pra (N/\Omega_0)^2$ is also a relevant parameter and its value is typically much smaller than one in main sequence massive and intermediate-mass stars. Taking a rotation period of $2.7$ days, we find that $\Pra (N/\Omega_0)^2 \leq 0.02$ for a $3 M_{\sun}$ main sequence star. The situation is different in the solar radiative zone where $\Pra (N/\Omega_0)^2$ is rather of the order of $1$ \citep{garaud2009}.
Another important parameter is the Ekman number, that compares the rotation and viscous time scales. Its very small value, $E_k = \nu/(R^2 \Omega) \sim 10^{-14}$, is out of reach in direct numerical simulations. We nevertheless intend to consider small enough Ekman numbers to respect the order of the characteristic times involved, if not the actual time scale ratio.

The typical conditions at large length scales $d$ in radiative zones of intermediate-mass and massive stars thus read $E_k \ll \Pra < \Pra (N/\Omega_0)^2 \ll 1$. It corresponds to the following time scale order : $t_\nu \gg t_{es} > t_\kappa \gg t_\Omega > t_N $
where $t_\nu = d^2/\nu$, $t_{es} = (d^2/\kappa) (N/\Omega_0)^2$, $t_\kappa = d^2/\kappa$, $t_\Omega = 1/\Omega_0$ and $t_N = 1/N$.
As shown in table 1, the simulations performed respect that order. 

The timescale associated with the initial poloidal field is  $t_{Ap}=d\sqrt{4\pi\rho}/B_0$, the poloidal Alfv\'en time. We have no direct constrain on the field intensity within stars, but we can use spectropolarimetric observations to get surface values. In particular, the lower limit of the dipolar field of Ap/Bp magnetic stars is close to $300$ Gauss and, for a rotation period of 5 days, this field corresponds to a Lorentz number $Lo = t_{\Omega}/t_{Ap}$ close to $1$ \citep{auriere2007}. This number is expected to decrease strongly towards the stellar interior as the variation of the Alfv\'en speed $v_{Ap}=B_p/\sqrt{ 4 \pi \rho}$ is dominated by the density increase. For example, at a radius $r=R_\star/3$, the Lorentz number would be $2.7 \times 10^{-3}$ assuming a density ratio of $10^8$ and a dipolar-like radial increase $B_p \propto 1/r^3$. Even lower Lorentz numbers are expected in Vega-like magnetic stars with $1$ Gauss surface field and $1$-day rotation period. In the radiative interior of intermediate-mass and massive stars, the magnetic field could also result from a convective core dynamo. Numerical simulations of A and B-type star convective cores \citep{brun2005, augustson2016} indicate that, in the low Rossby number regime characterizing
these convective motions, the generated fields have low Lorentz numbers. Indeed, in \citet{brun2005} simulation
of a 7-days rotating A star, a ratio $B_{rms}^2/(4 \pi \rho r_c^2 \Omega^2) \sim 2.10^{-4}$ is found at
mid-depth of the convective core.
Finally, the Lundquist number measures the ratio of the magnetic diffusion time scale to the poloidal Alfv\'en time and is expected to be very large, much larger than the value attainable in numerical simulations. \\

Table \ref{table_cases} lists the parameters used in our simulations. The cylindrical and radial cases are respectively labelled C and R. The effect of varying the profile of differential rotation is first studied, keeping the other parameters fixed (cases C1 and R1). Both for the cylindrical and radial cases, we vary $Pr$ (cases 2, 3 and 4), then $Pr$ and $N/\Omega_0$ while keeping $Pr N^2/\Omega_0^2$ fixed (cases 2, 5, 6 and R9 for the radial case). We also decrease the value of $Pr N^2/\Omega_0^2$ (cases 7 and 8) and finally consider lower $Lo$ (case C9 and R1, to be compared with C2 and R2). In all cases the Lundquist number is maintained equal to 50 and the magnetic Prandtl number to 1. 

\begin{table*}
 \caption{Parameters of the different simulations}
\label{table_cases} 
\centering 
\begin{tabular}{c c c c c c c c} 
\hline\hline 
Case & Rotation &$N/\Omega_0$ & Pr & $E_k$ &  Lo & $Pr N^2/\Omega_0^2$\\ 
\hline 
C1 & Cyl & 5 & $10^{-2}$ & $5\times 10^{-5}$ &  $2.5\times10^{-3}$ & $2.5\times10^{-1}$\\
C2 & Cyl & 5 & $10^{-2}$ & $10^{-4}$ &  $5\times10^{-3}$ & $2.5\times10^{-1}$\\ 
C3 & Cyl & 5 & $10^{-1}$ & $10^{-4}$ &  $5\times10^{-3}$& $2.5$\\
C4 & Cyl & 5 & $1$ & $10^{-4}$ &  $5\times10^{-3}$ & $25$\\
C5 & Cyl & 15.8 & $10^{-3}$ & $10^{-4}$  & $5\times10^{-3}$ & $2.5\times10^{-1}$\\
C6 & Cyl & 50 &  $10^{-4}$ & $10^{-4}$  & $5\times10^{-3}$ & $2.5\times10^{-1}$\\
C7 & Cyl & 5 &  $1.6\times10^{-3}$ & $10^{-4}$  & $5\times10^{-3}$ & $4\times10^{-2}$\\
C8 & Cyl & 50 &  $1.6\times10^{-5}$ & $10^{-4}$  & $5\times10^{-3}$ & $4\times10^{-2}$\\
C9 & Cyl & 5 &  $10^{-2}$ & $2\times10^{-4}$ &  $10^{-2}$ & $2.5\times10^{-1}$\\

R1 & Rad & 5 & $10^{-2}$ & $5\times 10^{-5}$ &  $2.5\times10^{-3}$ & $2.5\times10^{-1}$\\
R2 & Rad & 5 &  $10^{-2}$ & $2\times 10^{-5}$  & $10^{-3}$& $2.5\times10^{-1}$\\
R3 & Rad & 5 & $10^{-1}$ & $2\times 10^{-5}$  & $10^{-3}$& $2.5$\\
R4 & Rad & 5 & $1$ & $2\times 10^{-5}$ & $10^{-3}$ & $25$\\
R5 & Rad & 15.8 & $10^{-3}$ & $2\times 10^{-5}$  & $10^{-3}$ & $2.5\times10^{-1}$\\
R6 & Rad & 50 &  $10^{-4}$ & $2\times 10^{-5}$  & $10^{-3}$ & $2.5\times10^{-1}$\\
R7 & Rad & 5 &  $1.6\times10^{-3}$ & $2\times 10^{-5}$  & $10^{-3}$ & $4\times10^{-2}$\\
R8 & Rad & 50 &  $1.6\times10^{-5}$ & $2\times 10^{-5}$  & $10^{-3}$ & $4\times10^{-2}$\\
R9 & Rad & 2 & $6.25 \times10^{-2}$ & $2\times 10^{-5}$ & $10^{-3}$ & $2.5\times10^{-1}$\\

\hline 
\end{tabular}

\end{table*}

\subsection{Influence of a radial vs cylindrical initial differential rotation} 
\label{submag}

\begin{figure}[!h]
\begin{center}
  \includegraphics[width=0.45\textwidth]{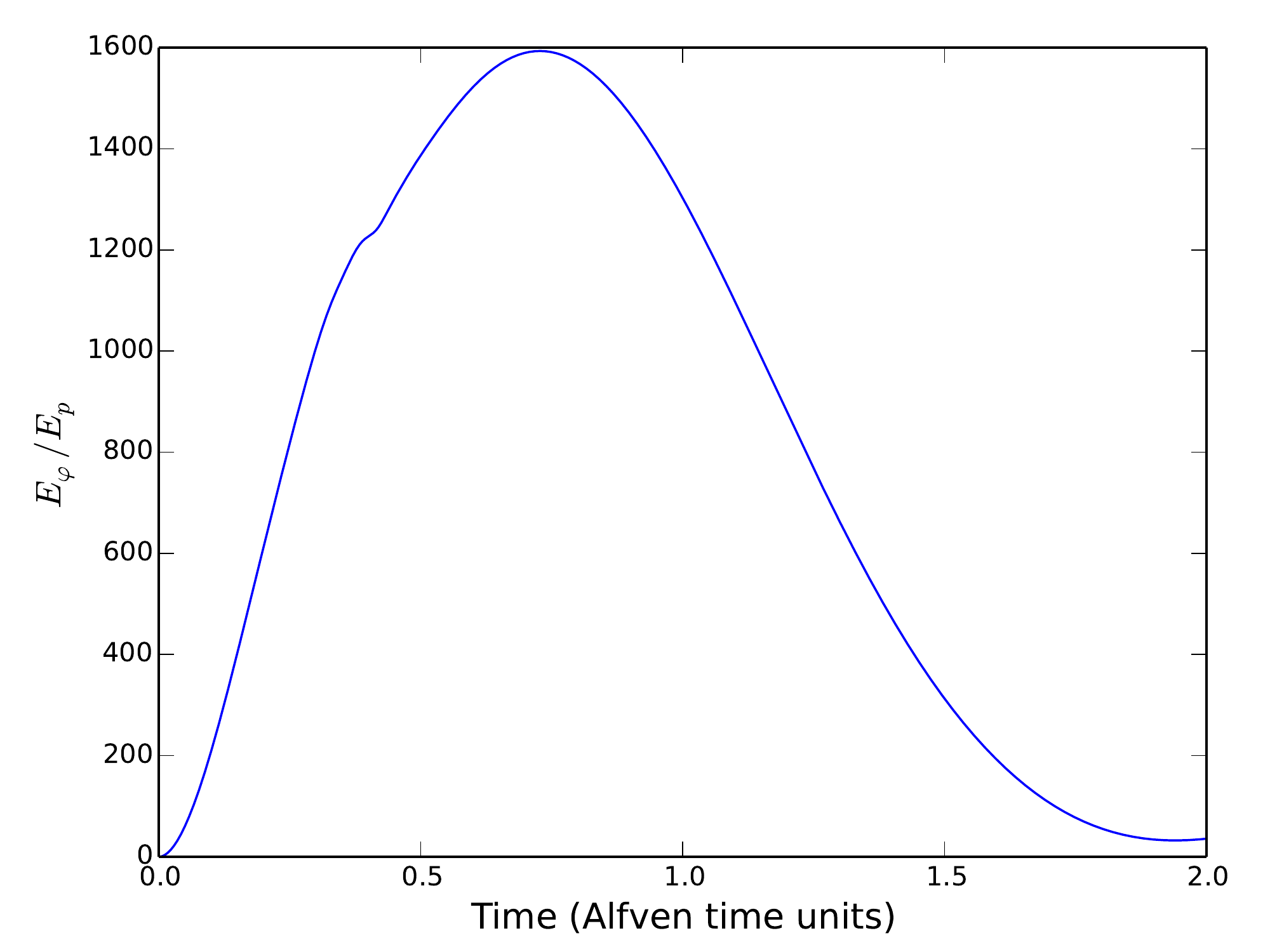}
  \includegraphics[width=0.45\textwidth]{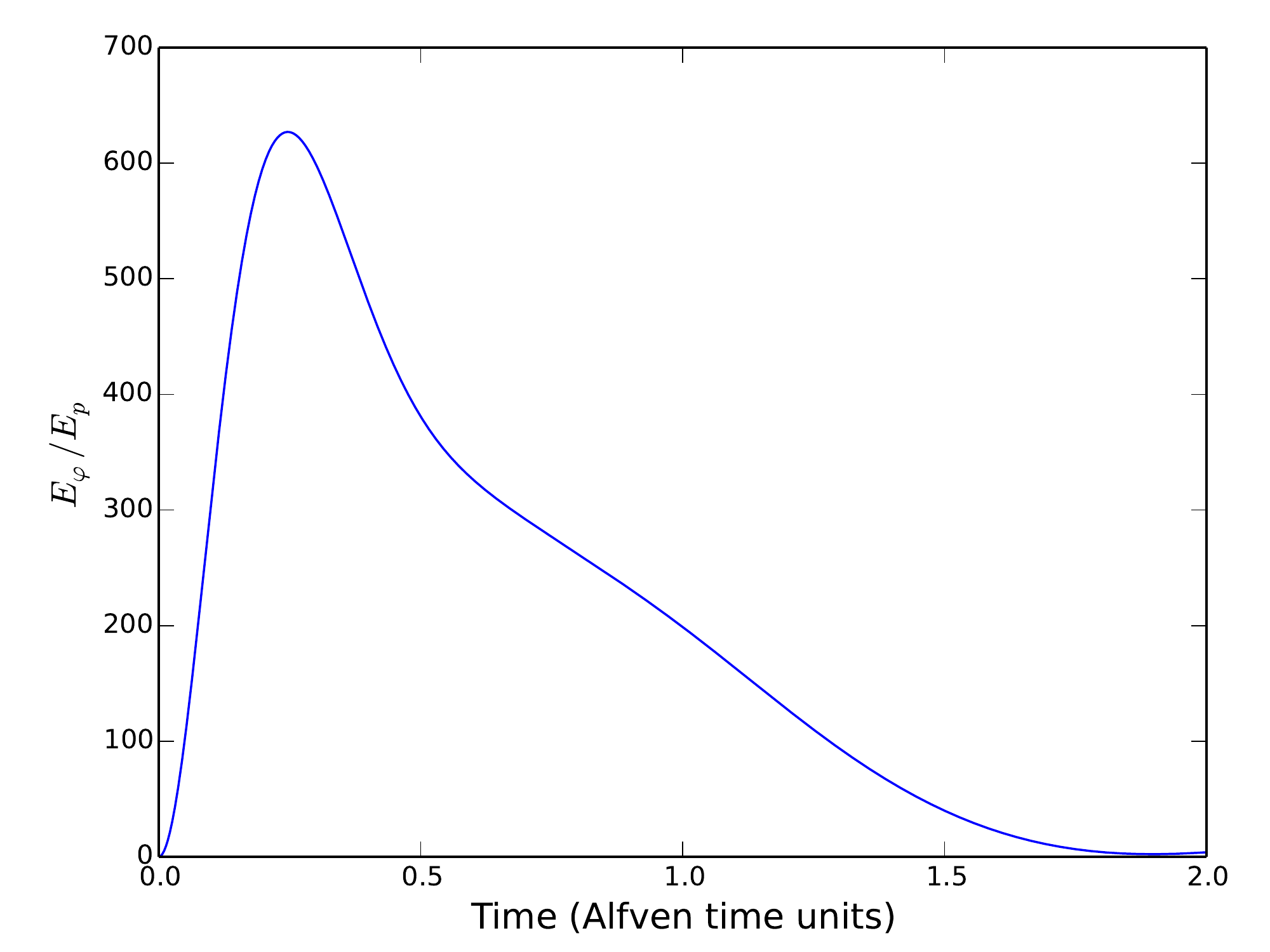}
  \caption{Evolution of the ratio between the toroidal and poloidal magnetic energies for cases C1 (left) and R1 (right). $E_\varphi$ seems much stronger in the cylindrical case but it's because $B_\varphi$ is more extended and located closer to the surface where $B_p$ is much smaller. If we look locally, we also find that $B_\varphi/B_p$ is stronger in the cylindrical case but mostly because of the location of $B_\varphi$ (where $B_p$ is small).}
 \label{fig_magener}
\end{center}
\end{figure}

Fig. \ref{fig_magener} displays the evolution of the ratio of the total (integrated over the whole spherical shell) dimensioned azimuthal magnetic energy $E_{m_\varphi}$ to the total dimensioned poloidal magnetic energy $E_{m_p}$ in two axisymmetric simulations in which the cylindrical differential rotation defined by Eq. (\ref{cylindrical}) (left panel) and the radial differential rotation defined by Eq. (\ref{radial}) (right panel) are used.
The other parameters of these simulations, denoted C1 and R1 in table \ref{table_cases}, are identical.
In both simulations, the ratio $E_{m_\varphi}/E_{m_p}$ initially increases quadratically before reaching a maximal value in a fraction of $t_{Ap}$, namely around $0.70 t_{Ap}$ for the cylindrical case and $0.35 t_{Ap}$ in the radial case. 
The maximal value of the quantity $E_{m_\varphi}/E_{m_p}$ is equal to $1600$ in the cylindrical case and to $620$ in the radial case, showing that
the magnetic configurations are dominated by the toroidal component.

This evolution of the magnetic field, namely a near-linear growth of $B_\varphi$ followed by a maximum reached at $t \sim t_{Ap}$, is the same as observed in simulations where only the coupled equations for the azimuthal magnetic and velocity fields were solved (for example \cite{charbonneau92} or \cite{gaurat2015} for exactly identical initial conditions). This means that the avdection by meridional flows and the diffusive decay of poloidal fields only have a weak effect in this case. The physical explanation is rather simple and well-known in this context of purely azimuthal dynamics: the toroidal magnetic field $B_\varphi$, initially set to zero, increases through the winding-up of the initial poloidal magnetic field by the initial differential rotation. This growth is linear as long as the differential rotation and the poloidal field are not modified by the back-reaction of the Lorentz force on the flow. When the Maxwell stress associated with the magnetic field becomes sufficiently strong to change the differential rotation, the $\Omega$-effect is modified accordingly and the growth of the toroidal field stops. This results in a maximum in the evolution of $B_\varphi$ and of the ratio $B_\varphi/B_p$, that occurs after a time of the order of the poloidal Alfv\'en time $t_{Ap}$. Locally, the maximum ratio should then be of the order of $(B_\varphi/B_p)_{max} \sim \Delta \Omega t_{Ap}$ where $\Delta \Omega$ is the initial differential rotation. The exact value depends on the shear along the poloidal field line during the linear growth phase (see a more detailed model in \citet{gaurat2015}) and this explains why the maxima reached for the cylindrical and radial differential rotation are different. 
On longer timescales after the maximum (not shown in Fig. \ref{fig_magener}), damped oscillations of the global magnetic energy are observed and are due to torsional Alfv\'en waves whose damping through the so-called phase-mixing mechanism \citep[see for example][]{ionson78,spruit99} finally leads to uniform rotation as nothing enforces differential rotation in our set-up. 
Note that we are not interested in the final uniformly rotating state of the axisymmetric evolution as we focus on MHD instabilities that are likely to be triggered before it is reached.
The simplified problem considered in \citet{gaurat2015} is still relevant to describe the global evolution of the present simulations even though we included meridional motions and stable stratification. However the detailed distributions of the differential rotation and the magnetic field, that are crucial for the occurrence and the nature of possible MHD instabilities, will depend on these processes.

\begin{figure*}[!h]
\begin{center}
  \includegraphics[width=0.3\textwidth]{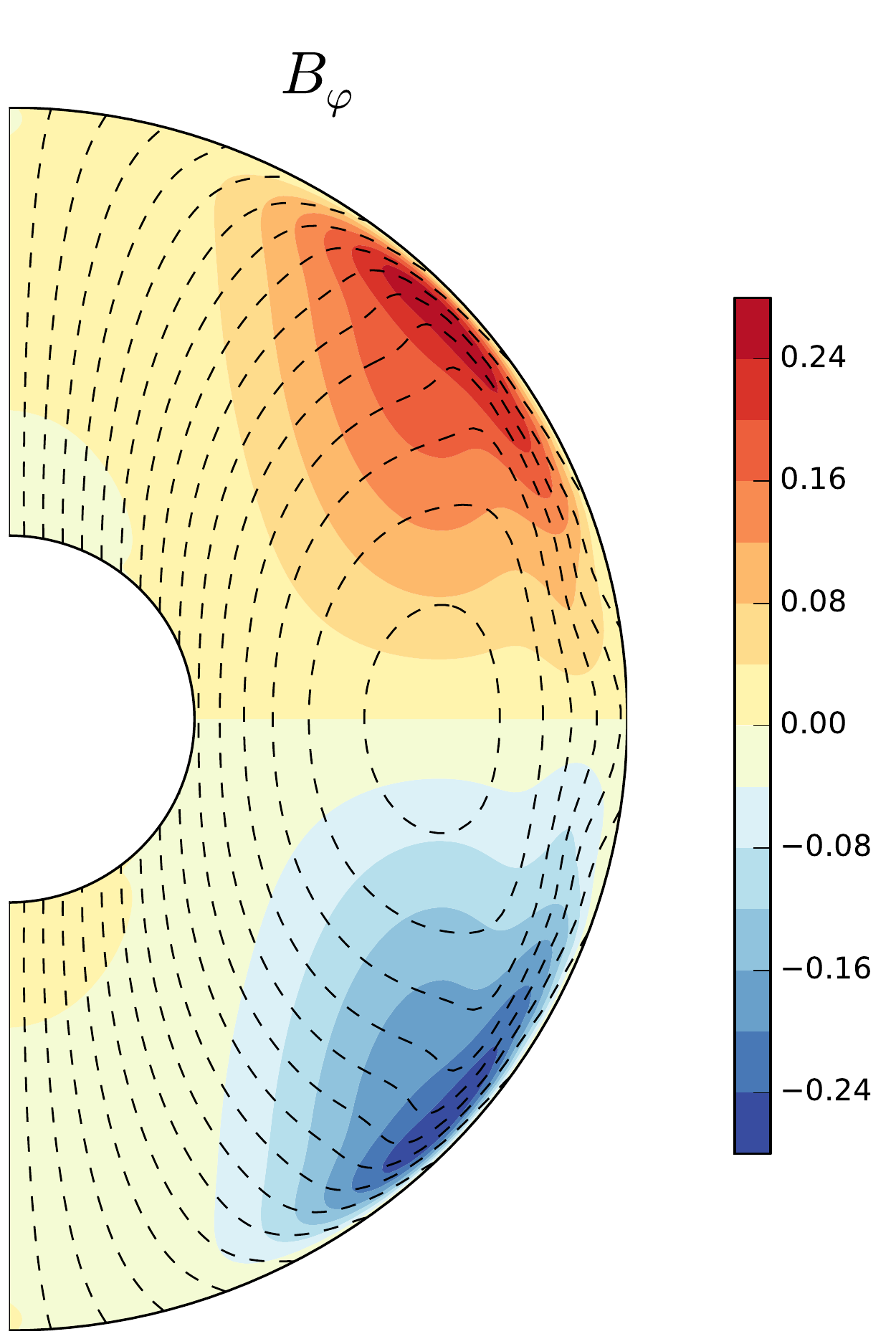}
  \includegraphics[width=0.3\textwidth]{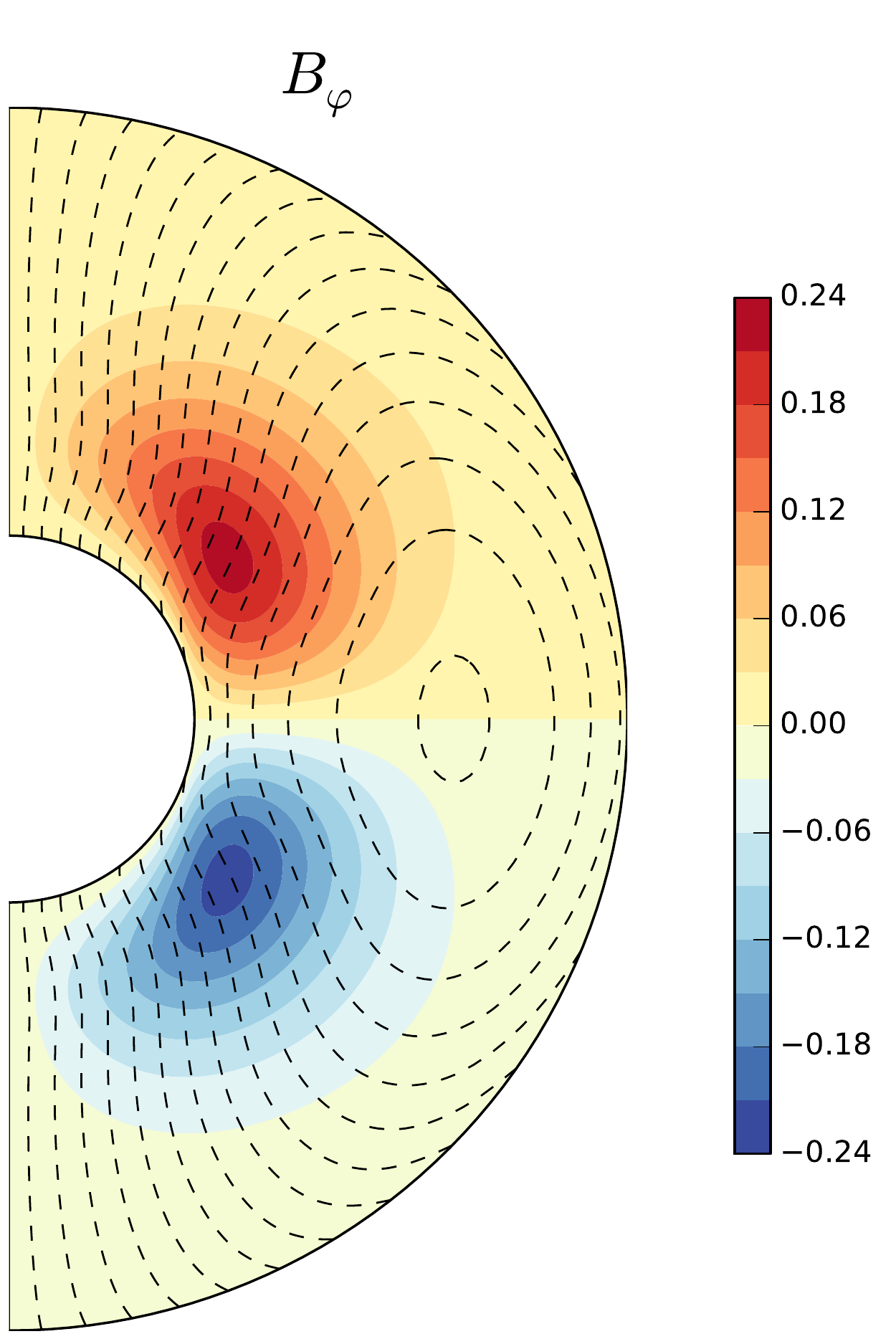}
  \caption{Contours of toroidal magnetic field (colors) for cases C1 (left) and R1 (right). Time is taken in the middle of the linear growth of $B_\varphi$, i.e. at $0.125 t_{ap}$ for case R1 and $0.25 t_{ap}$ for case C1. Superimposed are the poloidal magnetic field lines. Units are in $Lo_\varphi=\omega_{A\varphi}/\Omega$. Here, the strength of $B_\varphi$ compared to the global rotation is similar, we always have $Lo_\varphi < 1$. }
 \label{fig_bphi}
\end{center}
\end{figure*}

Figure \ref{fig_bphi} shows the distribution of the toroidal field (together with the contours (in black) of the poloidal field) during the linear growth of magnetic field, i.e. at $0.25 t_{ap}$ for the cylindrical case (left panel) and at $0.125 t_{ap}$ for the radial case (right panel). The value of $B_\varphi$ is normalized by $d\Omega_0\sqrt{4\pi\rho}$. We clearly see that the distribution of $B_\varphi$ is quite different in both cases, the field being mostly confined close to the upper boundary in the cylindrical case and to the bottom boundary in the radial case. This is due to the $\Omega$-effect which acts differently because the angle between the isocontours of $\Omega$ and the poloidal field lines is maximal at very different locations in the two cases.

Beside these spatial distributions, the ratio of the azimuthal Alfv\'en frequency, $\omega_{A_\varphi}=B_\varphi/r\sqrt{4\pi\rho}$ to the rotation rate $\Omega$, denoted $Lo_\varphi=\omega_{A_\varphi}/\Omega$, is another relevant quantity as it allows to distinguish between the two types of instabilities likely to be triggered : the Tayler instability (TI) or the azimuthal-magnetorotational instability (A-MRI)\citep{jouve2015}. For sufficiently large values of $Lo_\varphi$ (i.e. when the toroidal field dominates over rotation), the TI should be favoured because the A-MRI is suppressed when the magnetic field becomes too strong. On the contrary, the A-MRI is favoured for small values of $Lo_\varphi$ (when the rotation is fast compared to the toroidal Alfv\'en time) since the growth rate of the TI is strongly reduced by a fast rotation \citep{PT85}.

With the chosen normalization, the azimuthal field presented Figs. \ref{fig_bphi} thus provides the values of the Lorentz number $Lo_\varphi$, showing that it is everywhere lower than one.
More generally, in our simulations, the maximum value of $Lo_\varphi$ reached in the whole computational domain is always of the order of $0.2-0.3$. Indeed, arbitrarily strong values of $Lo_\varphi$ cannot be reached because only a certain amount of toroidal field can be built before the Lorentz force back-reacts on the differential rotation. As a consequence, we can already expect that the A-MRI will be the favored instability likely to develop in our simulations. This point will be addressed in Sect. \ref{sec_stab}.

\subsection{Influence of the stable stratification}

\label{submeri}

In the previous subsection, the initial growth of the toroidal field and the subsequent regime of damped oscillations have been explained by the winding-up of poloidal field induced by the initial differential rotation followed by the back-reaction of the magnetic field through Alfv\'en waves. While stable stratification does not seem to play a role in this process, we know from simulations performed in uniform density background \citep{jouve2015} that its presence is in fact crucial when the initial differential rotation is radial. Indeed, in the absence of stable stratification, radial differential rotation drives a fast meridional circulation (of time scale $1/\Delta \Omega$) that strongly redistributes the initial poloidal field and angular momentum,  before any coherent and therefore efficient winding-up can happen. As a consequence, the build-up of a magnetic configuration dominated by a toroidal component does not happen. 
With stable stratification, particularly in a $N/\Omega_0 \gg 1$ regime, such a fast adiabiatic meridional circulation is efficiently suppressed. Instead, on a larger thermal diffusion time scale, the latitudinal temperature perturbations generated by the differential rotation (through the so-called thermal wind equation) drive an Eddington-Sweet type circulation of time scale $t_{es}=(d^2/\kappa) (N/\Omega_0)^2$ (see for example \citet{Spiegel1992}) \footnote{Our boundary conditions do not enforce Ekman boundary layers, thus a circulation driven by Ekman pumping is not expected in our simulations}. As long as $t_{Ap}$ is of the same order or smaller than $t_{es}$, the Eddington-Sweet circulation should not prevent the winding-up process to occur. This is indeed what occurs in our numerical simulations as the ratio $t_{Ap}/t_{es} = Pm/(Lu Pr N^2/\Omega_0^2)$ varies between $8 \times 10^{-4}$ and $1$ (according to table \ref{table_cases}). In this regime, the detailed distribution of the angular momentum and of the azimuthal field can nevertheless be affected by the meridional circulation as we shall see below. 

\begin{figure*}[!h]
\begin{center}
  \includegraphics[width=0.32\textwidth]{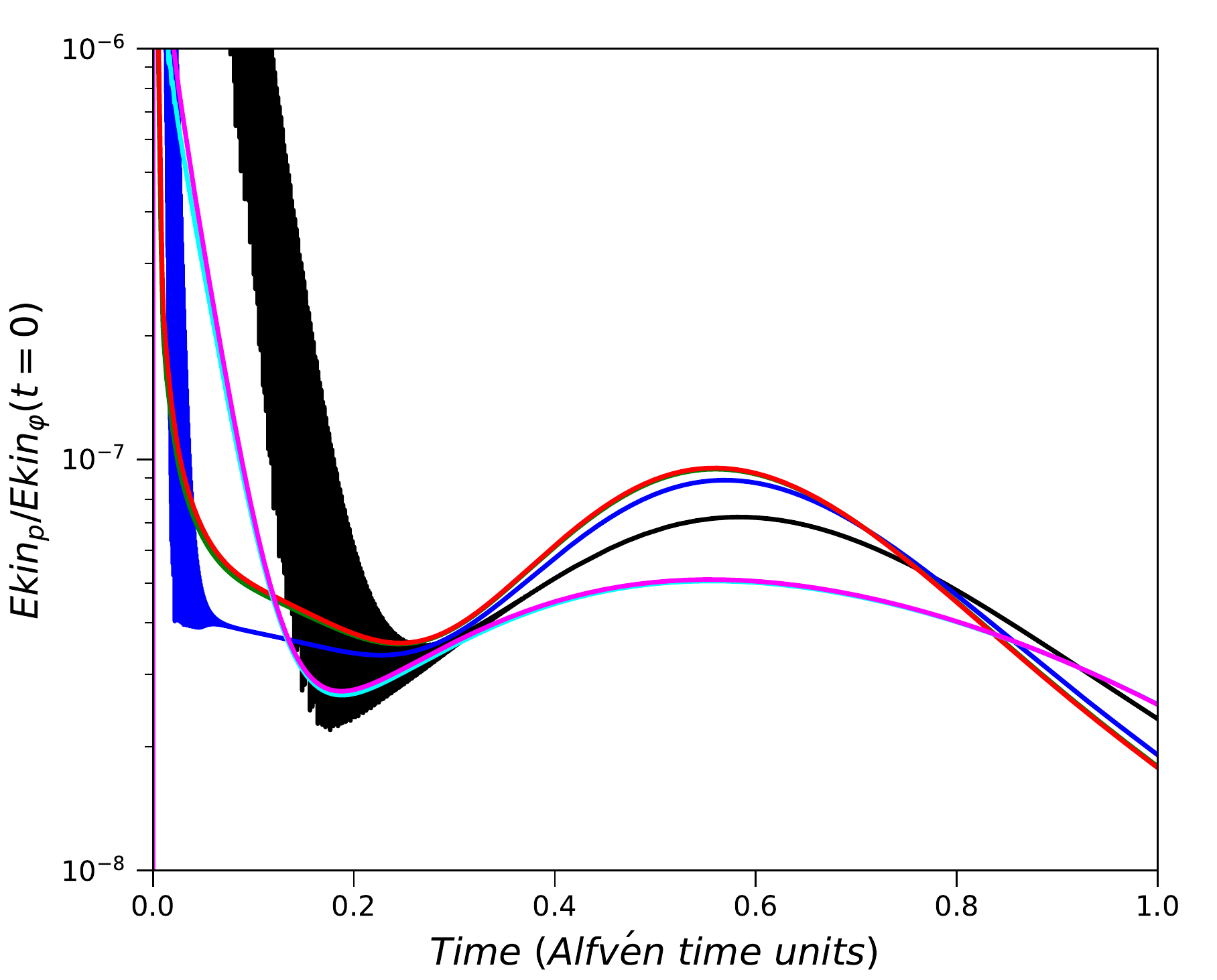}
 \includegraphics[width=0.32\textwidth]{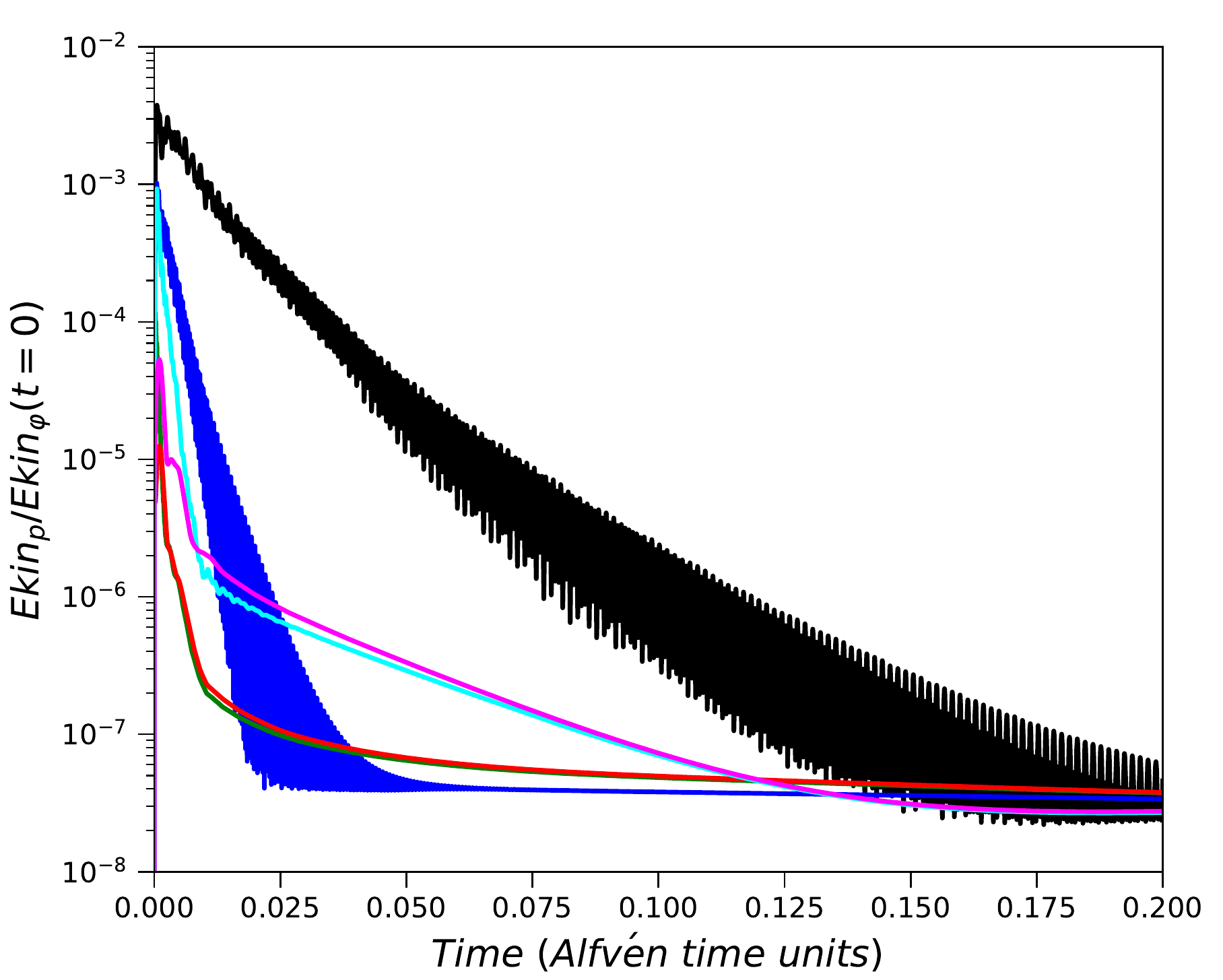}
   \includegraphics[width=0.32\textwidth]{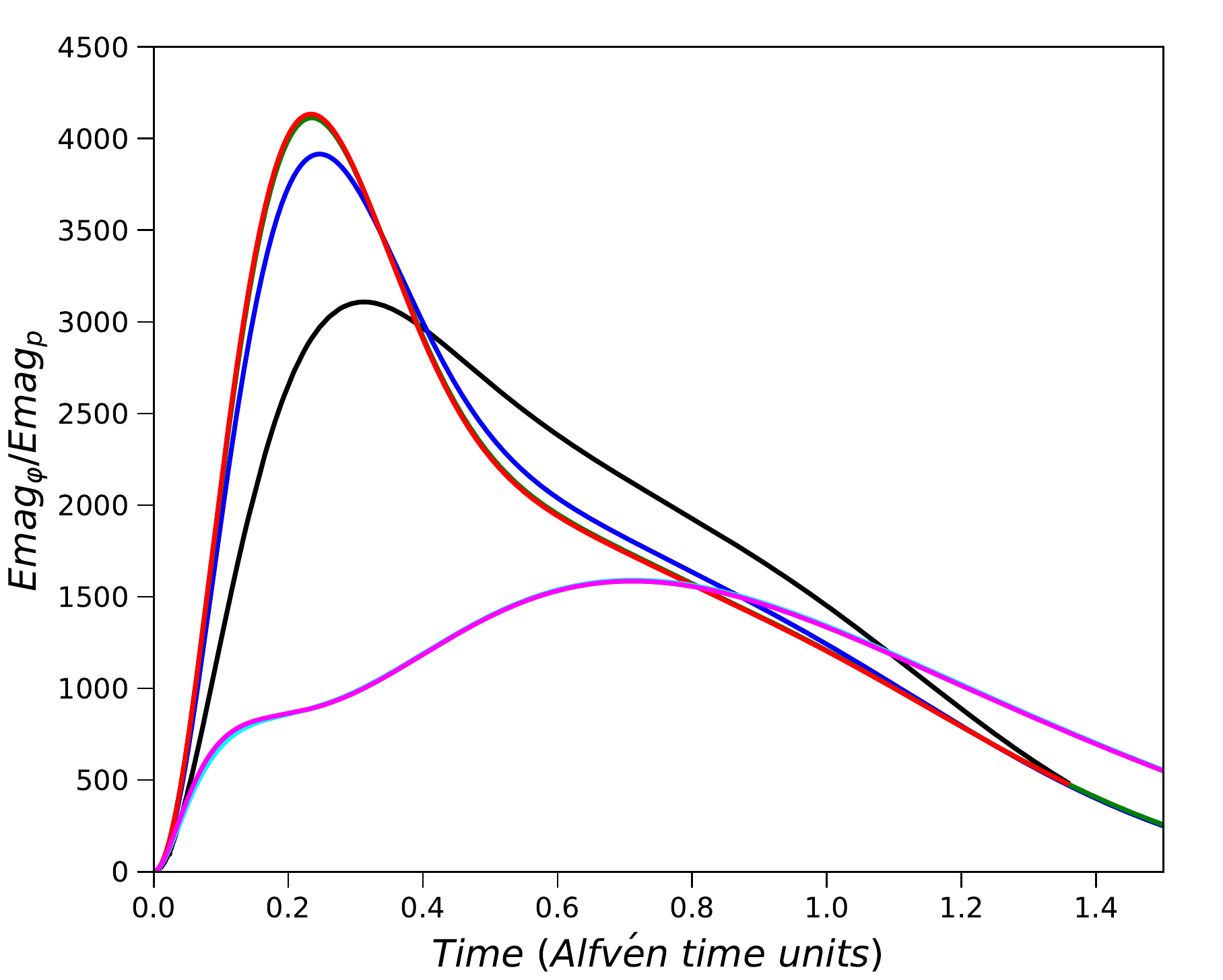}
  \caption{Temporal evolution of the kinetic energy (on a long timescale, left and zoomed in, mid-panel) and magnetic energy ratio (right) for 5 different cases: 3 cases with $PrN^2/\Omega_0^2=0.25$ (blue: R2, green: R5, red: R6 and black: R9) and 2 cases with $PrN^2/\Omega_0^2=0.04$ (cyan: R7 and magenta: R8). For the magnetic energy plot, the cyan and magenta curves are almost superimposed, as well as the red and green curves. The long term evolution is similar for the cases with the same $PrN^2/\Omega_0^2$, as long as $Pr$ is small enough, but different for different values of $PrN^2/\Omega_0^2$. In particular, the amount of toroidal field produced is much less in the case where $PrN^2/\Omega_0^2=0.04$.}
 \label{fig_enertot}
\end{center}
\end{figure*}

The effect of the stable stratification on the axisymmetric evolution has been analyzed by varying the parameters $Pr$ and $N/\Omega_0$. A striking feature is that, after a transient phase, most of our axisymmetric solutions are controlled by the product $P_rN^2/\Omega_0^2$, rather than by the two parameters $Pr$ and $N/\Omega_0$ independently. Indeed, as seen in Fig.\ref{fig_enertot}, the evolution of the kinetic and magnetic energies is very similar after about $t=0.2t_{Ap}$ for three simulations having the same $P_r N^2/\Omega_0^2 = 0.25$ but different $Pr$ and $N/\Omega_0$, namely $Pr=10^{-2}, N/\Omega_0=5$ (blue),  $Pr=10^{-3}, N/\Omega_0=15.8$ (green) and $Pr=10^{-4}, N/\Omega_0=50$ (red). The two cases $Pr=10^{-3}$ and $Pr=10^{-4}$ are even indistinguishable on this plot after about $t=10^{-3} t_{Ap}$. Figure \ref{fig_enertot} also displays the energy evolution of two simulations at $Pr N^2/\Omega_0^2=0.04$, the $Pr=10^{-3}$ (cyan) and $Pr=10^{-4}$ (magenta) cases, again showing a very similar behaviour. All the runs of Fig.\ref{fig_enertot} (R2, R5, R6, R7, R8) have been performed with the initial radial differential rotation but using the initial cylindrical profile (runs C2, C5 and C6) leads to the same conclusion regarding the dependence on $P_rN^2/\Omega_0^2$.

On a short timescale however, flows having the same $Pr N^2/\Omega_0^2$ can evolve differently. This is illustrated in the mid-panel of Fig. \ref{fig_enertot} where a zoom is made between $t=0$ and $t=0.2t_{Ap}$. Indeed, the initial conditions  generate gravity waves that propagate and oscillate until they are damped by thermal diffusion. The oscillations are clearly visible in the $Pr=10^{-2}$ case (run R2). The thermal damping is so efficient in the case $Pr=10^{-4}$ (for which thermal diffusivity is $100$ times larger than for the case $Pr=10^{-2}$) that the oscillations of the kinetic energy are not even visible. The invariance of the solution with $PrN^2/\Omega_0^2$ is thus relevant only after this initial transient phase. In addition, we observe that run R9 deviates also at late time from the other runs with $P_r N^2/\Omega_0^2 = 0.25$.

\begin{figure*}[!h]
\begin{center}
 \includegraphics[width=0.22\textwidth]{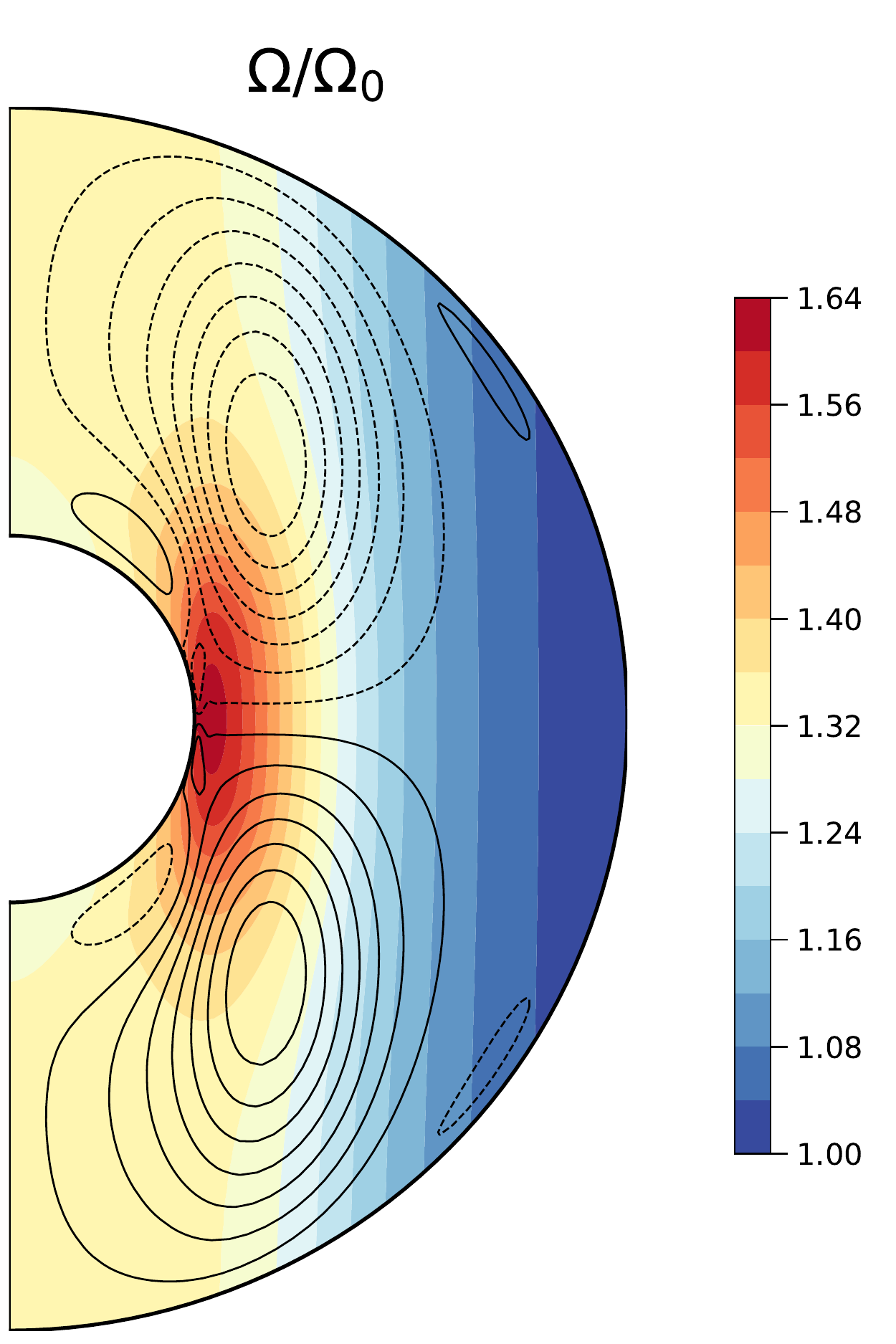}
 \includegraphics[width=0.22\textwidth]{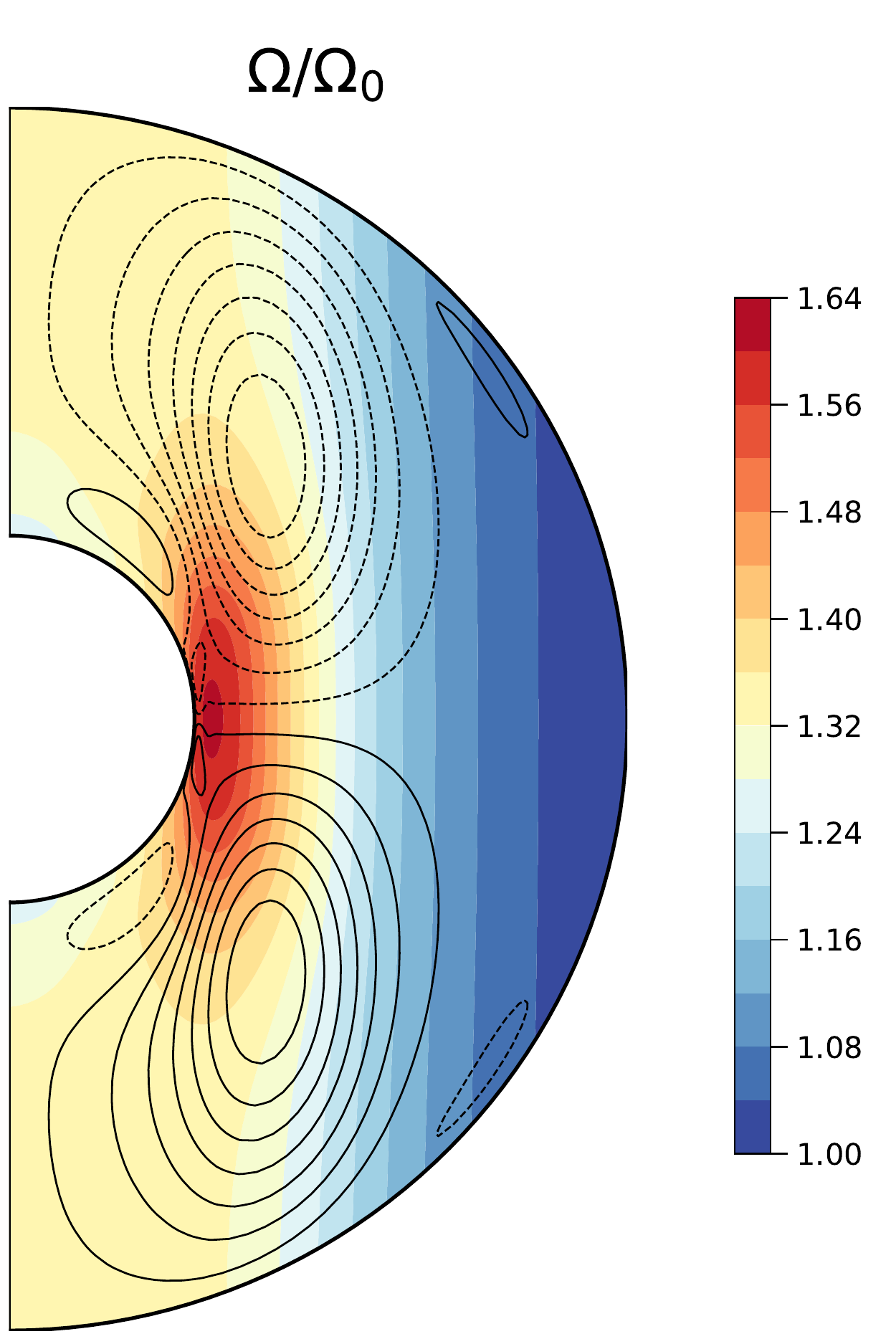}
  \includegraphics[width=0.22\textwidth]{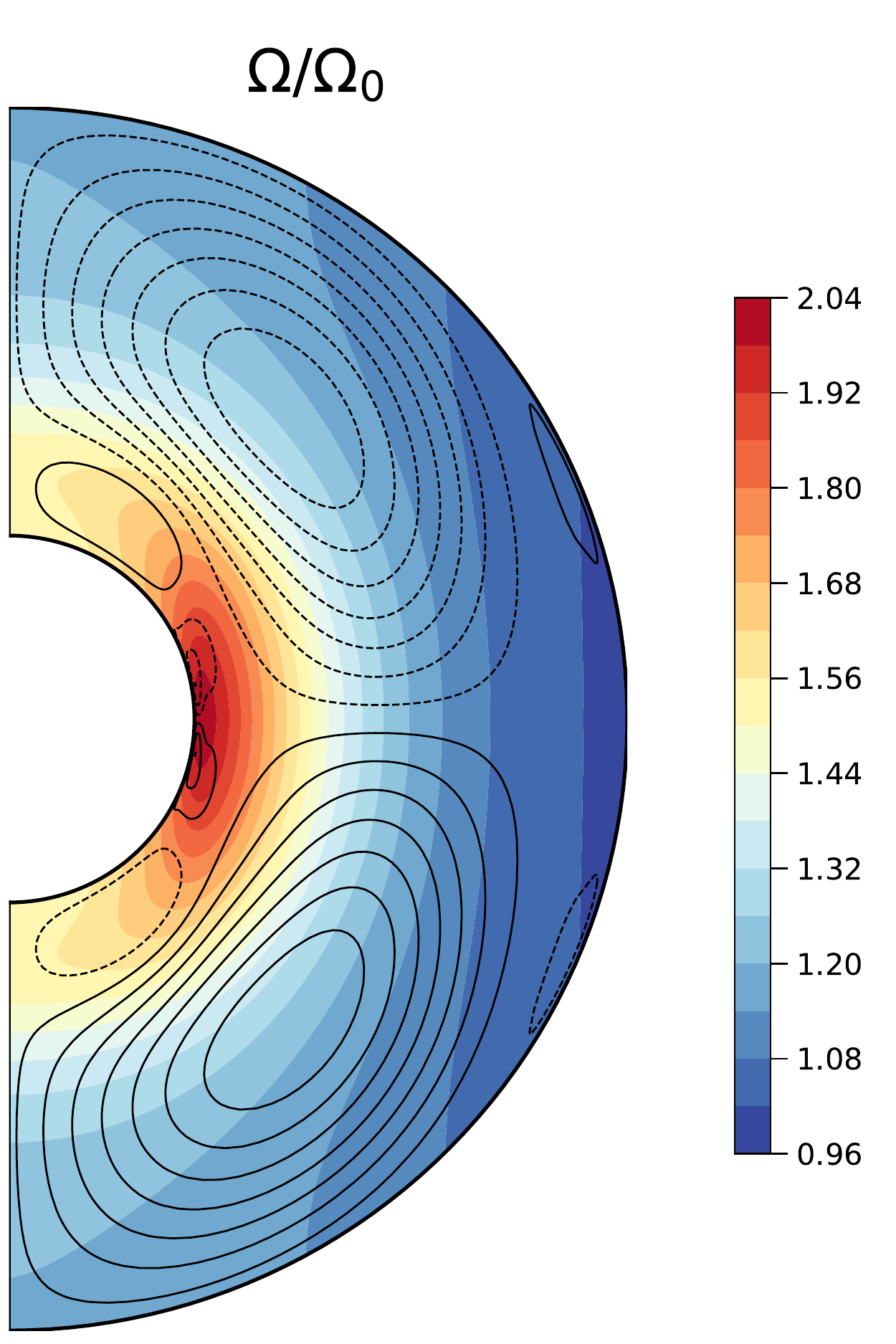}
  \includegraphics[width=0.22\textwidth]{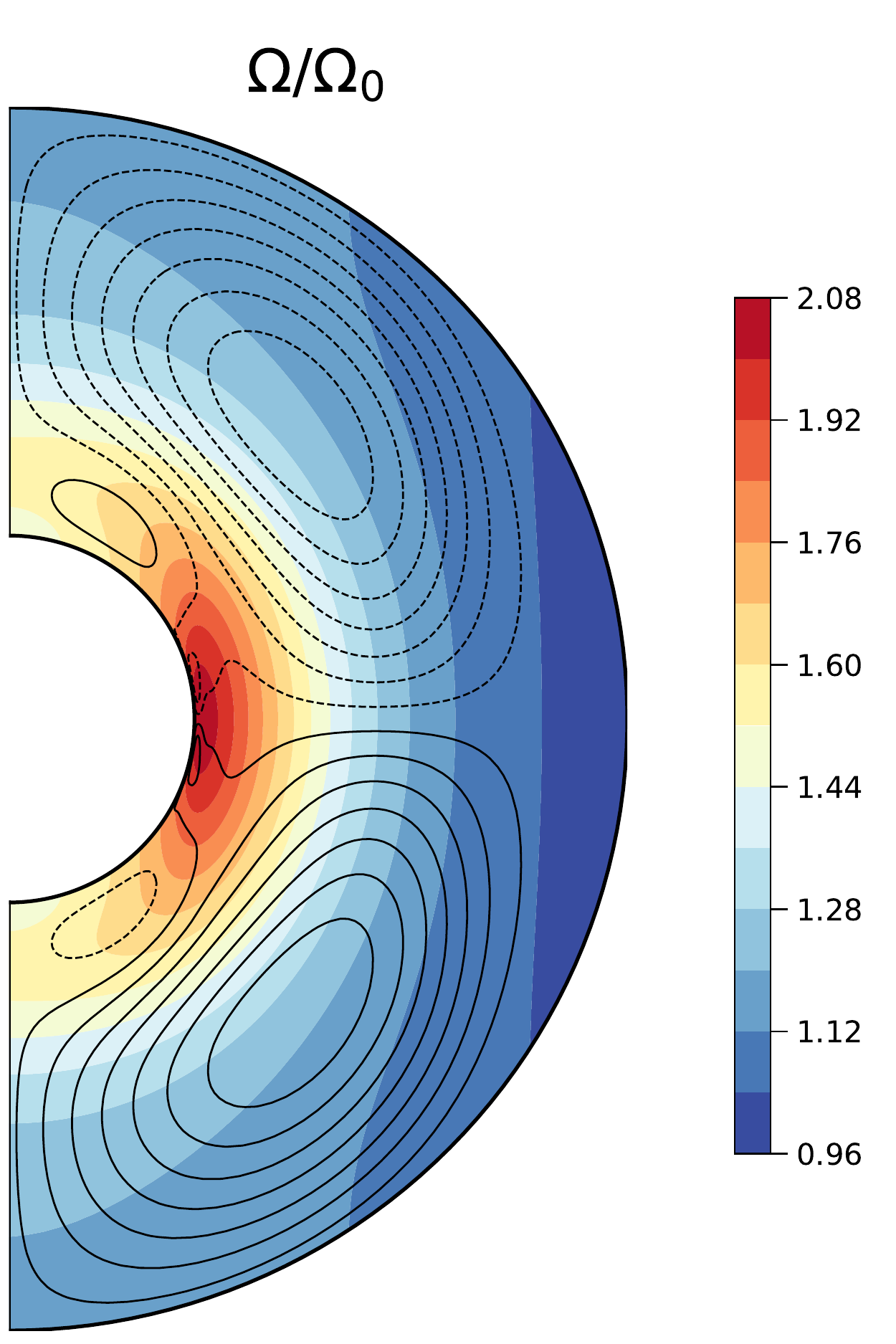}
   \includegraphics[width=0.22\textwidth]{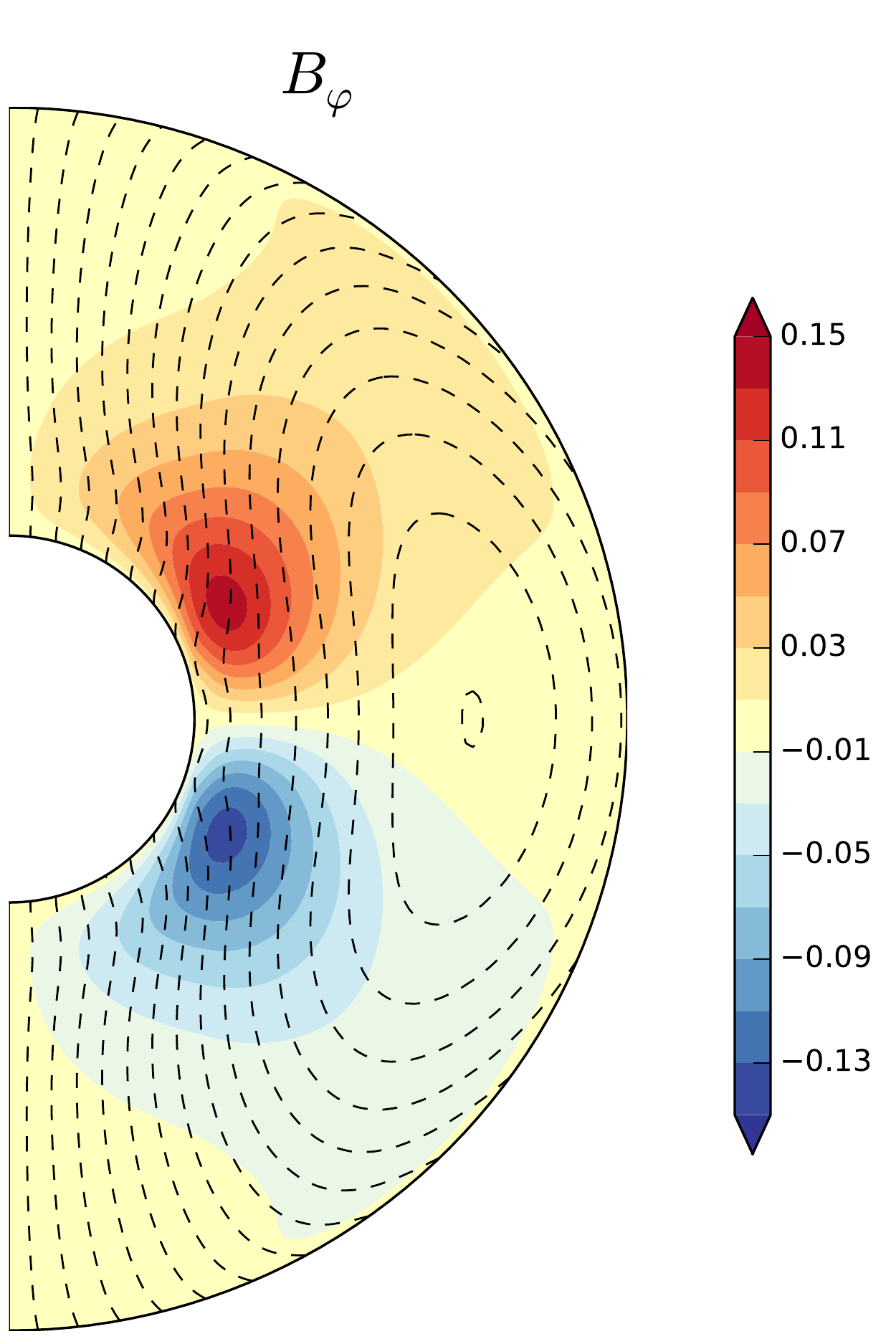}
   \includegraphics[width=0.22\textwidth]{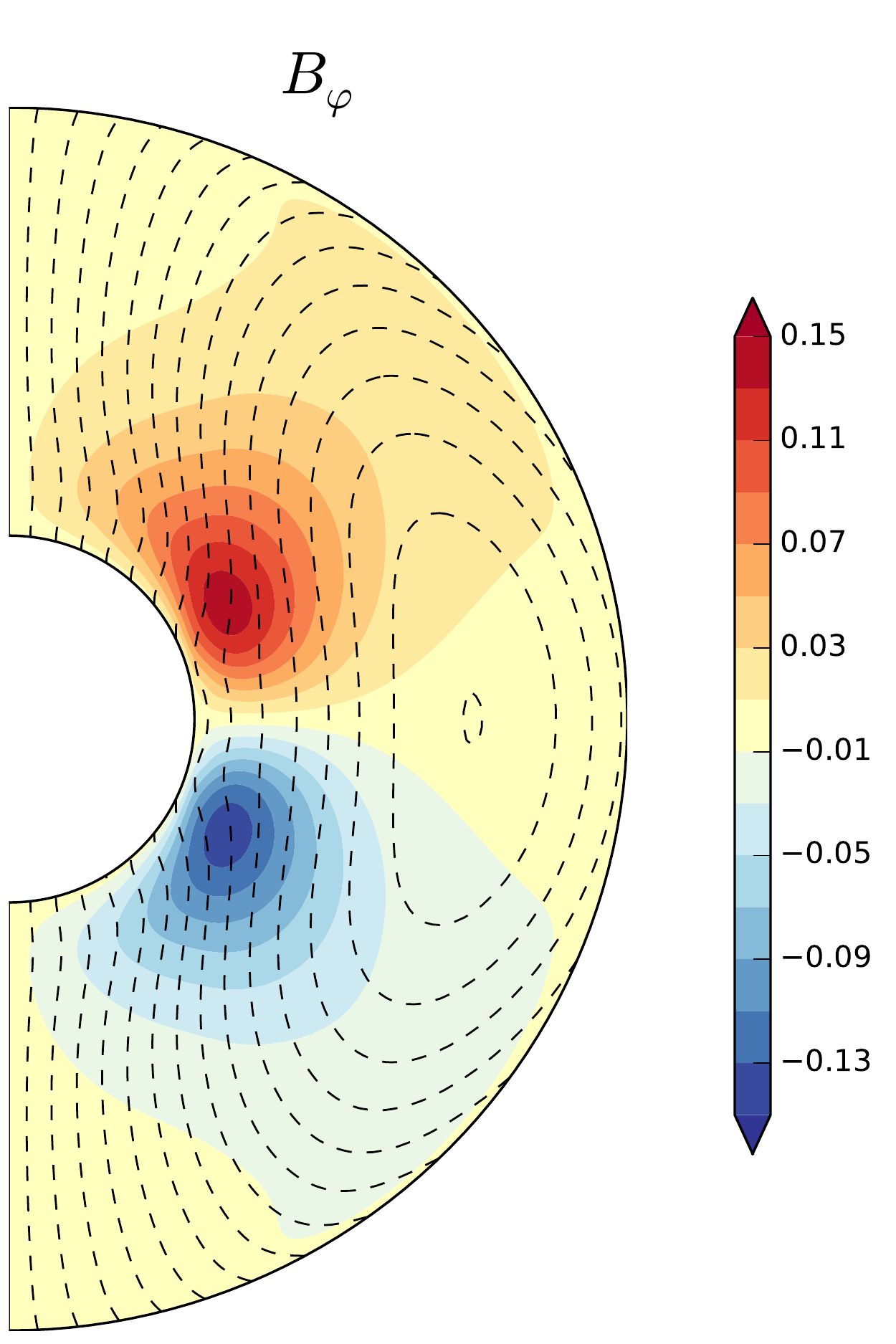}
  \includegraphics[width=0.22\textwidth]{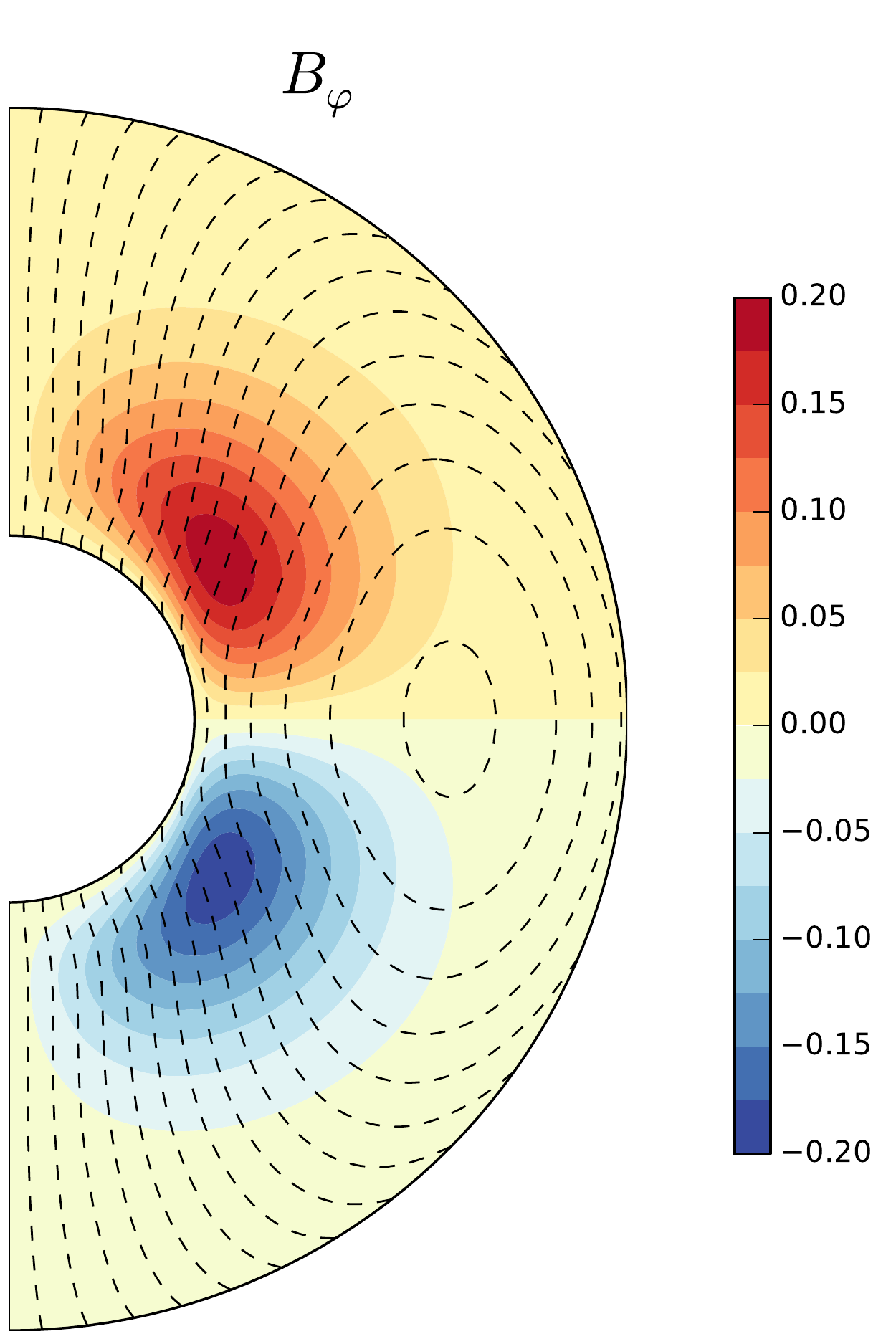}
  \includegraphics[width=0.22\textwidth]{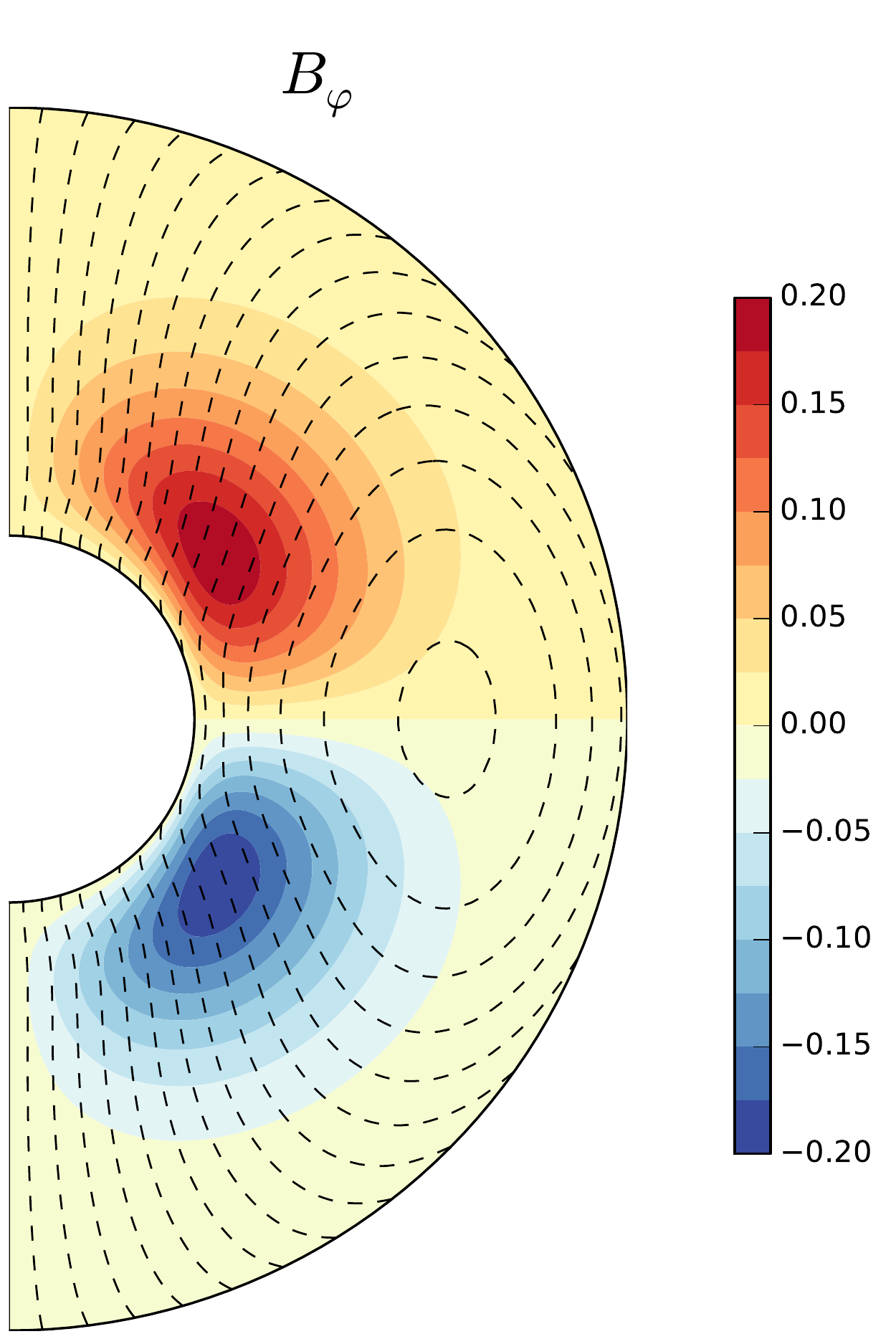}
  \caption{Structure of the flow (top panels: rotation rate in color, meridional flow contours in black lines) and of the magnetic field (bottom panels: toroidal field in color and poloidal field lines in dashed lines) at time $t=0.1 t_{ap}$, for R7 and R8 (two left panels) and for R2 and R6 (two right panels).}
 \label{fig_flowmag}
\end{center}
\end{figure*}

Then, not only the kinetic and magnetic energies evolution are close for identical value of $\Pra N^2/\Omega_0^2$ but in fact the whole solutions are very similar. This is illustrated in Fig.\ref{fig_flowmag} where the spatial structures of the flow and the magnetic field are shown at $t=0.1 t_{Ap}$ at $\Pra N^2/\Omega_0^2=0.04$ (two left panels) and $\Pra N^2/\Omega_0^2=0.25$ (two right panels), in both cases for 2 different values of $\Pra$, namely $1.6\times 10^{-3}$ and $1.6\times 10^{-5}$ for the left panels (cases R7 and R8) and $10^{-2}$ and $10^{-4}$ for the right panels (cases R2 and R6). The top panels show the rotation rate in color and the meridional flow contours while the bottom panels present the toroidal magnetic field in color and the poloidal field lines in dashed lines. 

In appendix~\ref{sec:edd}, we perform a scale analysis of the Boussinesq MHD equations in the parameter regime of the simulations. It shows that for a time scale ordering, $t_\nu \gg t_{Ap} \gg t_\kappa \gg t_\Omega \gg t_B$, 
the evolution of the system only depends on $\Pra N^2/\Omega_0^2$, $Lu$ and $Pm$ if time is scaled by $t_{Ap}$ and $B_\varphi$ by $d \Delta \Omega \sqrt{4 \pi \rho}$. The Lorentz number $Lo$ only appears as a scaling factor of the ratio  $B_{\varphi}/B_p = Lo^{-1} f(t/t_{Ap}, \vec{r}/d, \Pra N^2/\Omega_0^2,Lu, Pm)$. 
The simulations that verify the required time ordering do show
this $\Pra N^2/\Omega_0^2$ dependence. The deviations at small time due to the initially excited gravity waves are expected because gravity waves are filtered out by the scale analysis. The strong deviation observed at late time for run R9 is also expected since $\frac{t_\kappa}{t_{Ap}} = \frac{Lu \Pra }{Pm} = 3.125$ is larger than $1$ in this case (in appendix~\ref{sec:edd}, the regime $t_\kappa \gg t_{Ap} \gg t_\Omega \gg t_B$ is shown to be dominated by waves with negligible effect of the meridional circulation). 
Besides the $\Pra N^2/\Omega_0^2$ dependence, the expression of
$B_{\varphi}/B_p$
is fully compatible with the maximum toroidal to poloidal magnetic energy ratio found in simulations performed for the same $\Pra N^2/\Omega_0^2=0.25$ but two different Lorentz numbers (runs R1 and R2). The ratio indeed increased by a factor $\approx 2.5^2$, as $Lo$ was reduced from $2.5\times 10^{-3}$ (right panel of Fig.\ref{fig_magener}) to $Lo= 10^{-3})$ (right panel of Fig. \ref{fig_enertot}). In addition to a simplification in the physical interpretation, the scale analysis allows us to conduct the parametric study of the flow by varying only one non-dimensional number $\Pra N^2/\Omega_0^2$ instead of the three $Lo$, $N/\Omega_0$ and $\Pra$.

We thus consider how the rotation and magnetic configurations depend on the values of $Pr N^2/\Omega_0^2$. As shown on the right panel of Fig. \ref{fig_enertot}, the maximum ratio of toroidal to poloidal magnetic energy is $1500$ for $Pr N^2/\Omega_0^2=0.04$ while it is close to 4000 for $Pr N^2/\Omega_0^2=0.25$ (and is reached at an earlier time). The stable stratification is thus favorable to the creation of a magnetic field configuration more strongly dominated by its toroidal component. According to Fig.\ref{fig_flowmag}, in the most stably stratified case (right panels), the differential rotation remains close to its initial profile and then mostly dependent on radius, contrary to the less stably stratified case (left panels) for which the differential rotation is reduced and tends to become cylindrical. This tendency is expected because when stable stratification is less efficient the system can evolve more freely towards a flow satisfying the Taylor-Proudman constraint, valid for unstratified systems: $\frac{\partial \Omega}{\partial z} =0$, where the $z$-direction parallel to the rotation axis. The reduced level of differential rotation can also be explained by an efficient meridional transport of angular momentum in the less stratified case. The ratio $\Omega_i/\Omega_o$ indeed decreases from $2$ at $t=0$ to $1.6$ at $t=0.1 t_{Ap}$ in the less stratified case whereas it remains close its initial value in the more stratified case. 

This difference in the level of differential rotation then naturally explains why a weaker toroidal magnetic field is produced in the less stratified case. This is visible on the bottom left panels of Fig.\ref{fig_flowmag} where the maximum value of $Lo_\varphi$ only reaches $0.15$ compared to the other cases where it is already close to $0.2$. In addition, we observe that the poloidal field configuration has been significantly altered in the less stratified cases compared to the initial condition. The poloidal field tends to align on the cylindrical isocontours of $\Omega$ at mid-latitudes, again preventing a strong $\Omega$-effect to be at play. This significant change of the poloidal field is due to its advection by the meridional circulation which is more efficient in the less stratified case.

From the axisymmetric numerical simulations performed for different stable stratifications, we conclude that in the regime considered,
the effect of the stable stratification is controlled by the product $P_rN^2/\Omega_0^2$ and
that stable stratification favors the creation of magnetic field configurations more strongly dominated by their toroidal component.

\section{Stability of the magnetic configurations}
\label{sec_stab}

We now turn to investigate the stability of the axisymmetric magnetic configurations determined in the first part of this work. We perturb the magnetic field by adding a random noise on the axisymmetric poloidal field and then follow the temporal evolution of the various non-axisymmetric modes $m\neq 0$, in the same way as was done in \citet{jouve2015}. We first consider the stability of the system with radial and cylindrical differential rotations and a fixed  $PrN^2/\Omega_0^2=0.25$ (cases R2 and C2) and argue that the observed instability is of the MRI type, this is presented in Sect.\ref{sub_r2c2}. In Sect.\ref{sub_thermal}, the effect of varying the thermal diffusivity on the instability is then studied (cases C3, C4 and R3, R4). To help us understand the characteristics of the instabilities, we compare our results with a local stability analysis in Sect.\ref{sub_local}. While the Lorentz number $Lo$ has been fixed to a small value (namely $Lo=5\times10^{-3}$ for the cylindrical case and $Lo=10^{-3}$ for the radial case) to maximize the possibility for a non-axisymmetric instability to fully develop (see \citet{jouve2015}), we investigate in Sect.\ref{sub_lo} the effects of increasing $Lo$ (cases C9 and R1).  

\subsection{Radial VS cylindrical differential rotation}
\label{sub_r2c2}

\begin{figure*}[!h]
\begin{center}
 \includegraphics[width=0.5\textwidth]{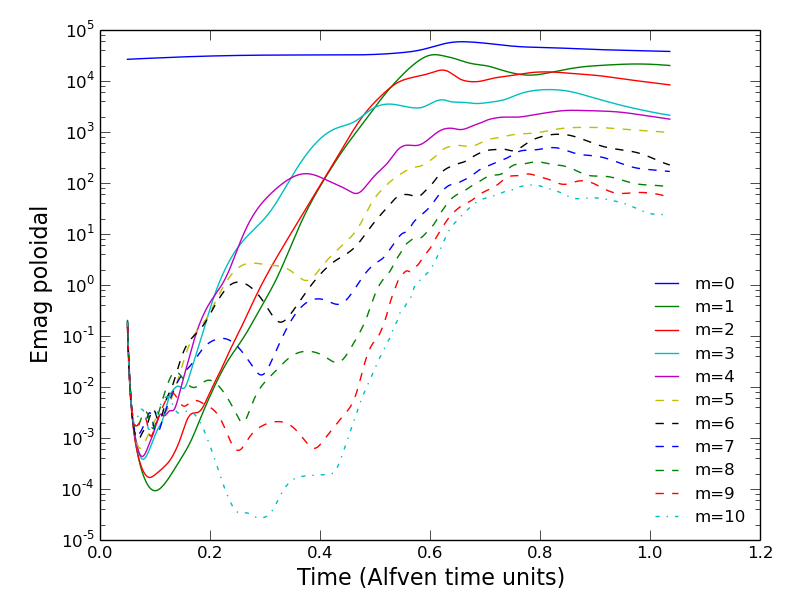}
 \includegraphics[width=0.18\textwidth]{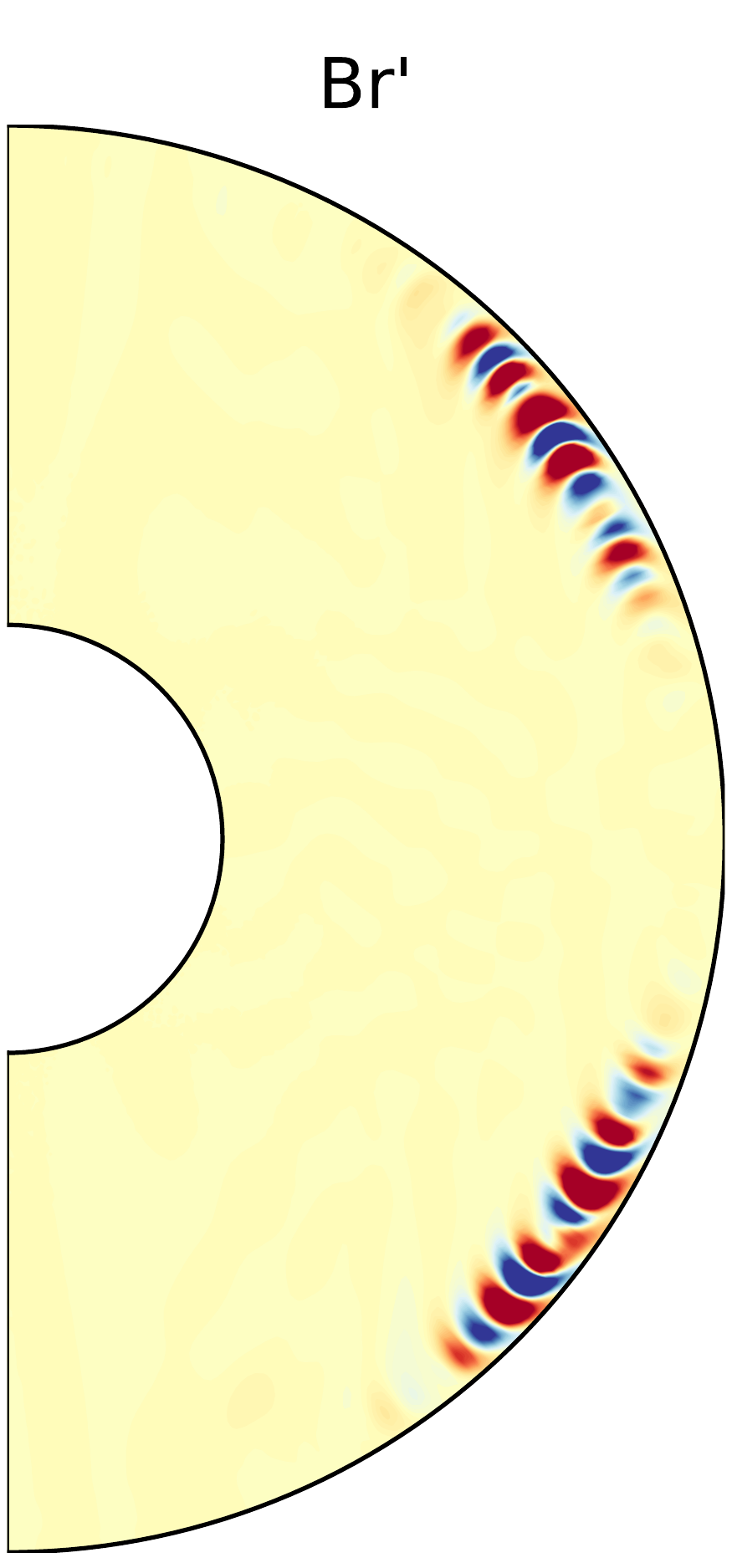}
 \includegraphics[width=0.25\textwidth]{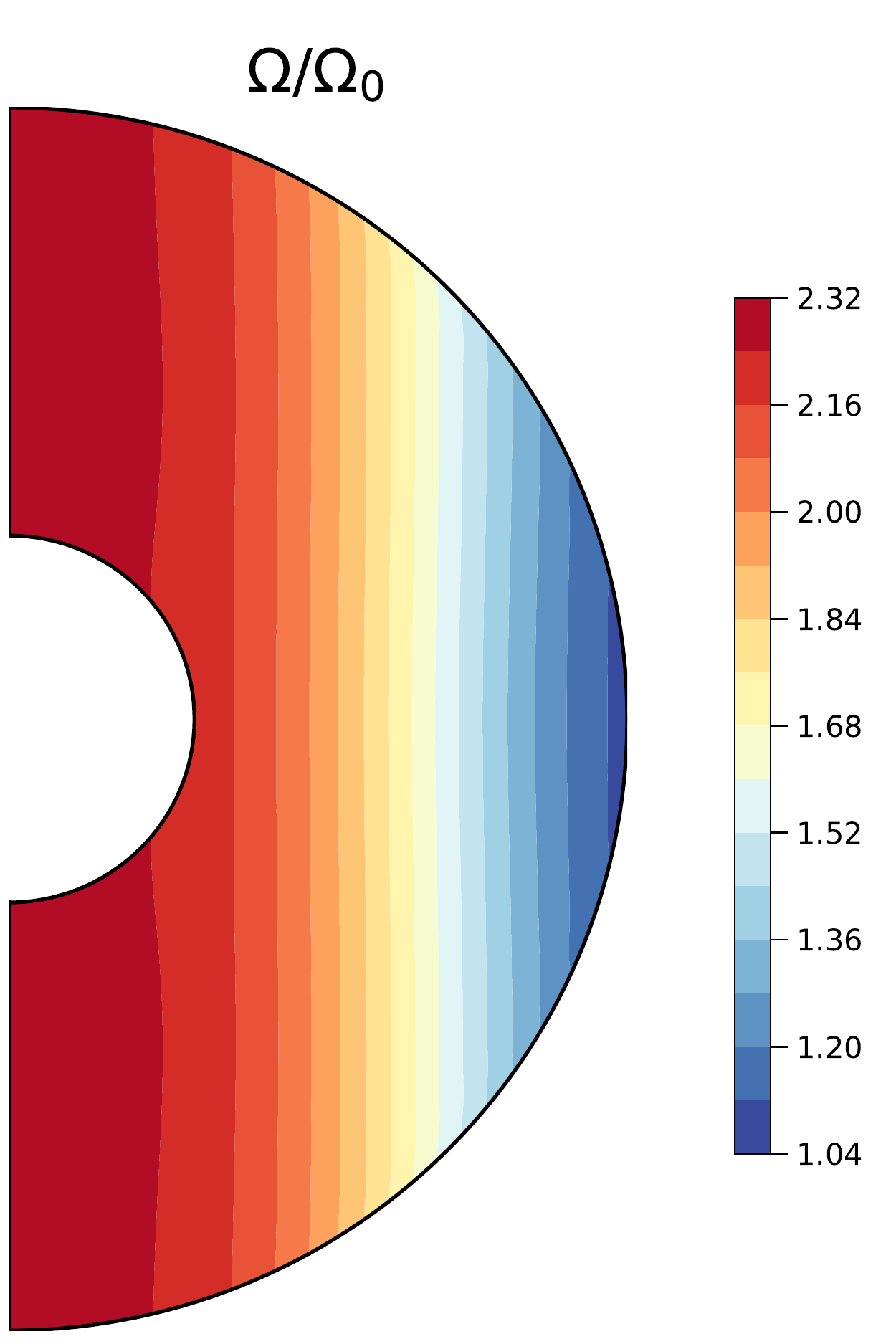}
 \includegraphics[width=0.5\textwidth]{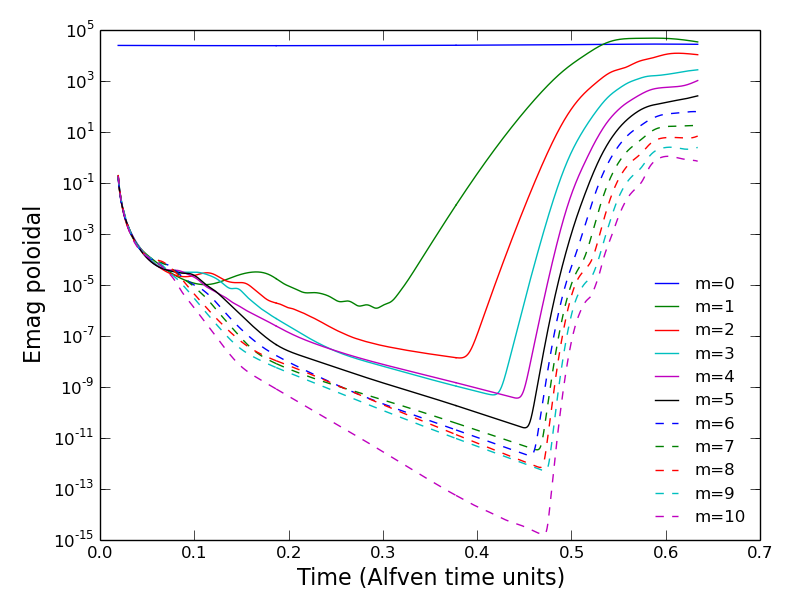}
 \includegraphics[width=0.18\textwidth]{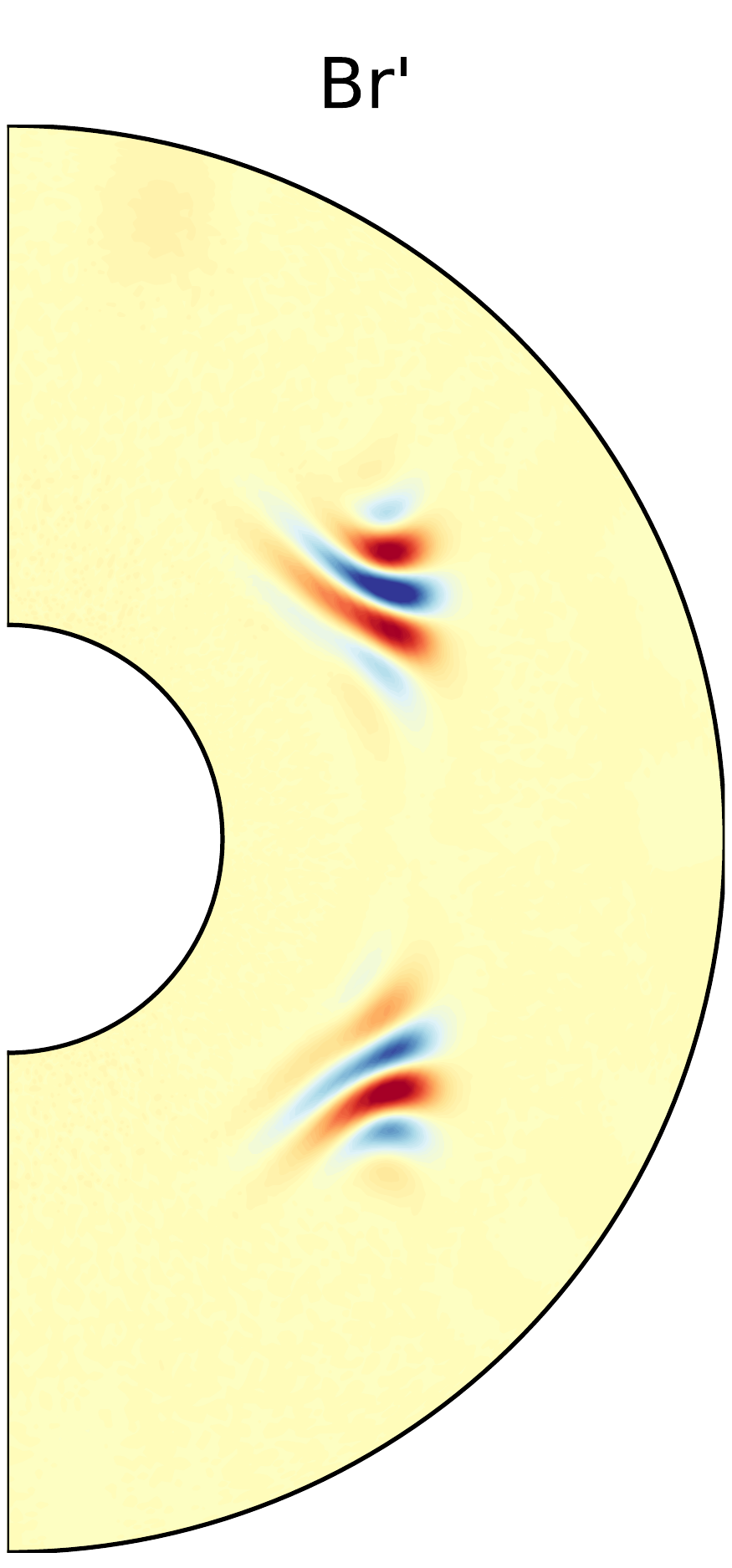}
 \includegraphics[width=0.25\textwidth]{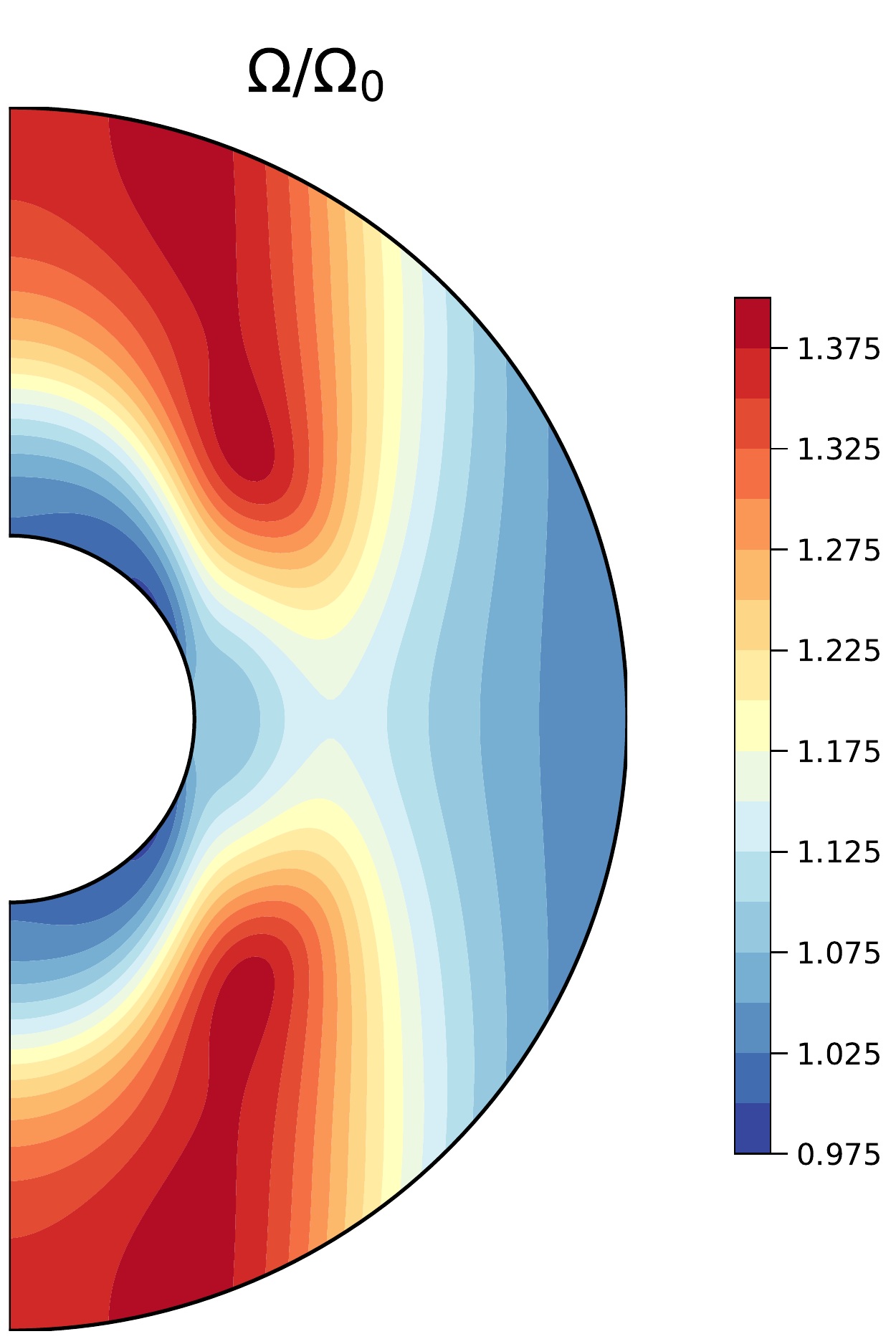}
  \caption{Instability in case C2 (top) and R2 (bottom). Shown are the temporal evolution of the poloidal magnetic energy in the first 11 azimuthal wavenumbers (averaged in $r$ and $\theta$) (left), the fluctuations of the radial component of the magnetic field at a particular longitude (mid) and the rotation profile (right) when the instability starts to develop, i.e. at $t=0.15 t_{ap}$ for C2 and $t=0.35t_{ap}$ for R2. The colorbar applies to the rotation rate.}
 \label{fig_instab}
\end{center}
\end{figure*}

\begin{figure*}[!h]
\begin{center}
 \includegraphics[width=0.8\textwidth]{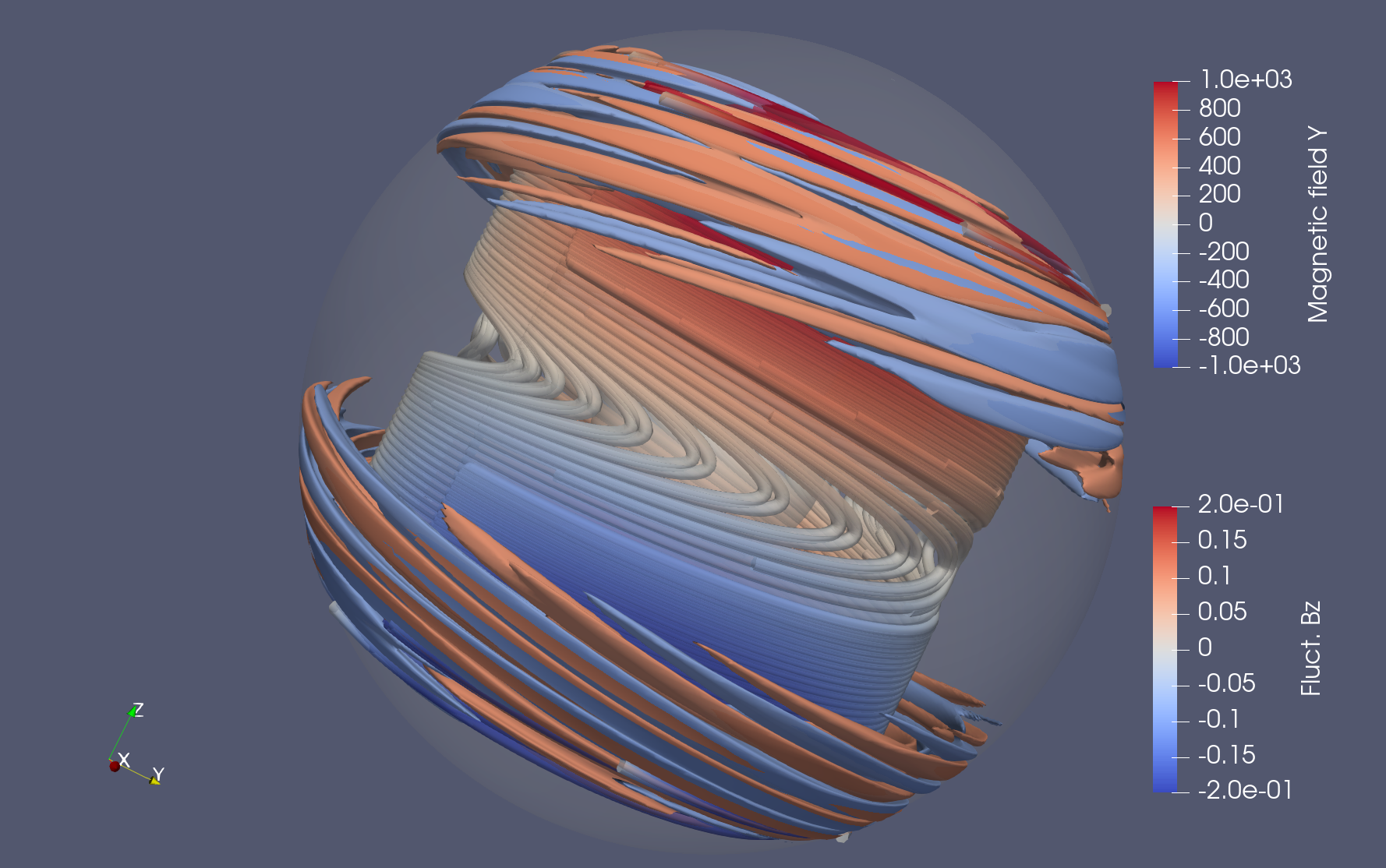}
  \includegraphics[width=0.8\textwidth]{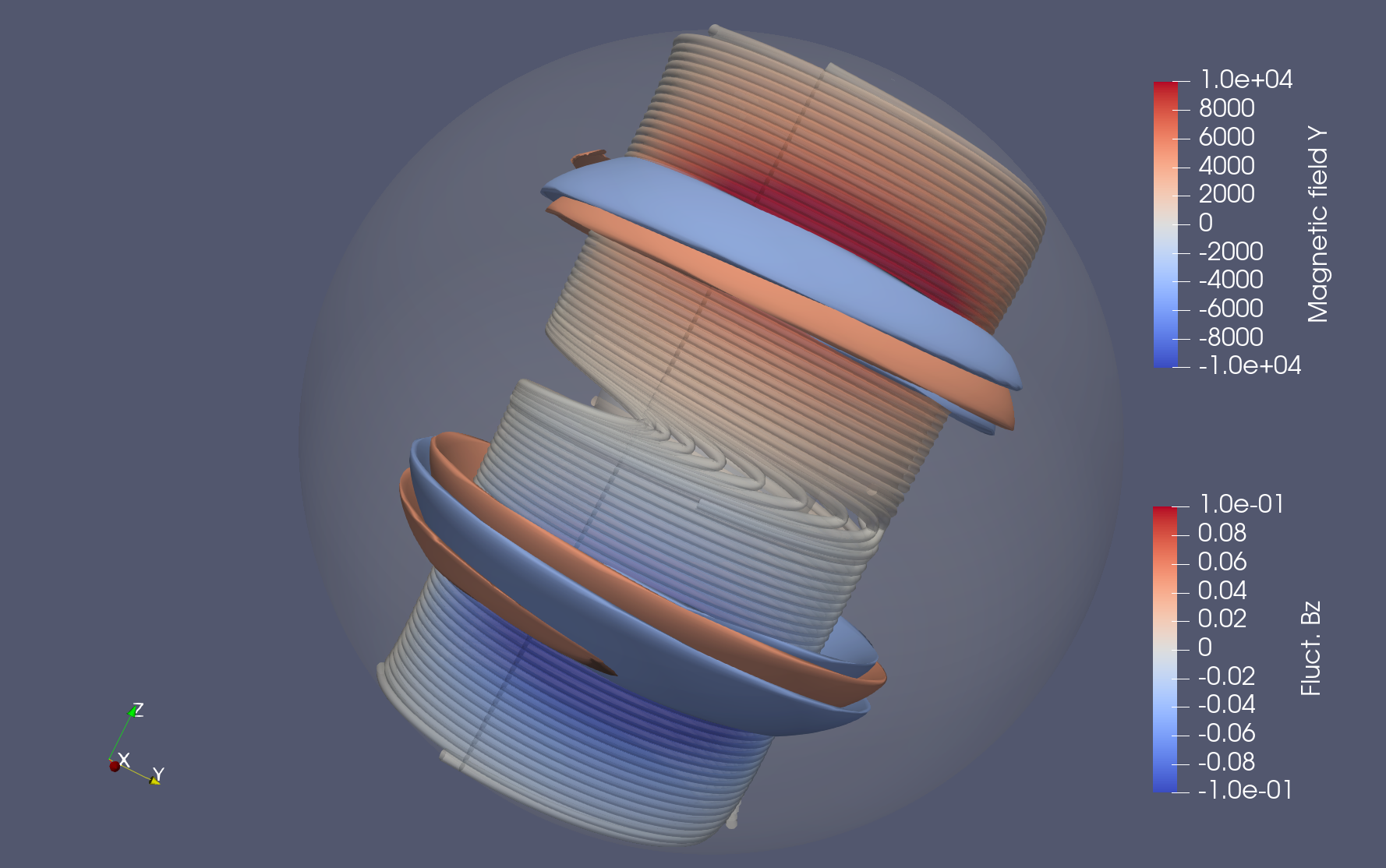}
  \caption{3D views of the instability in the 2 cases: Case C2 on top and case R2 below. The magnetic field lines of the background axisymmetric magnetic field are plotted around the location of the instability and colored with the values of the toroidal magnetic field (in the $Y$-direction in the Cartesian frame shown at the bottom left) and isosurfaces of the axial component of the fluctuating magnetic field are overplotted.}
 \label{fig_3D}
\end{center}
\end{figure*}

We first investigate the typical evolution of an unstable situation and compare the behaviour of the simulations initialized with the cylindrical and radial differential rotation profiles. Figure \ref{fig_instab} shows the evolution of the poloidal magnetic energy contained in the first 11 azimuthal wavenumbers, including the axisymmetric $m=0$ mode, which is approximately steady during the time considered. In both cases, a non-axisymmetric instability grows exponentially in a fraction of Alfv\'en time to quickly reach the level of the axisymmetric energy at about $0.6 t_{ap}$. The right panels of Fig.\ref{fig_instab} enable us to visualize the location and structure of the unstable modes, by showing the amplitude of the fluctuations of the radial component of the magnetic field. In both cases, the instability develops preferentially where the azimuthal magnetic field is maximum (see Fig. \ref{fig_bphi} for the axisymmetric configuration which was perturbed) and where and when a significant amount of differential rotation exists. In both cases, the growth rate of the most unstable mode is a fraction of the rotation rate. It is approximately equal to $5\times10^{-2}\Omega$ for the $m=4$ mode in the cylindrical case and $2.5\times10^{-2}\Omega$ for the $m=1$ mode in the radial case.

We now emphasize the differences between the 2 cases. First, the time at which the instability starts to grow is quite different. Indeed, in the cylindrical case, the axisymmetric equilibrium which is perturbed is already unstable as soon as the perturbation is introduced, leading to the exponential growth of the non-axisymmetric modes from approximately $t=0.1 t_{ap}$. At this stage, the axisymmetric evolution is still in its linear growth of azimuthal magnetic field, as shown in Fig.\ref{fig_magener}. On the contrary, in the radial case, the instability develops only later, at about $t=0.3 t_{ap}$, approximately when the maximum of axisymmetric $B_\varphi$ is reached and thus when a strong back-reaction of the Lorentz force on the differential rotation profile has acted. To illustrate this, the right panels of Fig.\ref{fig_instab} show the profiles of differential rotation at the time when the unstable non-axisymmetric modes start to grow. It is clear that the cylindrical differential rotation is still mostly identical to its initial condition whereas the radial case has been significantly modified by the back-reaction of the Lorentz force. In particular, a latitudinal differential rotation appears here, which was not present initially since the rotation rate was dependent on radius only. It is exactly at the location where the latitudinal shear is the strongest that the unstable modes are confined.

Another major difference between the case R2 and C2 lies in the structure of the unstable modes. From the mid panels of Fig.\ref{fig_instab}, it is clear that the displacement of the perturbations is not in the same direction in both cases. Let us express the perturbation as proportional to: 
$$
\exp \left[\mi (\kr r+\kt \theta+m \varphi-\sigma\, t)\right]
$$
\noindent with $\kr$, $\kt$ and $m$ the radial, latitudinal and azimutal wavenumbers and $\sigma$ the complex growth rate. Then in the cylindrical case, the latitudinal wavenumber $\kt$ is large compared to $\kr$ and the displacement is thus mainly in the radial direction, parallel to gravity. We argue that the radial extent of the unstable mode is in fact mostly due to the structure of the axisymmetric background and not due to the effect of stable stratification. Indeed, in our previous study where the effects of stable stratification were not included \citep{jouve2015}, the structure of the unstable modes and the growth rates in the equivalent of case C2 where very close to the ones found here.

On the other hand, in case R2, $\kr$ is now dominant compared to $\kt$, so that the displacement is mainly in the latitudinal direction, perpendicular to gravity. We thus anticipate that in this simulation, the stable stratification, which is much less effective if the displacement is horizontal, only affects the geometry of the unstable mode and not its growth rate. This is investigated in the next section, where the effect of the stable stratification is increased for the two initial differential rotations.    

In both cases, we argue that the instability found here is of MRI type. First, we checked that the flow is hydrodynamically stable by perturbing the flow when the magnetic field is set to 0 at the time where the instability develops in the MHD case. The instability could nevertheless be a current-driven instability of the Tayler type since the magnetic configuration contains current and is strongly dominated by the toroidal component, as we can clearly see on Fig.\ref{fig_3D}. This figure shows the magnetic field lines of the background axisymmetric magnetic field traced around the location of the instability. This 3D view enables to clearly see the dominance of the toroidal component of the field and also allows us to see that the maximum amplitude of the unstable modes (shown on Fig.\ref{fig_3D} by the isosurfaces of the fluctuating axial component of the field) are mainly located where the toroidal field is maximum. The location of the maximum $B_\varphi$ naturally corresponds to the region where the shear is also maximum since the shear is responsible for the generation of toroidal field through the $\Omega$-effect. We thus find here that the instability develops mainly where the shear is concentrated. This would be different if the instability was of the Tayler type, because then the location of the unstable modes would be correlated with the gradients of toroidal field, where the currents are maximal. Moreover, the most unstable mode in the cylindrical case is not the $m=1$ as expected for the Tayler instability. It is however the $m=1$ mode which is the most unstable in the radial case as seen on the figure but this is not incompatible with an MRI instability in the fast thermal diffusion case. In \citet{Acheson78}, a detailed theoretical description is made of all the various MHD instabilities which can arise in stellar radiative zones. In this seminal paper, the MRI is not explicitly quoted but an instability associated with a shear and which necessitates the presence of a magnetic field is studied, when the thermal diffusivity is high and in the limit where $(\omega_{A_\varphi}/\Omega)^2 E_k^{-1} Pm$ is also high. In this situation, he argues that the most favored unstable mode is precisely the $m=1$ mode. The values in our simulations of the parameter $(\omega_{A_\varphi}/\Omega)^2 E_k^{-1} Pm$ when the instability develops in the radial case is of the order of $2\times 10^3$ so that the limit studied by \citet{Acheson78} does apply here. In both cases started with a cylindrical or a radial differential rotation, we thus observe the presence of a MRI which is driven mostly by the initial radial differential rotation in the cylindrical case and driven by the latitudinal shear that is produced by the back-reaction of the Lorentz force in the radial case. The instability is allowed to exist in both cases with a relatively high thermal diffusivity (Prandtl number of $10^{-2}$). We now wish to investigate the effect of varying the thermal diffusion on the instability.

\subsection{Effect of the thermal diffusivity}
\label{sub_thermal}

\begin{figure*}[!h]
\begin{center}
 \includegraphics[width=0.95\textwidth]{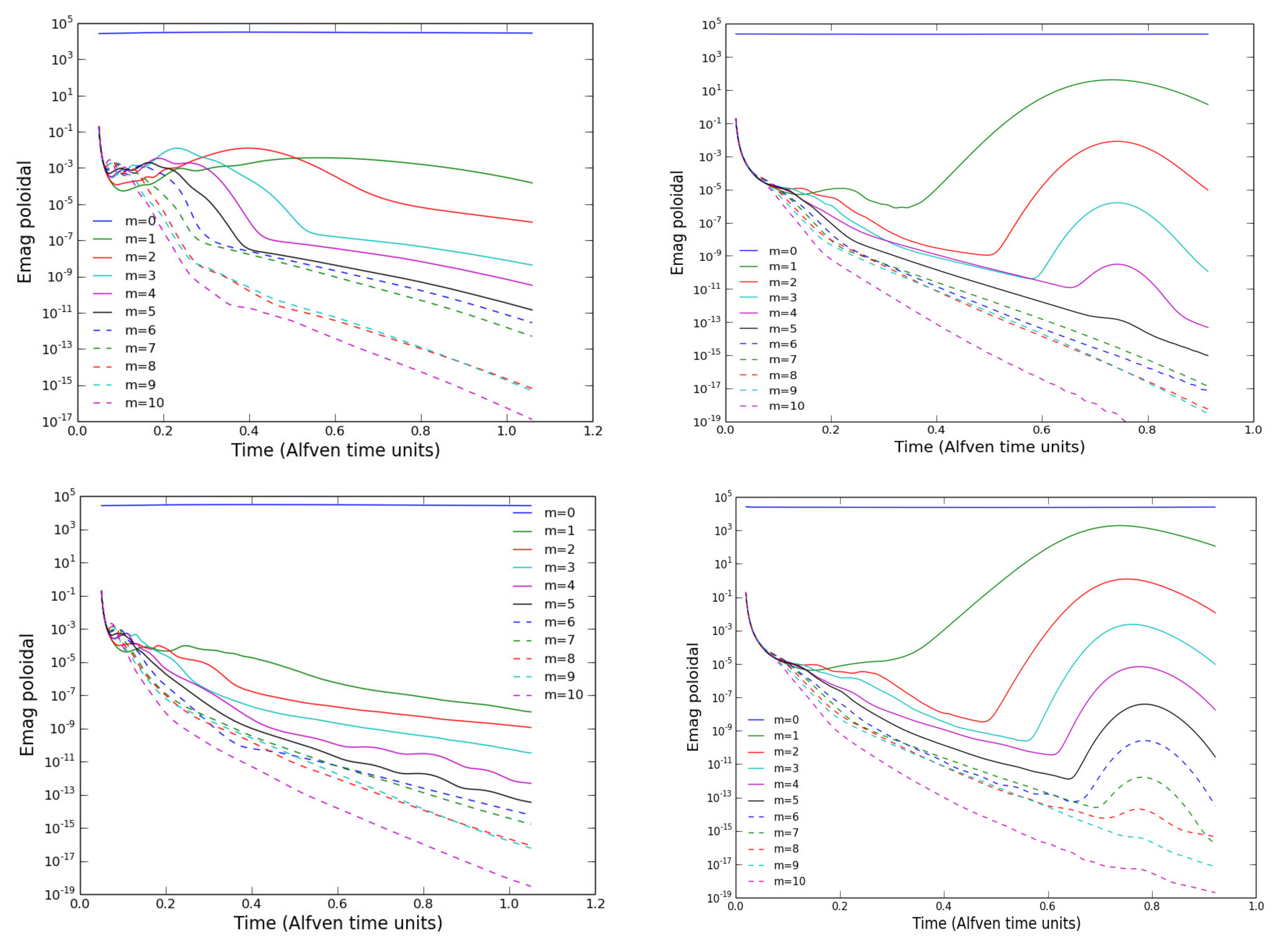}
  \caption{Temporal evolution of the poloidal magnetic energy in the first 11 azimuthal wavenumbers for cases C3 (top left), C4 (bottom left), R3 (top right) and R4 (bottom right).}
 \label{fig_pr_ener}
\end{center}
\end{figure*}

The stable stratification has the tendency to strongly reduce the development of non-axisymmetric instabilities, as shown for example in \citet{spruit99}. In particular, as the stable stratification limits radial displacements, it will strongly affect instabilities that require them to develop. By damping temperature deviations, thermal diffusion diminishes the amplitude of the restoring buoyancy force and thus the effect of the stable stratification. In order to study these effects, we therefore decrease the thermal diffusivity, and thus increase the Prandtl number $Pr$, keeping the same value for $N/\Omega_0$. These correspond to cases C3, C4 and R3, R4. Figure \ref{fig_pr_ener} shows the temporal evolution of the poloidal magnetic energy decomposed into the first 11 azimuthal wavenumbers in cases where $Pr=0.1$ and $Pr=1$. The left panels correspond to cases C3 and C4 and the right panels cases R3 and R4. Compared to the magnetic energy evolution of Fig. \ref{fig_instab} where we had $Pr=10^{-2}$, it is clear that for the cylindrical case, the instability is largely suppressed by the increase of the stable stratification effect. In particular, for $Pr=1$, the axisymmetric solution becomes completely stable to any non-axisymmetric perturbation. In other words, the preferentially radial displacements that were unstable at $Pr=10^{-2}$ are inhibited at $Pr=1$. The transition 
between $Pr=10^{-2}$ and $Pr=1$ can be linked to the value of the critical lengthscale above which the effects of the stable stratification are not diminished by thermal diffusion. This critical lengthscale $l_c$ is determined by equating the buoyancy and the thermal diffusion time scales : 

$$
\frac{l_c^2}{\kappa} = \frac{1}{N} \,\,\,\,\,\, \mbox{and thus} \,\,\,\,\,\, l_c=\sqrt{\frac{\kappa}{N}}
$$

With the dimensionless parameters used in our calculations, the critical lengthscale reads :

$$
\frac{l_c}{d}=\sqrt{\frac{E_k}{Pr}\frac{\Omega}{N}}
$$

The computation of this quantity gives a value ranging from $4\%$ to $0.4\%$ when $Pr$ goes from $10^{-2}$ to $Pr=1$, for these cases where $Pr N^2/\Omega_0^2=0.25$. The unstable radial lengthscale seen in Fig.\ref{fig_instab} being of the order of a few percent of the computational domain, we argue that this case is only marginally affected by the stratification. This is consistent with the fact that, as mentioned above, a very similar unstable mode was found in the corresponding unstratified simulation by \citet{jouve2015}. However, the reduction of the critical lengthscale causes the instability to disappear in the $Pr=1$ case. Such a behaviour where increasing the stable stratification removes the instability is reminiscent of the vertical shear instability in a vertically stratified medium \citep{dudis74,lignieres99b}.

\begin{figure*}[!h]
\begin{center}
 \includegraphics[width=0.23\textwidth]{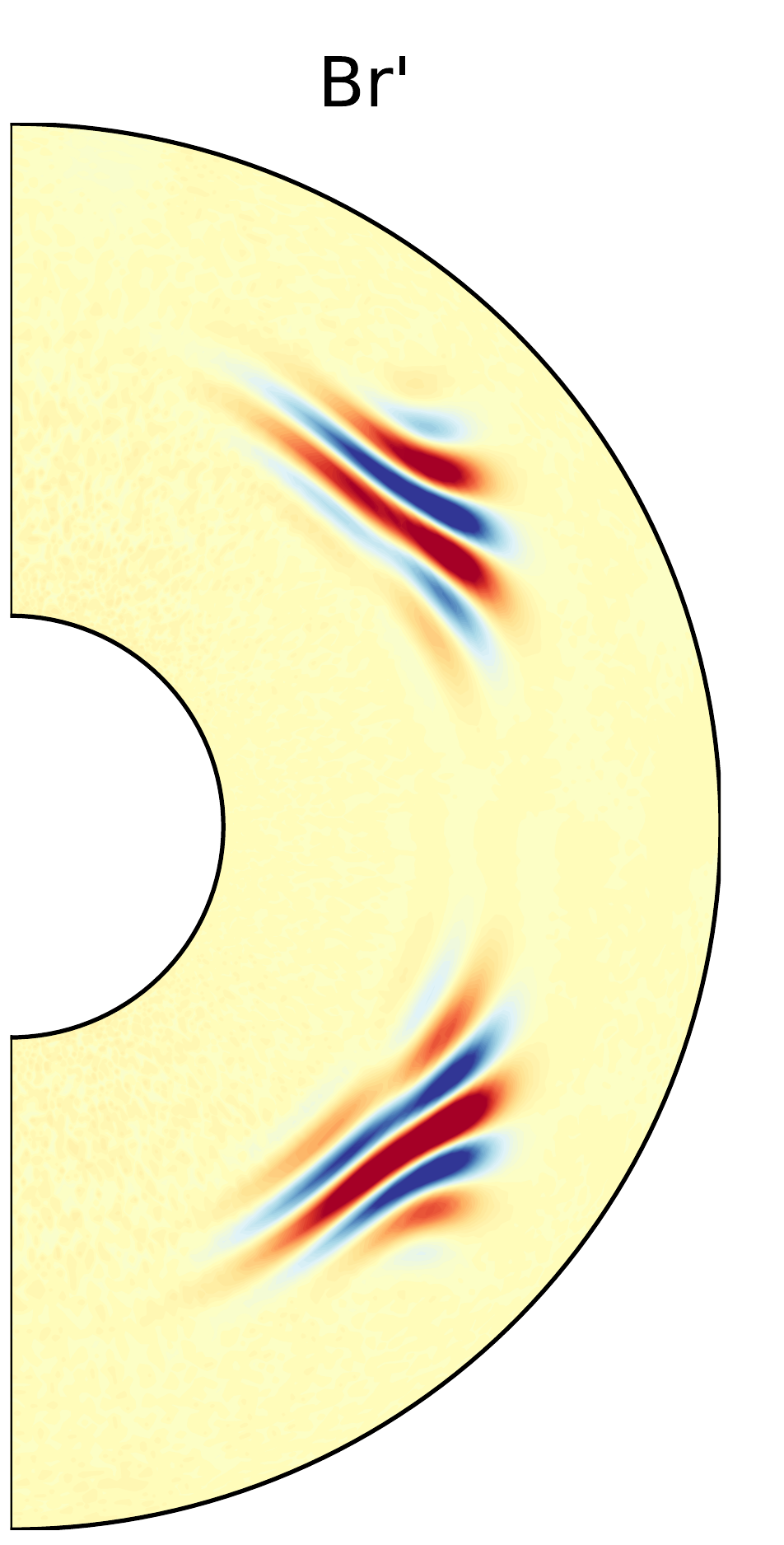}
  \includegraphics[width=0.3\textwidth]{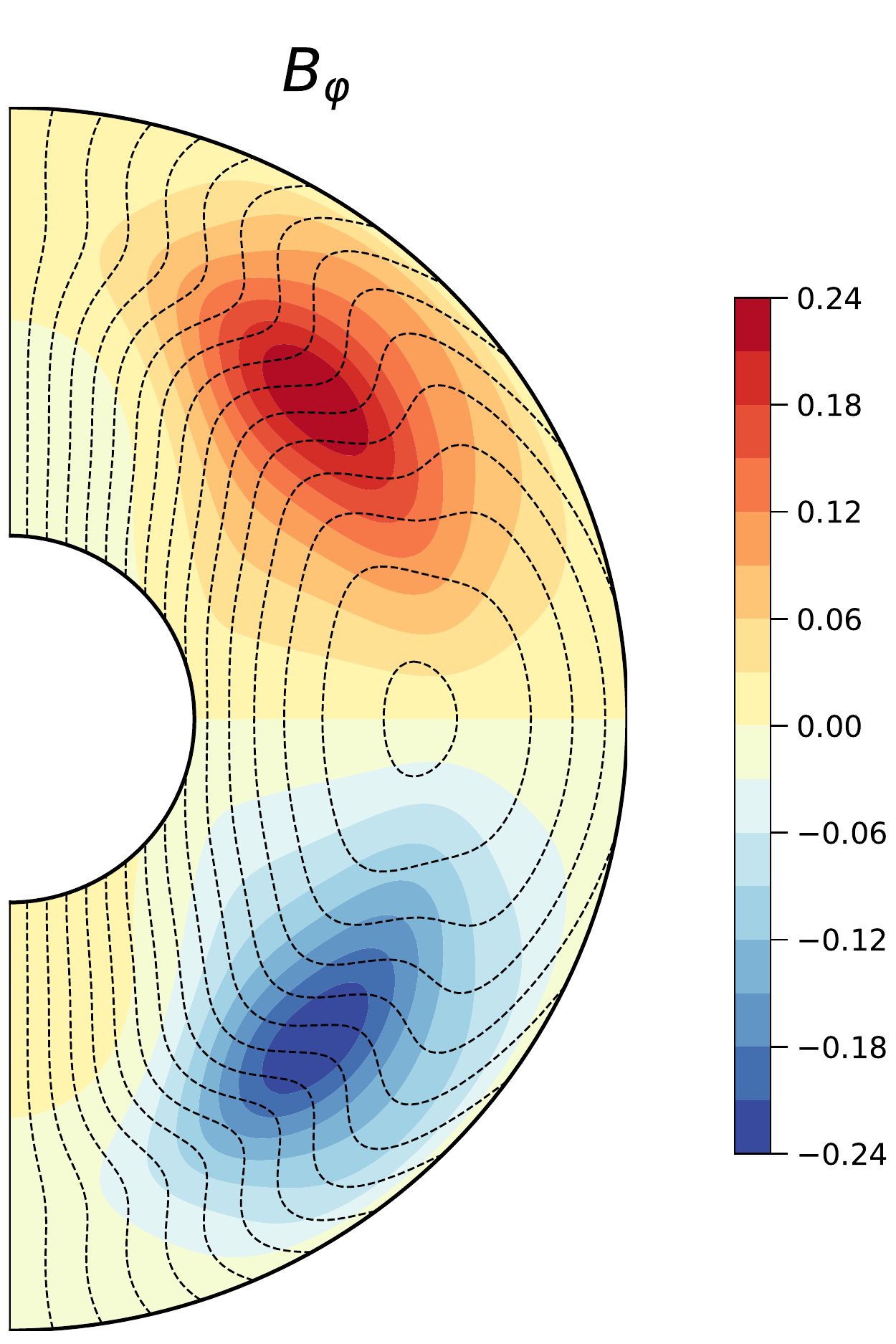}
   \includegraphics[width=0.3\textwidth]{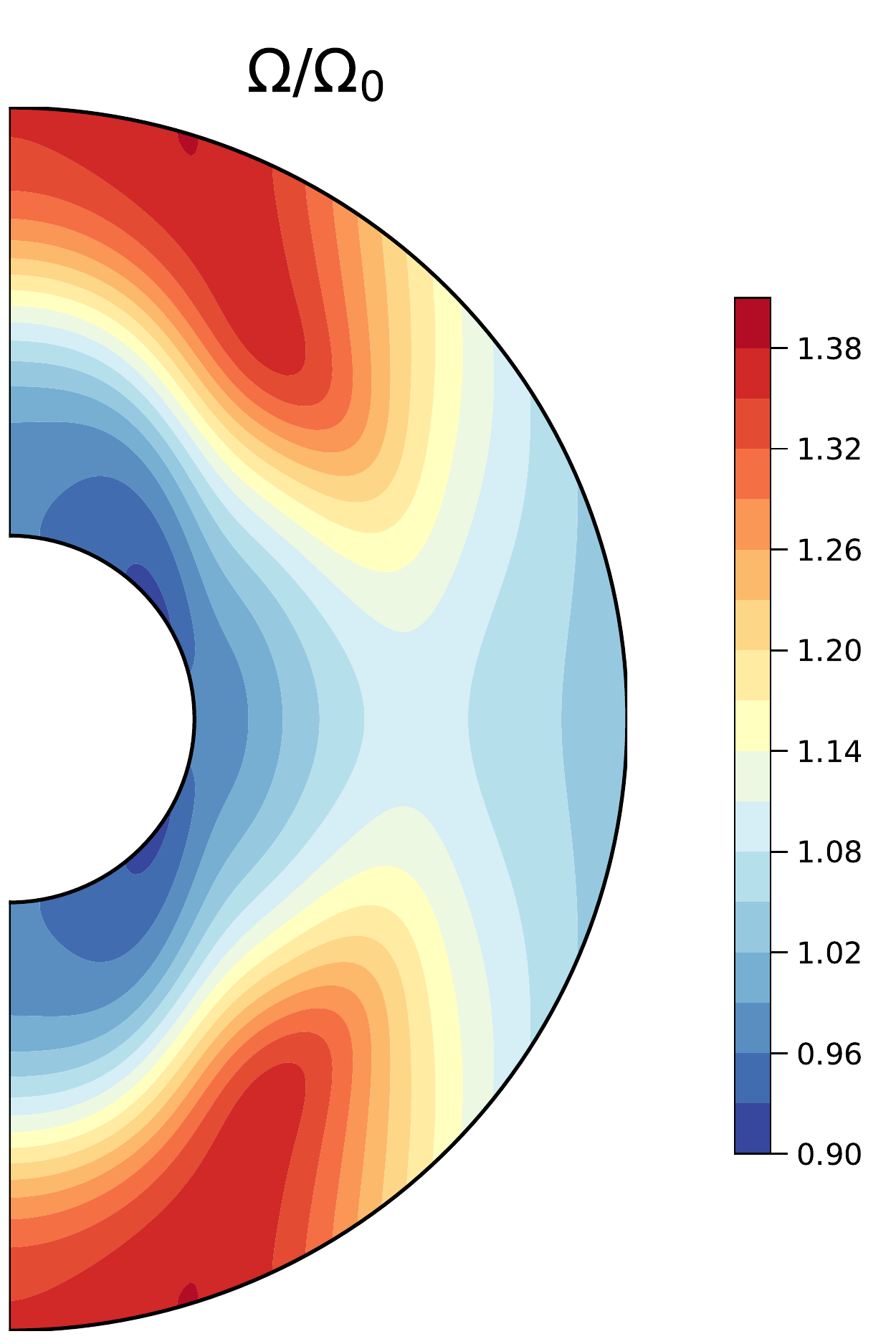}
  \includegraphics[width=0.23\textwidth]{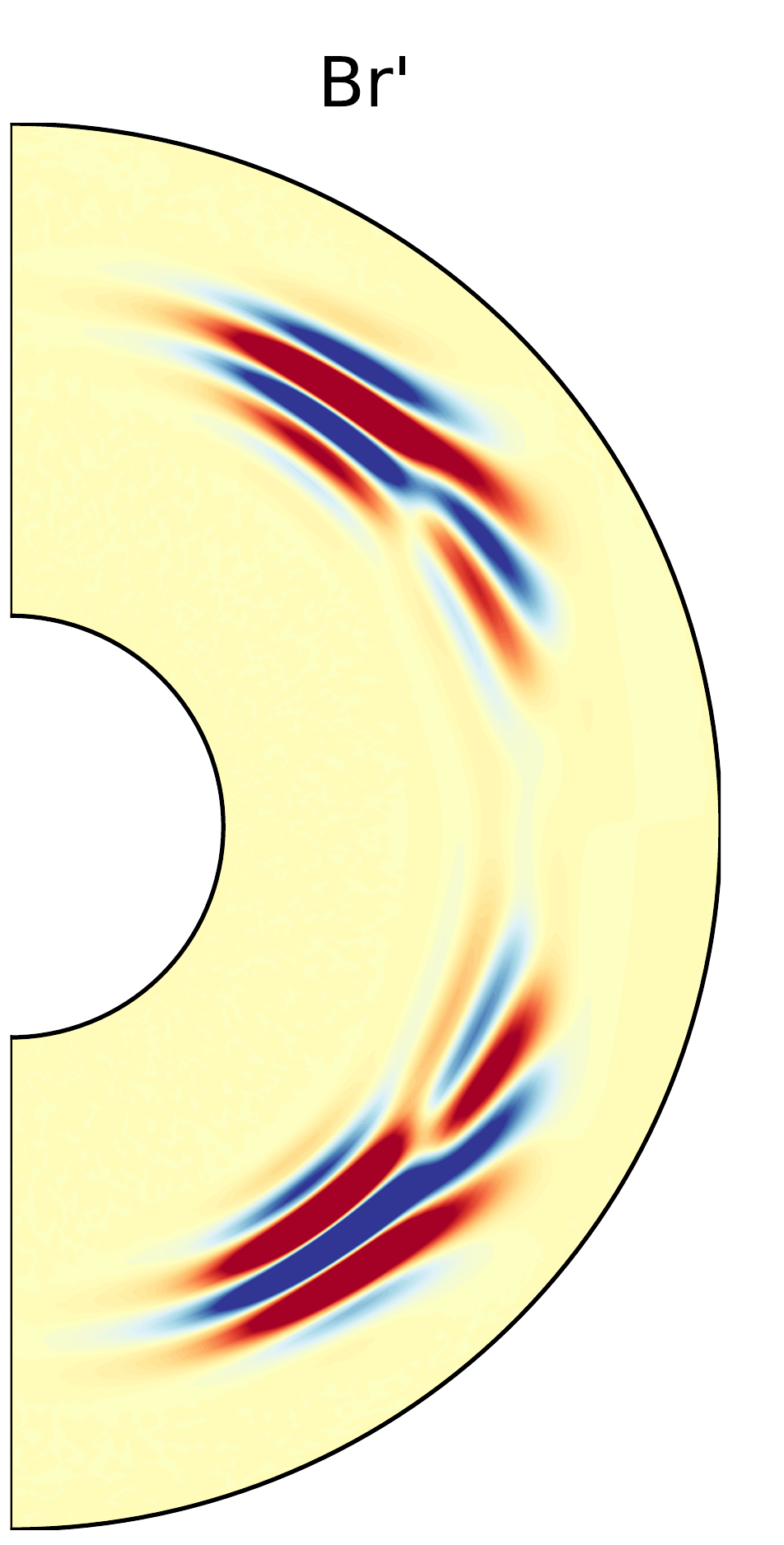}
   \includegraphics[width=0.3\textwidth]{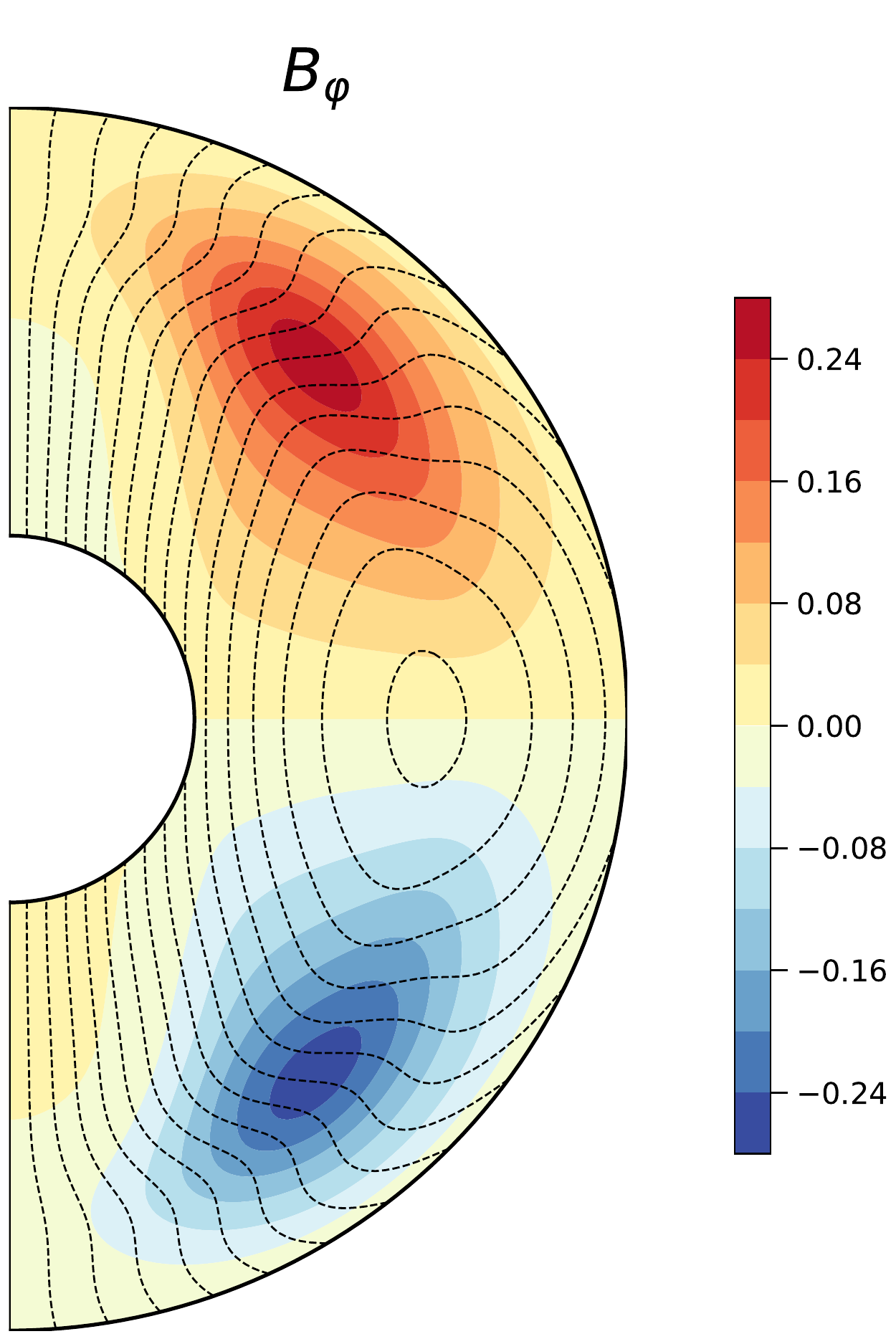}
  \includegraphics[width=0.3\textwidth]{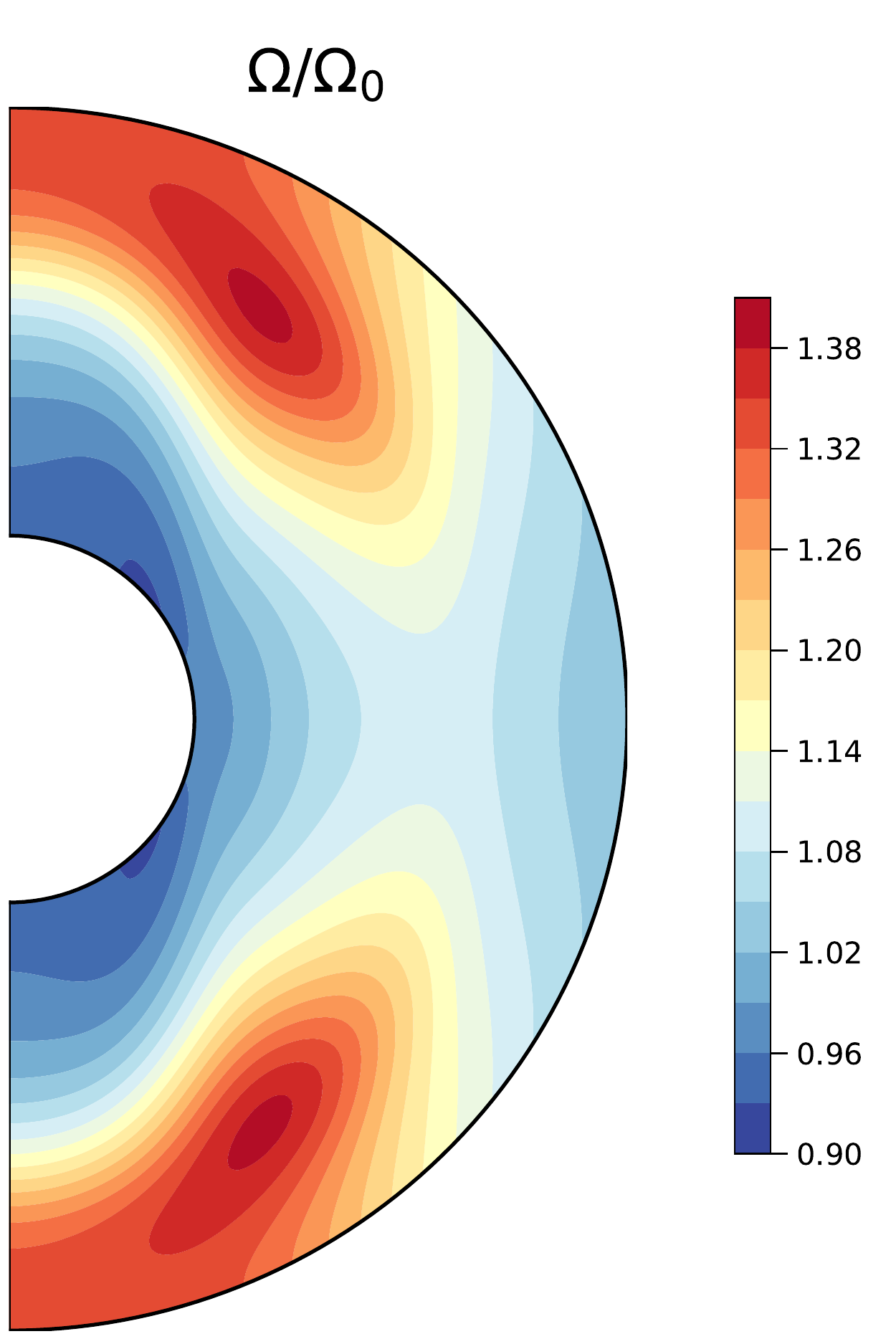}
  \caption{Fluctuations of the radial component of the magnetic field at a particular longitude, axisymmetric magnetic field (toroidal field in color and poloidal field lines) and rotation rate at approximately time $t=0.5 t_{ap}$ (during the linear phase of the instability) for cases R3 ($Pr=0.1$, top) and R4 ($Pr=1$, bottom). We clearly see that the perturbation direction is orthogonal to gravity, especially for the most stratified case R4. Note that the background flow and field are very similar in both cases.}
 \label{fig_pr_br}
\end{center}
\end{figure*}

The situation is quite different in the radial case (left panels of Fig.\ref{fig_pr_ener}). Now, the instability survives even with the increase of the effects of stable stratification, and grows on time scales similar as in the $Pr=10^{-2}$ case. As we show below, this comes with the fact that the unstable displacements become more and more horizontal. On figure \ref{fig_pr_br}, we show the structure of the unstable mode for cases R3 and R4 at two different longitudes and we plot the profile of the magnetic field and the rotation rate, averaged in longitude. In both cases, the background flow and field are quite similar even if the value of $Pr N^2/\Omega_0^2$ differs. This is also true for the cylindrical case (not shown here), which also confirms that the absence of an instability in cases C3 and C4 is mostly due to the effect of stable stratification on the characteristics of the instability (namely the lenghtscale) and not on the background flow and field. We recover the fact that the displacement is indeed mostly horizontal, with a latitudinal lengthscale extremely dominant in comparison to the radial scale in case R4 where $Pr N^2/\Omega_0^2=25$. The location of the instability is still mostly where the latitudinal gradient of $\Omega$ lie, as seen on the right panels. It is thus clear here that the effect of the strong stratification is to force the unstable modes to become more horizontal and since their origin is the latitudinal gradient of $\Omega$, the instability survives even when the degree of stratification is increased. We note that the most unstable azimuthal wavenumber is still $m=1$ so that the strong stratification does not seem to significantly affect the azimuthal scale.

It is quite striking here that the growth rates of the unstable modes do not seem to be strongly affected by the stratification. Indeed, the growth rates of the cases R2, R3 and R4 are similar. Meanwhile, the ratio of the radial to the latitudinal wavenumbers increases with the increased stratification. This can be understood by the fact that the instability here is driven by the latitudinal (or horizontal) gradient of $\Omega$. 
Indeed, this behaviour is reminiscent of previously studied hydrodynamical instabilities driven by an horizontal shear in a vertically stratified medium. In the case of the centrifugal (or inertial) instability studied by \citet{kloosterziel2008}, the dispersion relation of the unstable modes clearly shows that 
the growth rate of a mode with given latitudinal and azimuthal wavenumbers can be made invariant to a stratification increase by
adapting (i.e. increasing) the vertical wavenumber accordingly. The possibility to adapt the vertical lengthscale to get the same growth rate also exists when the shear instability of the inflectional type \citep{deloncle2007}. 
The effect of the stable stratification on a vertical shear instability is very different. In an inviscid and adiabatic case there is simply no instability
when the Richardson number exceeds $1/4$, while a high thermal diffusivity can potentially destabilize predominantly horizontal perturbations but then the growth rates are vanishingly small \citep{lignieres99b}. A simple physical interpretation is
that the most unstable modes of a vertical shear necessarily involve vertical motions, such as for example in Kelvin-Helmholtz billows. Thus, by opposing vertical motions, stable stratification either kills the instability or reduces it strongly. On the contrary, for an horizontal shear, the stable stratification may affect the preferred vertical wavelength of the perturbation but this does not prevent the unaffected horizontal motions to efficiently draw energy from the horizontal shear.

While these purely hydrodynamical cases help interpret the effect of the stratification, both the centrifugal and the inflectional instabilities are absent from our simulations since the differential rotation does not fulfill the inviscid and unstratified criteria for these instabilities. We thus expect that the observed instability is a magnetorotational instability due to the latitudinal shear and supported by the magnetic field.
In the next section, we check the consistency of our interpretation using a local linear stability analysis in the MHD case.

\subsection{Comparison to the Acheson dispersion relation}
\label{sub_local}

We now wish to analyze our numerical results at the light of a local linear instability analysis, strongly inspired by the work of \cite{Acheson78} where various types of MHD instabilities in different regimes were investigated, as already quoted at the end of Section \ref{sub_r2c2}. We are particularly interested in the impact of stable stratification on our instabilities and on the differences found between the cylindrical and radial cases. We recall here the various steps of the establishment of the Acheson dispersion relation of interest in our case \citep[eq. 3.20 in][]{Acheson78} without indicating all the details, which can be found in Appendix \ref{sec_acheson}.

First, the MHD equations governing the system with thermal, viscous and magnetic diffusion are linearised
around the background axisymmetric state in cylindrical geometry (which is assumed to be purely toroidal both for the magnetic and the velocity fields) and, by considering small amplitude harmonic
perturbations in space and time of the form
\begin{equation}
\exp \left[\mi (\ks s+\kz z+m \varphi-\sigma\, t)\right]
\end{equation}

Here $\ks=2\pi/\wls$ ($\kz=2\pi/\wlz$) is the radial (axial) wavenumber
of the instability and $m$ its azimuthal order which is an $O(1)$ integer.
When the imaginary part of $\sigma$ is positive, the applied perturbation
is unstable and grows exponentially at a rate $\gr=\Im(\sigma)$.
Then, we assume here that the thermal diffusivity $\kappa$ is much higher than the magnetic diffusivity $\eta$, which is the case in our setup where $Pm=1$ and $Pr \ll 1$. In this situation, the dispersion relation of Acheson is reduced to a simpler expression: a polynomial equation of degree 4 in the dimensionless frequency  $\omegaND=\omega/\Omega_0$. 

We solve numerically that polynomial equation \ref{a:AchesonND} by choosing as background axisymmetric profiles our numerical solutions $\Omega(r,\theta)$ and $B_\varphi(r,\theta)$ at the time where the instability develops in the simulations. The various parameters defined in appendix B and which play a role in the calculation of the instability growth rate are the ratio of poloidal wavenumbers $\beta$, the azimuthal wavenumber $m$, the shear parameter $q$, a parameter $b$ quantifying the gradient of $B_\varphi$, the azimuthal Lorentz number $Lo_\varphi$, the stratification parameter $Pr N^2/\Omega_0^2$ and the Reynolds numbers $Re$, $Rm$ and $Rt$. For all these parameters, we take the values estimated or calculated from the simulations. With this procedure, we obtain a 2D map of the theoretical growth rate $\sigma (r,\theta)$ at the time where the instability starts to grow in the simulation. The aim is then to compare the location and the value of the maximum theoretical growth rate in the 2D domain with the location of the unstable mode and the growth rate estimated from the simulation.

\begin{figure*}[!h]
\begin{center}
 \includegraphics[width=0.4\textwidth]{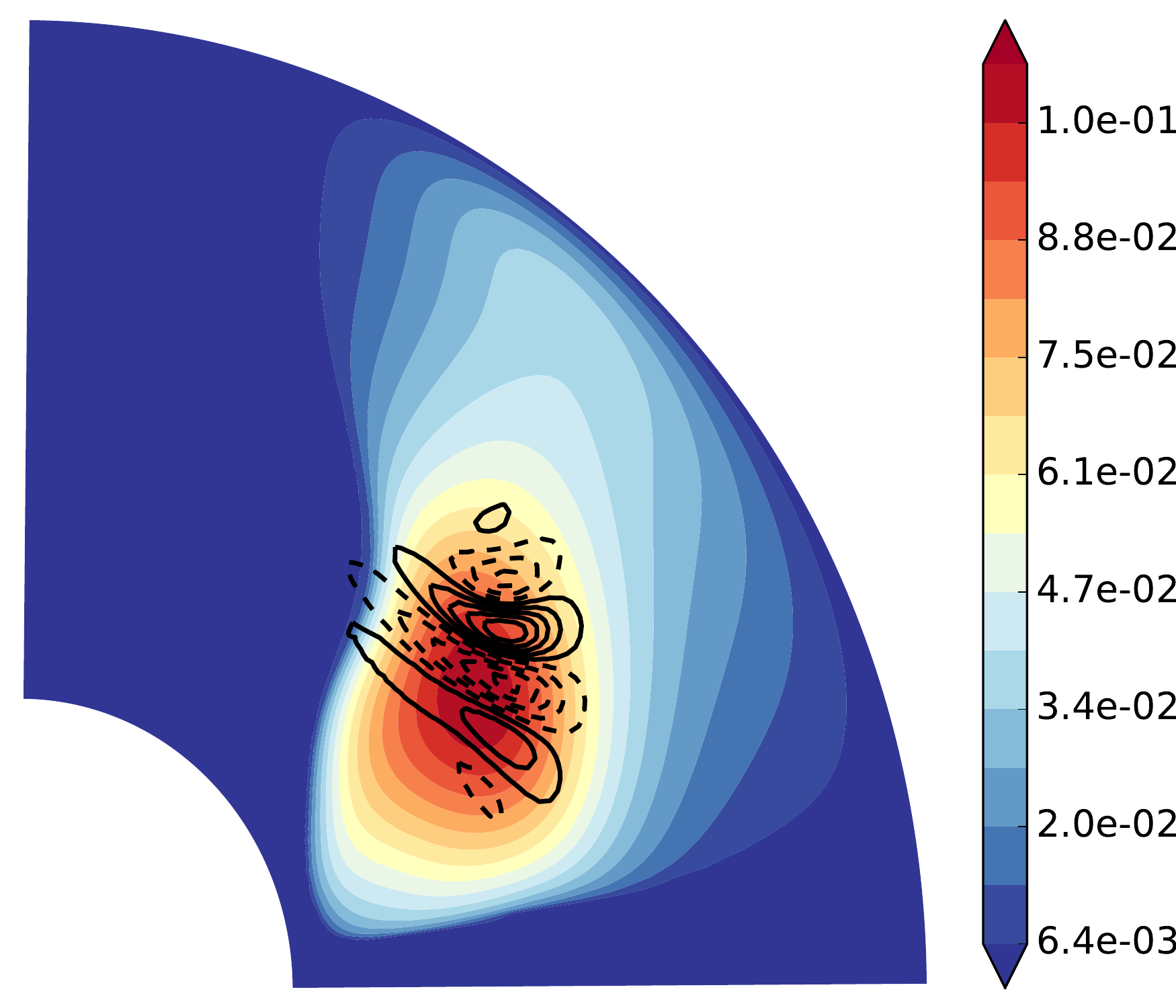}
  \includegraphics[width=0.4\textwidth]{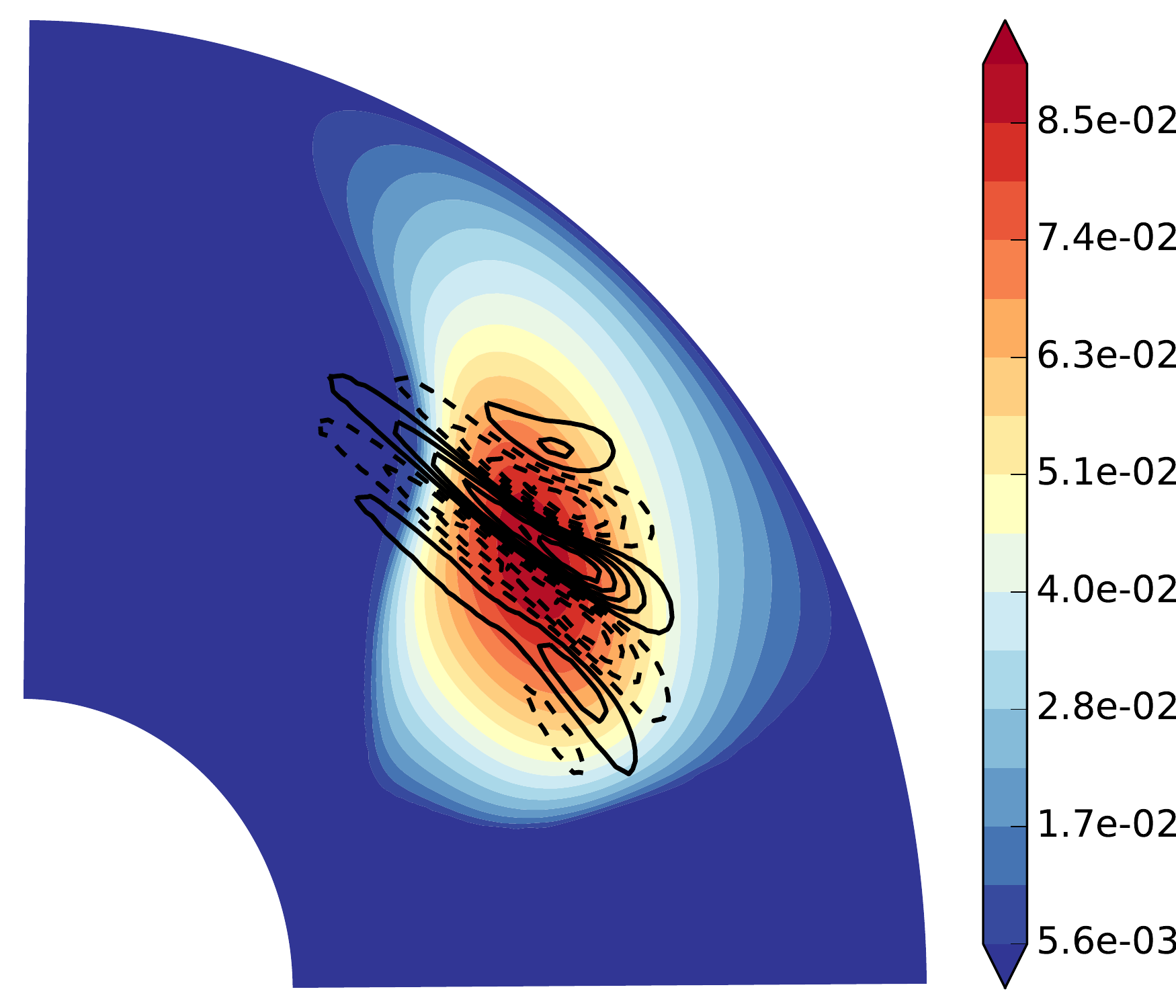}
  \caption{Cases R2 (left) and R3 (right): contours of the growth rate $\sigma/\Omega_0$ of the $m=1$ mode obtained through the Acheson dispersion relation (colors) and superimposed in black lines are the contours of the fluctuating radial magnetic field coming from the 3D simulation. The agreement for the location of the instability is quite satisfactory in both cases.}
 \label{fig_da_pr1e-2}
\end{center}
\end{figure*}

An example of such a map is given in Figure \ref{fig_da_pr1e-2}, where the azimuthal wavenumber was chosen to be $m=1$ and the ratio of poloidal wavenumbers such that $k_\theta \ll k_r$. This case corresponds to case R2 where the instability was clearly present in the numerical simulation and mostly on the $m=1$. On this map of $\sigma/\Omega_0$, we superimpose the isocontours of the fluctuating component of the radial magnetic field coming from the 3D simulation. We find that the location of the unstable mode coincides well with the theoretical location of the maximum growth rate. The value of the maximum growth rate reaches $\sigma/\Omega_0\approx 0.1$, compared to $2.5 \times 10^{-2}$ in the simulation. We do not expect to recover exactly the same growth rates because of the various assumptions underlying the derivation of the dispersion relation, which might not be entirely fulfilled in our simulations. In particular, the analysis of \citet{Acheson78} is local and we are comparing it here with global numerical simulations, with possible effects of the boundary conditions, especially for the cylindrical case where the instability develops very close to the top boundary of our computational domain. Then, as already pointed out in a previous work \citep{meduri2019}, the use of the short wavelength approximation (meridional perturbation wavelength much smaller than the typical scale of variation of the background) could also be questioned here, in particular for the radial direction. Anyhow, we do not try here to understand in detail the discrepancy in the values of the growth rate obtained in the local analysis and in the numerical simulations, we just aim at gaining some insight from the local analysis on the possible causes for instability observed in the simulations.

\begin{figure*}[!h]
\begin{center}
 \includegraphics[width=0.4\textwidth]{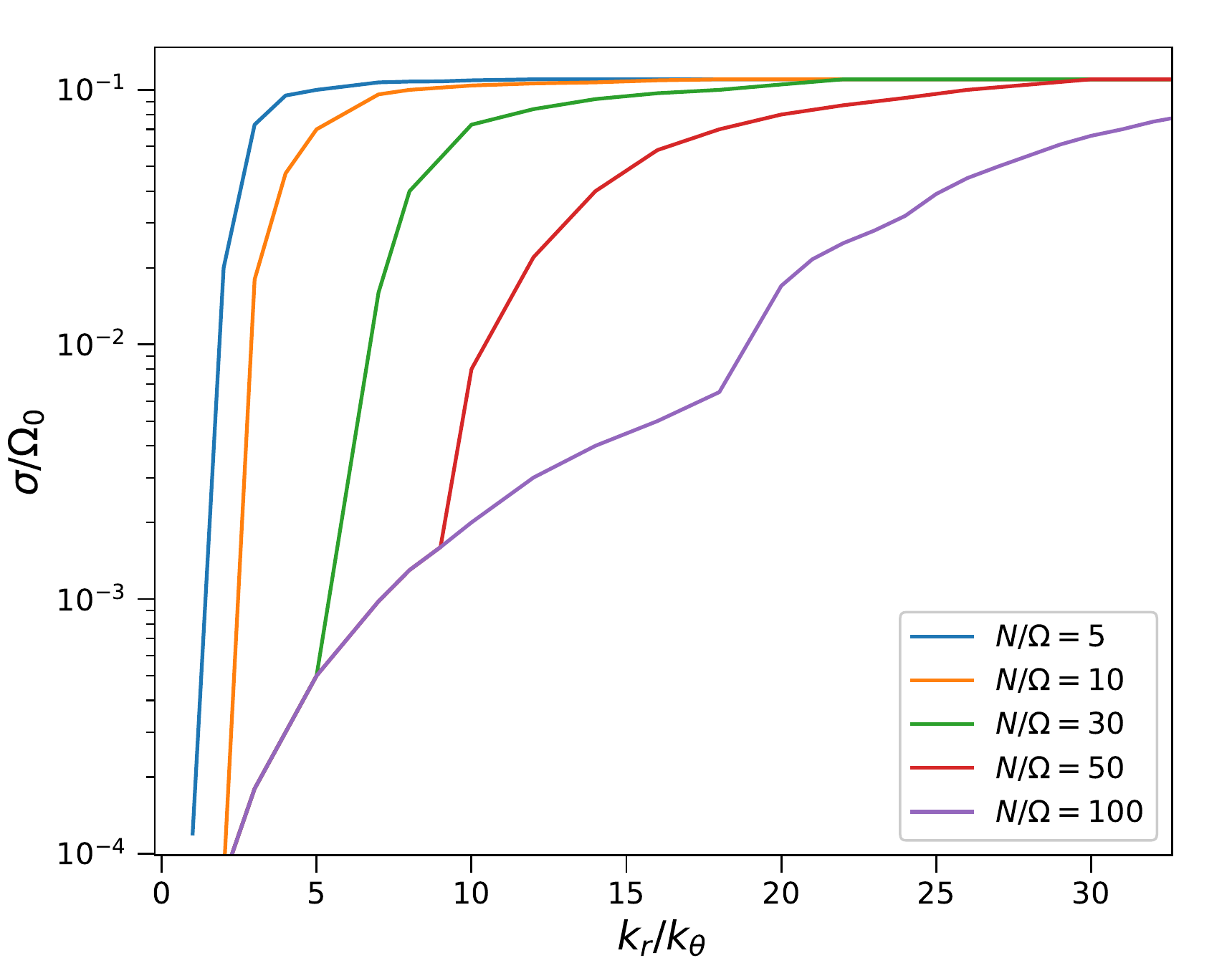}
  \caption{Spatial maximum of the instability growth rate for the $m=1$ mode calculated from equation \ref{a:AchesonND}, as a function of the ratio of poloidal wavenumbers, for different values of $N/\Omega_0$, keeping $Pr=10^{-2}$ and for the background azimuthal velocity and magnetic fields of case R2.}
 \label{fig_sigmabeta}
\end{center}
\end{figure*}

In the radial cases, the local dispersion relation helps us to determine that it is mostly the latitudinal gradient of $\Omega$ which is responsible for the instability and that the presence of the background magnetic field is needed. Indeed, the local analysis predicts that the background flow is hydrodynamically stable (the growth rate is  negative when the magnetic field is set to 0). Moreover, when the latitudinal gradient of $\Omega$ is set to 0, the growth rate drops dramatically while it stays around the same maximum value of $\sigma/\Omega_0\approx0.1$ when the radial gradient is set to 0. We thus confirm here the argument developed in the previous section: the instability is here driven by the gradient of rotation in the $\theta$-direction, i.e. orthogonal to the stable stratification. As discussed above, this is also probably the reason for the persistence of the instability when the stability of the stratification is increased. The right panel of Fig.\ref{fig_da_pr1e-2} shows the location of the instability when the effect of the stable stratification is increased, i.e. with $Pr=10^{-1}$ instead of $Pr=10^{-2}$. The local analysis still predicts a significant growth rate, again located around the maximum latitudinal gradient of rotation. The value of the growth rate itself is reduced by about 20\% but the instability still exists and indeed also observed in the 3D simulation at approximately the same location for the $m=1$ mode. In fact, in this case where we chose the poloidal wavenumber $\beta$ such that the displacement is nearly horizontal ($k_\theta \ll k_r$), it is expected that the local dispersion relation predicts a small effect of the stable stratification on the growth rate. Indeed, if we look at the coefficients of equation \ref{a:AchesonND}, we see that all the terms involving $Pr N^2/\Omega_0^2$ are multiplied by the quantity $\sin \theta - \beta \cos \theta$. In the limit case where $k_\theta$ is 0 and $k_r$ is large (but finite), $\beta$ reduces to $\tan \theta$ and the term multiplying $Pr N^2/\Omega_0^2$ vanishes. Of course, we are not really in this limit here but there can be a factor of at least 10 between the poloidal wavenumbers such that the effect of the stable stratification becomes very weak on the value of the growth rates. 

On Figure \ref{fig_sigmabeta}, we illustrate the fact that the linear analysis also predicts that the geometry of the unstable mode in the radial case should change as the stable stratification increases. The figure shows the maximum growth rate reached in the $(r,\theta)$ plane for the  background flow and field of case R2 as a function of the poloidal wavenumber ratio when the value of $N/\Omega_0$ is varied from 5 to 100. We clearly see that when the level of stratification is increased, the most unstable mode adapts its radial to horizontal wavenumber ratio: the most unstable mode becomes more and more horizontal when the stratification is increased, as also seen in the 3D simulation and as observed in the hydrodynamical studies discussed in the previous section. We also note that the maximum growth rate always tends to the same value as the level of stratification is increased, so that, theoretically, all modes with a sufficiently large radial to horizontal wavenumber ratio should be equally unstable. The unstable mode seen in the 3D simulation of course possesses a finite $k_r/k_\theta$, probably chosen to minimize the stable stratification effects while fitting in the extension of the background field. 

\begin{figure*}[!h]
\begin{center}
 \includegraphics[width=0.4\textwidth]{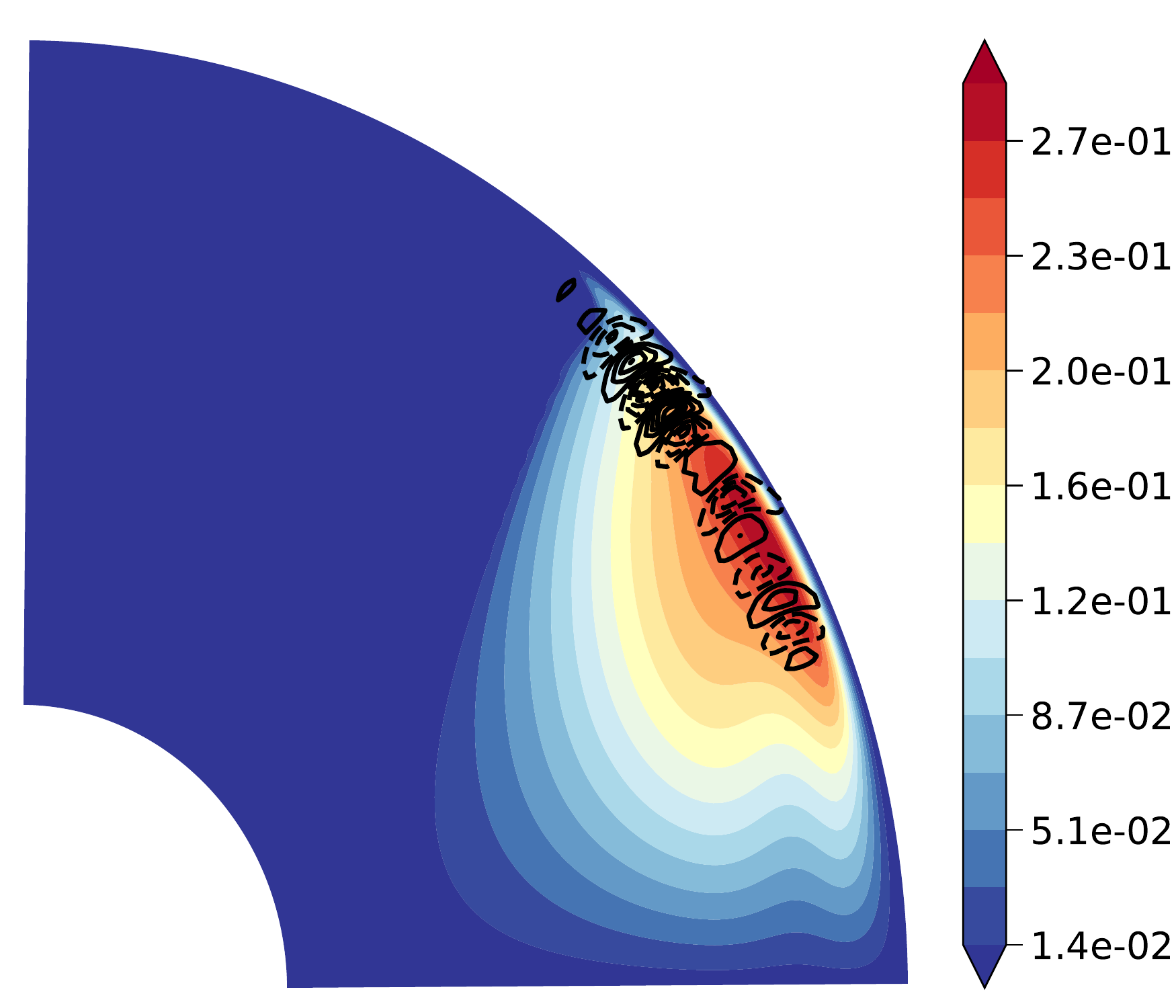}
  \caption{Same as Fig.\ref{fig_da_pr1e-2} but for Case C2 and for the $m=4$ mode. Again, the location of the instability in the simulation corresponds quite well with the position of the expected maximum growth rate from the local analysis. Case C3 is not shown since the local analysis does not predict any instability in this case, in agreement with the 3D simulation.}
 \label{fig_pr_cyl}
\end{center}
\end{figure*}

The situation is different in the cylindrical cases C2 and C3 where we chose, on the contrary, to calculate the growth rate as a function of $r$ and $\theta$ but using a ratio $\beta$ such that $k_\theta \gg k_r$, as seen in the 3D simulation. And we now choose to focus on the mode $m=4$ which is one of the most unstable ones. The results would be similar for the equally unstable $m=2$ and $m=3$ modes. Figure \ref{fig_pr_cyl} shows the map of the theoretical growth rate obtained from the dispersion relation for the $m=4$ mode, together with the contours of the radial component of the magnetic fluctuations coming from the 3D simulation. Again, the location of the maximum growth rate coincides quite well with the position where the instability is observed in the simulation but the expected growth rate is larger ($2.7\times 10^{-1}$ compared to $5\times 10^{-2}$ in the simulation). In this case, our procedure enables us to attest that it is now the radial gradient of $\Omega$ which is responsible for the instability found here, the growth rates (value and location) being very similar when the $\frac {\partial \Omega}{\partial \theta}$ is set to 0. When the Prandtl number is now increased to $Pr=0.1$, the instability completely vanishes, showing the very strong effect of the stable stratification on the instability in this cylindrical case, as observed in the simulation.

\subsection{Effect of $Lo$}
\label{sub_lo}

\begin{figure*}[!h]
\begin{center}
 \includegraphics[width=0.95\textwidth]{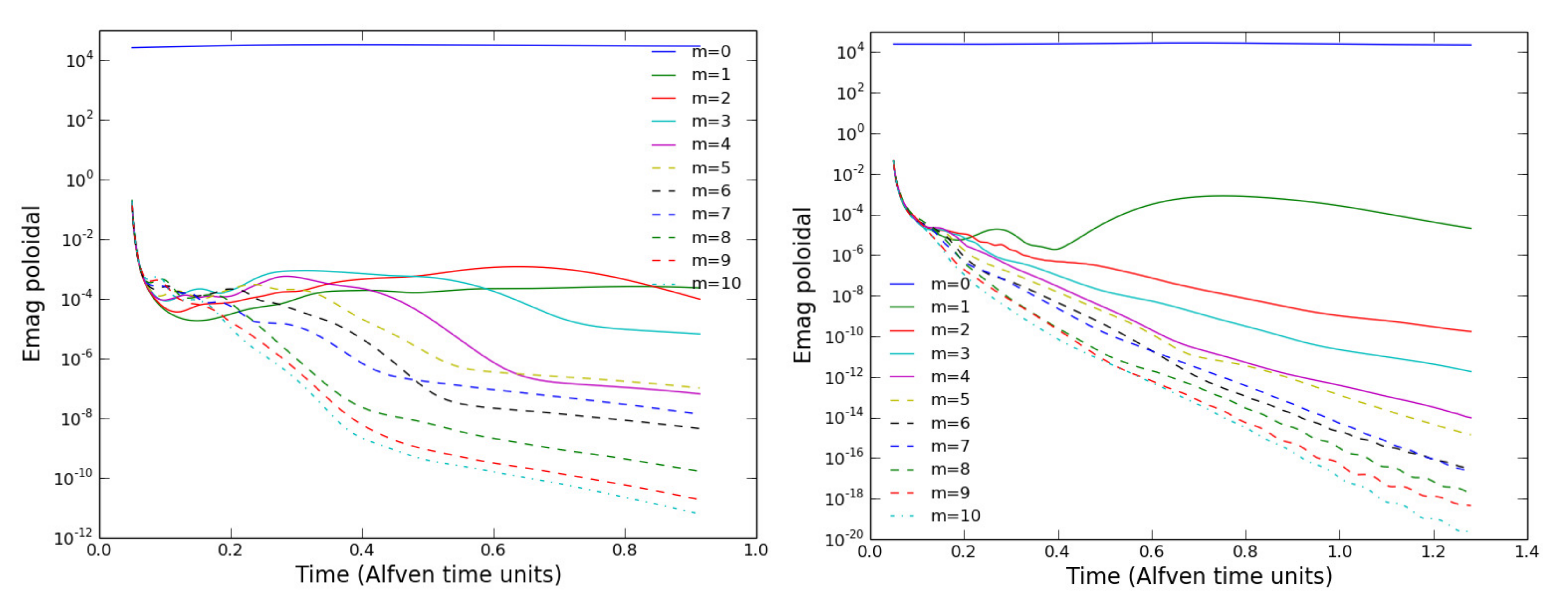}
  \caption{Temporal evolution of the poloidal magnetic energy in the first 11 azimuthal wavenumbers for case C9 (left) and R1 (right): compared to cases C2 and R2, the value of $Lo$ was increased to $Lo=10^{-2}$ for the cylindrical case and $Lo=2.5\times10^{-3}$ for the radial case.}
 \label{fig_lo}
\end{center}
\end{figure*}

In this last section, we investigate the effect of varying the Lorentz number, which measures the ratio between the dynamical timescales of interest in this study: the rotation time scale to the poloidal Alfv\'en time scale. From our previous study \citep{jouve2015}, we know that this parameter is crucial to the full development of the instability. Indeed, since we identify our instability here of the magneto-rotational type with a typical growth rate of the order of the rotation frequency, the Lorentz number quantifies the time it takes for the instability to grow compared to the typical lifetime of the background toroidal magnetic field on which it grows. The optimal case for the full development of the instability is consequently when the Lorentz number is small. To test this argument with the simulations performed in this work, we increased the Lorentz number both in the radial and the cylindrical cases. The growth in time of the magnetic energy contained in the first 11 azimuthal modes for an increased Lorentz number is shown in Figure \ref{fig_lo}, both for the cylindrical (left panel) and the radial case (right panel).

As expected, the main effect of increasing $Lo$ in both cases is to suppress the instability in the cylindrical case and drastically decrease its impact on the axisymmetric field in the radial case. To be more precise, the Lorentz number was increased here by decreasing the rotation rate and thus increasing the rotation time. The evolution of the axisymmetric magnetic field is then similar to what is shown in Fig.\ref{fig_magener} but with a smaller value for the ratio between toroidal and poloidal magnetic energies. In particular, the maximum toroidal field will still peak at approximately $t=0.7 t_{Ap}$ for the cylindrical case and at $t=0.3 t_{Ap}$ for the radial case, but the growth rates are divided by approximately 2 since the Lorentz number was doubled in case C9 and multiplied by 2.5 in case R1 compared to C2 and R2 respectively. As a consequence, the instability does not have time to sufficiently develop to reach the level of the axisymmetric field. The magnetic field after a few Alfv\'en times will remain mostly unaffected by the presence of non-axisymmetric components. 

The conclusion here is similar to the unstratified case \citep{jouve2015} and is still valid for the radial case. This is not surprising since the instability is also of MRI type and thus the ratio between the instability growth time and the background magnetic field lifetime will still control the ability of the non-axisymmetric unstable modes to reach the energy of the axisymmetric field. We then anticipate that for the mean axisymmetric field to be significantly modified by the development of the instability, the system must be at low Lorentz number, i.e. a relatively weak poloidal magnetic field embedded in a fastly rotating environment. To be more quantitative, in the cylindrical case, $Lo$ must be weaker than $5\times10^{-3}$ while in the radial case, it must be even weaker, of the order of $10^{-3}$. Both values are compatible with the values expected to be found in stellar interiors (see Sec.\ref{subpara}), especially for Vega-like stars which possess a weak surface field and a rapid rotation. The difference between the radial and cylindrical cases can be understood by the fact that for the instability to be triggered in the radial case, we first need to wait for the magnetic field to back-react on the flow to produce the latitudinal shear. The instability starts to develop when the toroidal field has already reached its maximum value and begins its decay. The instability thus needs to grow quite fast so that the toroidal field keeps approximately its maximum value during the whole development of the instability. The cylindrical case is different since the instability is able to grow right away on the existing radial shear and consequently while the toroidal field is building up. In the cylindrical case, the instability is thus allowed more time to grow and the range of Lorentz numbers allowing the instability to fully develop is thus extended.

\section{Conclusion}
\label{sec_conclu}

In this work, we studied the effects of the stable stratification in the non-adiabatic case on instabilities which can develop when an initial poloidal field is wound up by an initial differential rotation. Two different profiles for the differential rotation were considered, both likely to exist in stellar radiative zones: one, cylindrical, which satisfies the Taylor-Proudman constrain and the other, shellular, which corresponds to what could be expected in a strongly stably stratified layer. 

The axisymmetric solutions of this initial value problem were first investigated. We showed that, for fixed $Lu$ and $Pm$, the axisymmetric evolution depends only on one dimensionless parameter $Pr N^2/\Omega_0^2$ measuring the level of stratification, instead of the 3 independent parameters $L_o$, $Pr$ and $N/\Omega_0$. This result is found to be consistent with a scale analysis of the Boussinesq MHD equations performed for a time ordering $t_\nu \gg t_{Ap} \gg t_\kappa \gg t_\Omega \gg t_B$. In this simplified form, the gravity waves are filtered out, and the system evolves through Alfv\'en waves and an Eddington Sweet circulation prescribed by a magneto-thermal wind equilibrium and a thermal equilibrium. The parameter $Lo = t_\Omega/t_{Ap}$ only controls the ratio between toroidal and poloidal field. An interesting feature of the strongly stably stratified cases (with relatively high values of  $Pr N^2/\Omega_0^2$) is that the toroidal to poloidal field ratio becomes higher since the transport of angular momentum through meridional flows is inhibited. Indeed, in this situation, the initial differential rotation is not modified before the Lorentz force starts to back-react on the flow and the $\Omega$-effect is more efficient at producing a toroidal field component. 

In stars, the ordering in time-scales given above may not apply since the Alfv\'en time-scale may be small compared to the thermal diffusion time-scale. Then, the Eddington-Sweet circulation becomes negligible and the system evolution is dominated by Alfv\'en waves as in \cite{gaurat2015} where only the coupled evolution of $v_\varphi$ and $B_\varphi$ was analysed. In our calculations, we considered a situation where $t_{Ap} \sim t_\kappa$ and found that the gravity waves existing in the transient phase do persist during the whole winding-up process and significantly perturb the flow and field. However, this initial gravity wave transient is a direct consequence of our initial condition which is far from an equilibrium and such a transient is not likely to be present in stars. 

When axisymmetric solutions strongly dominated by their toroidal component exist, there are expected to be unstable. This is indeed what was found in \citet{jouve2015} in the non-stratified case. We tested here the effects of the stable stratification on the instability. It turns out that the situations involving two different initial differential rotation profiles respond quite differently to perturbations. When non-adiabatic effects are important, i.e. when a large thermal diffusivity is considered, both cases are unstable to a magneto-rotational instability. However, when the thermal diffusivity is reduced and thus when the effects of the stable stratification are increased, the instability disappears in the cylindrical case while the unstable displacements become more and more horizontal in the radial case, with similar growth rates. We argue that this is due to the fact that the radial shear is responsible for the instability in the cylindrical case while it is driven by the latitudinal shear in the other. This latitudinal shear does not exist initially, it is produced by the back-reaction of the magnetic field on the flow. The situation may appear quite specific since it is here the magnetic field itself which creates the conditions for its own instability. However, such phenomena could occur in stellar radiative zones where angular momentum is permanently redistributed by meridonal flows or Aflv\'en waves. In our case, the level of latitudinal shear which produces the instability does not need to be very high ($\Delta\Omega/\Omega < 1$) and can be localized in space. If such a gradient appears in a stellar radiative zone and persists for a few hundreds of rotation periods, we predict that an instability could develop and strongly modify the axisymmetric magnetic field despite the stable stratification. 

As far as Ap and Bp stars are concerned, we predict here that the instability could appear in stars for which the Lorentz number is less than $10^{-3}$, meaning that the Alfv\'en frequency should be 1000 times smaller than the rotation frequency. As argued in Sect. \ref{subpara}, small $Lo$ are indeed expected in stellar interiors especially for Vega-like stars which rotate rapidly and exhibit a small surface magnetic field. The Lorentz number is even smaller when deep layers of the stars are considered, where latitudinal shears could be locally generated and likely to be unstable. The existence of an instability for low $Lo$ stars would then possibly explain why strong fields are observed only for about $10\%$ of intermediate-mass and massive stars, these stars having potentially sufficiently high magnetic frequency compared to their rotation frequency so that the instability does not reach the level of the axisymmetric field. The present study also potentially applies to the angular momentum transport in evolved stars. Although the turbulent transport associated with the MRI is not quantified here, various studies \citep{rudiger2014, rudiger2015, jouve2015} have shown that the MRI of a toroidal field as seen here could produce a significant transport of angular momentum, which could possibly help to reconcile models and observations of the differential rotation of sub-giant and red giant stars observed with Kepler.

\acknowledgements

{The authors acknowledge financial support from the
Agence Nationale de la Recherche (ANR) through
the project IMAGINE (Investigating MAGnetism of
INtErmediate-mass and massive stars).
This work was granted access to the HPC resources of CALMIP supercomputing center under the allocation P1118. LJ acknowledges funding by the Institut Universitaire de France.}

\bibliographystyle{aa.bst}
\bibliography{jlg_2020} 

\begin{thebibliography}{61}
\expandafter\ifx\csname natexlab\endcsname\relax\def\natexlab#1{#1}\fi

\bibitem[{{Acheson}(1978)}]{Acheson78}
{Acheson}, D.~J. 1978, Philosophical Transactions of the Royal Society of
  London Series A, 289, 459

\bibitem[{{Augustson} {et~al.}(2016){Augustson}, {Brun}, \&
  {Toomre}}]{augustson2016}
{Augustson}, K.~C., {Brun}, A.~S., \& {Toomre}, J. 2016, \apj, 829, 92

\bibitem[{Auri{\`e}re {et~al.}(2007)Auri{\`e}re, Wade, Silvester,
  Ligni{\`e}res, {et~al.}}]{auriere2007}
Auri{\`e}re, M., Wade, G.-A., Silvester, J., Ligni{\`e}res, F., {et~al.} 2007,
  A\&A, 475, 1053

\bibitem[{{Balbus} \& {Hawley}(1992)}]{BH92}
{Balbus}, S.~A. \& {Hawley}, J.~F. 1992, ApJ, 400, 610

\bibitem[{{Blaz{\`e}re} {et~al.}(2016{\natexlab{a}}){Blaz{\`e}re}, {Neiner}, \&
  {Petit}}]{blazere2016a}
{Blaz{\`e}re}, A., {Neiner}, C., \& {Petit}, P. 2016{\natexlab{a}}, MNRAS, 459,
  L81

\bibitem[{{Blaz{\`e}re} {et~al.}(2016{\natexlab{b}}){Blaz{\`e}re}, {Petit},
  {Ligni{\`e}res}, {Auri{\`e}re}, {Ballot}, {B{\"o}hm}, {Folsom}, {Gaurat},
  {Jouve}, {Lopez Ariste}, {Neiner}, \& {Wade}}]{blazere2016b}
{Blaz{\`e}re}, A., {Petit}, P., {Ligni{\`e}res}, F., {et~al.}
  2016{\natexlab{b}}, A\&A, 586, A97

\bibitem[{{Braithwaite}(2006)}]{braithwaite2006b}
{Braithwaite}, J. 2006, A\&A, 449, 451

\bibitem[{{Brun} {et~al.}(2005){Brun}, {Browning}, \& {Toomre}}]{brun2005}
{Brun}, A.~S., {Browning}, M.~K., \& {Toomre}, J. 2005, \apj, 629, 461

\bibitem[{{Cantiello} {et~al.}(2014){Cantiello}, {Mankovich}, {Bildsten},
  {Christensen-Dalsgaard}, \& {Paxton}}]{cantiello2014}
{Cantiello}, M., {Mankovich}, C., {Bildsten}, L., {Christensen-Dalsgaard}, J.,
  \& {Paxton}, B. 2014, ApJ, 788, 93

\bibitem[{{Ceillier} {et~al.}(2013){Ceillier}, {Eggenberger}, {Garc{\'{\i}}a},
  \& {Mathis}}]{ceillier2013}
{Ceillier}, T., {Eggenberger}, P., {Garc{\'{\i}}a}, R.~A., \& {Mathis}, S.
  2013, A\&A, 555, A54

\bibitem[{{Chandrasekhar}(1960)}]{chandra60}
{Chandrasekhar}, S. 1960, Proceedings of the National Academy of Science, 46,
  253

\bibitem[{{Charbonneau} \& {MacGregor}(1992)}]{charbonneau92}
{Charbonneau}, P. \& {MacGregor}, K.~B. 1992, \apj, 387, 639

\bibitem[{{Deheuvels} {et~al.}(2014){Deheuvels}, {Do{\u g}an}, {Goupil},
  {Appourchaux}, {Benomar}, {Bruntt}, {Campante}, {Casagrande}, {Ceillier},
  {Davies}, {De Cat}, {Fu}, {Garc{\'{\i}}a}, {Lobel}, {Mosser}, {Reese},
  {Regulo}, {Schou}, {Stahn}, {Thygesen}, {Yang}, {Chaplin},
  {Christensen-Dalsgaard}, {Eggenberger}, {Gizon}, {Mathis},
  {Molenda-{\.Z}akowicz}, \& {Pinsonneault}}]{deheuvels2014}
{Deheuvels}, S., {Do{\u g}an}, G., {Goupil}, M.~J., {et~al.} 2014, A\&A, 564,
  A27

\bibitem[{{Deheuvels} {et~al.}(2012){Deheuvels}, {Garc{\'{\i}}a}, {Chaplin},
  {Basu}, {Antia}, {Appourchaux}, {Benomar}, {Davies}, {Elsworth}, {Gizon},
  {Goupil}, {Reese}, {Regulo}, {Schou}, {Stahn}, {Casagrande},
  {Christensen-Dalsgaard}, {Fischer}, {Hekker}, {Kjeldsen}, {Mathur}, {Mosser},
  {Pinsonneault}, {Valenti}, {Christiansen}, {Kinemuchi}, \&
  {Mullally}}]{deheuvels2012}
{Deheuvels}, S., {Garc{\'{\i}}a}, R.~A., {Chaplin}, W.~J., {et~al.} 2012, ApJ,
  756, 19

\bibitem[{{Deloncle} {et~al.}(2007){Deloncle}, {Chomaz}, \&
  {Billant}}]{deloncle2007}
{Deloncle}, A., {Chomaz}, J.-M., \& {Billant}, P. 2007, Journal of Fluid
  Mechanics, 570, 297

\bibitem[{{Dudis}(1974)}]{dudis74}
{Dudis}, J.~J. 1974, Journal of Fluid Mechanics, 64, 65

\bibitem[{{Eggenberger} {et~al.}(2019){Eggenberger}, {den Hartogh}, {Buldgen},
  {Meynet}, {Salmon}, \& {Deheuvels}}]{eggenberger2019b}
{Eggenberger}, P., {den Hartogh}, J.~W., {Buldgen}, G., {et~al.} 2019, \aap,
  631, L6

\bibitem[{{Eggenberger} {et~al.}(2012{\natexlab{a}}){Eggenberger},
  {Haemmerl{\'e}}, {Meynet}, \& {Maeder}}]{eggenberger2012a}
{Eggenberger}, P., {Haemmerl{\'e}}, L., {Meynet}, G., \& {Maeder}, A.
  2012{\natexlab{a}}, A\&A, 539, A70

\bibitem[{{Eggenberger} {et~al.}(2012{\natexlab{b}}){Eggenberger},
  {Montalb{\'a}n}, \& {Miglio}}]{eggenberger2012b}
{Eggenberger}, P., {Montalb{\'a}n}, J., \& {Miglio}, A. 2012{\natexlab{b}},
  A\&A, 544, L4

\bibitem[{{Fuller} {et~al.}(2019){Fuller}, {Piro}, \& {Jermyn}}]{fuller2019}
{Fuller}, J., {Piro}, A.~L., \& {Jermyn}, A.~S. 2019, \mnras, 485, 3661

\bibitem[{{Garaud} \& {Acevedo Arreguin}(2009)}]{garaud2009}
{Garaud}, P. \& {Acevedo Arreguin}, L. 2009, \apj, 704, 1

\bibitem[{{Garaud} {et~al.}(2015){Garaud}, {Medrano}, {Brown}, {Mankovich}, \&
  {Moore}}]{garaud2015}
{Garaud}, P., {Medrano}, M., {Brown}, J.~M., {Mankovich}, C., \& {Moore}, K.
  2015, \apj, 808, 89

\bibitem[{{Gastine} \& {Wicht}(2012)}]{Gastine12}
{Gastine}, T. \& {Wicht}, J. 2012, \icarus, 219, 428

\bibitem[{{Gaurat} {et~al.}(2015){Gaurat}, {Jouve}, {Ligni{\`e}res}, \&
  {Gastine}}]{gaurat2015}
{Gaurat}, M., {Jouve}, L., {Ligni{\`e}res}, F., \& {Gastine}, T. 2015, A\&A,
  580, A103

\bibitem[{{Gilman} \& {Glatzmaier}(1981)}]{GG81}
{Gilman}, P.~A. \& {Glatzmaier}, G.~A. 1981, ApJS, 45, 335

\bibitem[{{Guerrero} {et~al.}(2019){Guerrero}, {Del Sordo}, {Bonanno}, \&
  {Smolarkiewicz}}]{Guerrero2019}
{Guerrero}, G., {Del Sordo}, F., {Bonanno}, A., \& {Smolarkiewicz}, P.~K. 2019,
  \mnras, 490, 4281

\bibitem[{{Guervilly} \& {Cardin}(2010)}]{guervilly2010}
{Guervilly}, C. \& {Cardin}, P. 2010, Geophysical and Astrophysical Fluid
  Dynamics, 104, 221

\bibitem[{{Hale}(1908)}]{hale08}
{Hale}, G.~E. 1908, ApJ, 28, 315

\bibitem[{Ionson(1978)}]{ionson78}
Ionson, J.-A. 1978, ApJ, 226, 650

\bibitem[{{Jouve} {et~al.}(2015){Jouve}, {Gastine}, \&
  {Ligni{\`e}res}}]{jouve2015}
{Jouve}, L., {Gastine}, T., \& {Ligni{\`e}res}, F. 2015, A\&A, 575, A106

\bibitem[{{Kitchatinov} \& {R{\"u}diger}(2008)}]{Kitchatinov08}
{Kitchatinov}, L. \& {R{\"u}diger}, G. 2008, \aap, 478, 1

\bibitem[{{Kloosterziel} \& {Carnevale}(2008)}]{kloosterziel2008}
{Kloosterziel}, R.~C. \& {Carnevale}, G.~F. 2008, Journal of Fluid Mechanics,
  594, 249

\bibitem[{{Knobloch} \& {Spruit}(1982)}]{knobloch82}
{Knobloch}, E. \& {Spruit}, H.~C. 1982, \aap, 113, 261

\bibitem[{{Ligni{\`e}res} {et~al.}(1999){Ligni{\`e}res}, {Califano}, \&
  {Mangeney}}]{lignieres99b}
{Ligni{\`e}res}, F., {Califano}, F., \& {Mangeney}, A. 1999, \aap, 349, 1027

\bibitem[{Ligni{\`e}res {et~al.}(2009)Ligni{\`e}res, Petit, B{\"o}hm, \&
  Auri{\`e}re}]{lignieres2009}
Ligni{\`e}res, F., Petit, P., B{\"o}hm, T., \& Auri{\`e}re, M. 2009, A\&A, 500,
  L41

\bibitem[{{Marcotte} \& {Gissinger}(2016)}]{marcotte2016}
{Marcotte}, F. \& {Gissinger}, C. 2016, Physical Review Fluids, 1, 063602

\bibitem[{{Markey} \& {Tayler}(1973)}]{markey73}
{Markey}, P. \& {Tayler}, R.~J. 1973, \mnras, 163, 77

\bibitem[{{Marques} {et~al.}(2013){Marques}, {Goupil}, {Lebreton}, {Talon},
  {Palacios}, {Belkacem}, {Ouazzani}, {Mosser}, {Moya}, {Morel}, {Pichon},
  {Mathis}, {Zahn}, {Turck-Chi{\`e}ze}, \& {Nghiem}}]{marques2013}
{Marques}, J.~P., {Goupil}, M.~J., {Lebreton}, Y., {et~al.} 2013, A\&A, 549,
  A74

\bibitem[{{Meduri} {et~al.}(2019){Meduri}, {Ligni{\`e}res}, \&
  {Jouve}}]{meduri2019}
{Meduri}, D.~G., {Ligni{\`e}res}, F., \& {Jouve}, L. 2019, \pre, 100, 013110

\bibitem[{{Moffatt}(1978)}]{moffatt78}
{Moffatt}, H.~K. 1978, {Magnetic field generation in electrically conducting
  fluids}

\bibitem[{{Parker}(1955)}]{parker55}
{Parker}, E.~N. 1955, \apj, 122, 293

\bibitem[{{Philidet} {et~al.}(2019){Philidet}, {Gissinger}, {Ligni{\`e}res}, \&
  {Petitdemange}}]{Philidet2019}
{Philidet}, J., {Gissinger}, C., {Ligni{\`e}res}, F., \& {Petitdemange}, L.
  2019, arXiv e-prints, arXiv:1910.04092

\bibitem[{Pitts \& Tayler(1985)}]{PT85}
Pitts, E. \& Tayler, R.-J. 1985, MNRAS, 216, 139

\bibitem[{{R{\"u}diger} {et~al.}(2018){R{\"u}diger}, {Gellert}, {Hollerbach},
  {Schultz}, \& {Stefani}}]{rudiger2018}
{R{\"u}diger}, G., {Gellert}, M., {Hollerbach}, R., {Schultz}, M., \&
  {Stefani}, F. 2018, \physrep, 741, 1

\bibitem[{R{\"u}diger {et~al.}(2014)R{\"u}diger, Gellert, Schultz, Hollerbach,
  \& Stefani}]{rudiger2014}
R{\"u}diger, G., Gellert, M., Schultz, M., Hollerbach, R., \& Stefani, F. 2014,
  MNRAS, 438, 271

\bibitem[{{R{\"u}diger} {et~al.}(2015){R{\"u}diger}, {Gellert}, {Spada}, \&
  {Tereshin}}]{rudiger2015}
{R{\"u}diger}, G., {Gellert}, M., {Spada}, F., \& {Tereshin}, I. 2015, A\&A,
  573, A80

\bibitem[{{R{\"u}diger} \& {Kitchatinov}(2010)}]{Rudiger10}
{R{\"u}diger}, G. \& {Kitchatinov}, L.~L. 2010, Geophysical and Astrophysical
  Fluid Dynamics, 104, 273

\bibitem[{{R{\"u}diger} {et~al.}(2016){R{\"u}diger}, {Schultz}, \&
  {Kitchatinov}}]{Rudiger16}
{R{\"u}diger}, G., {Schultz}, M., \& {Kitchatinov}, L.~L. 2016, \mnras, 456,
  3004

\bibitem[{{Schaeffer}(2013)}]{Schaeffer13}
{Schaeffer}, N. 2013, Geochemistry, Geophysics, Geosystems, 14, 751

\bibitem[{{Spiegel} \& {Zahn}(1992)}]{Spiegel1992}
{Spiegel}, E.~A. \& {Zahn}, J.~P. 1992, \aap, 265, 106

\bibitem[{Spruit(1999)}]{spruit99}
Spruit, H.-C. 1999, A\&A, 349, 189

\bibitem[{Spruit(2002)}]{spruit2002}
Spruit, H.-C. 2002, A\&A, 381

\bibitem[{{Szklarski} \& {Arlt}(2013)}]{Szklarski13}
{Szklarski}, J. \& {Arlt}, R. 2013, \aap, 550, A94

\bibitem[{{Talon} \& {Charbonnel}(2008)}]{talon2008}
{Talon}, S. \& {Charbonnel}, C. 2008, \aap, 482, 597

\bibitem[{Tayler(1973)}]{tayler73}
Tayler, R.-J. 1973, MNRAS, 161, 365

\bibitem[{{Townsend}(1958)}]{townsend58}
{Townsend}, A.~A. 1958, Journal of Fluid Mechanics, 4, 361

\bibitem[{{Vallis}(2006)}]{Vallis2006}
{Vallis}, G.~K. 2006, {Atmospheric and Oceanic Fluid Dynamics} (Cambridge
  University Press)

\bibitem[{Velikhov(1959)}]{velikhov59}
Velikhov, E.~P. 1959, Sov. Phys. JETP, 36, 1398

\bibitem[{{Wicht}(2002)}]{Wicht2002}
{Wicht}, J. 2002, Physics of the Earth and Planetary Interiors, 132, 281

\bibitem[{Zahn(1992)}]{zahn92}
Zahn, J.-P. 1992, A\&A, 265, 115

\bibitem[{{Zahn} {et~al.}(2007){Zahn}, {Brun}, \& {Mathis}}]{zahn07}
{Zahn}, J.-P., {Brun}, A.~S., \& {Mathis}, S. 2007, \aap, 474, 145

\end{thebibliography}

\appendix
\onecolumn

\section{Full set of non-axisymmetric MHD Boussinesq equations}
\label{sec_eqnonaxi}

We give in this appendix the full set of non-axisymmetric Boussinesq equations solved in this work, using the adimensionalisation detailed in the main text in Section \ref{sec_model}.

For the 3 components of the velocity field, separating the toroidal and poloidal dynamics, the equations read:

    \begin{equation}
     \begin{split}
     \label{adim_vr}
     \vphantom{\left(\frac{B^2}{1}\right)}\frac{\partial \widetilde{v}_r}{\partial \widetilde{t}}&+\widetilde{\vec{v}}_{\vec{m}}\cdot\widetilde{\vec{\nabla}}\widetilde{v}_r+ \frac{1}{Lo}\frac{\widetilde{v}_\varphi}{\widetilde{r}\sin\theta}\frac{\partial \widetilde{v}_r}{\partial \varphi}-\frac{\widetilde{v}^2_\theta}{\widetilde{r}}-\frac{1}{Lo^2}\frac{\widetilde{v}^2_\varphi}{\widetilde{r}}=\frac{2}{Lo^2}\sin\theta\,\widetilde{v}_\varphi\\
     &+\frac{1}{Lo^2}\left(\frac{N}{\Omega_0}\right)^2\widetilde{T}_1-\frac{1}{Lo^2}\frac{\partial\widetilde{p}_1}{\partial\widetilde{r}}+\frac{\widetilde{B}_\theta}{\widetilde{r}}\left(\frac{\partial\widetilde{B}_r}{\partial\theta}-\frac{\partial\,(\widetilde{r}\widetilde{B}_\theta)}{\partial\widetilde{r}}\right)\\
     &-\frac{1}{Lo^2}\frac{\widetilde{B}_\varphi}{\widetilde{r}}\frac{\partial\,(\widetilde{r}\widetilde{B}_\varphi)}{\partial\widetilde{r}}+\frac{1}{Lo}\frac{\widetilde{B}_\varphi}{\widetilde{r}\sin\theta}\frac{\partial \widetilde{B}_r}{\partial \varphi}+\frac{Pm}{Lu}\left.\widetilde{\Delta}\widetilde{\vec{v}}\right|_r\ \ \ ,\\
     \end{split}     
    \end{equation}
    
    \begin{equation}
     \begin{split}
     \label{adim_vtheta}
     \vphantom{\left(\frac{B^2}{1}\right)}\frac{\partial \widetilde{v}_\theta}{\partial \widetilde{t}}&+\widetilde{\vec{v}}_{\vec{m}}\cdot\widetilde{\vec{\nabla}}\widetilde{v}_\theta+ \frac{1}{Lo}\frac{\widetilde{v}_\varphi}{\widetilde{r}\sin\theta}\frac{\partial \widetilde{v}_\theta}{\partial \varphi}+\frac{\widetilde{v}_\theta\widetilde{v}_r}{\widetilde{r}}-\frac{1}{Lo^2}\cot\theta\frac{\widetilde{v}^2_\varphi}{\widetilde{r}}=\frac{2}{Lo^2}\cos\theta\,\widetilde{v}_\varphi\\
     &-\frac{1}{Lo^2}\frac{1}{\widetilde{r}}\frac{\partial\widetilde{p}_1}{\partial\theta}+\frac{\widetilde{B}_r}{\widetilde{r}}\left(\frac{\partial\,(\widetilde{r}\widetilde{B}_\theta)}{\partial\widetilde{r}}-\frac{\partial\widetilde{B}_r}{\partial\theta}\right)\\
     &-\frac{1}{Lo^2}\frac{\widetilde{B}_\varphi}{\widetilde{r}\sin\theta}\frac{\partial\,(\sin\theta\widetilde{B}_\varphi)}{\partial\theta}+\frac{1}{Lo}\frac{\widetilde{B}_\varphi}{\widetilde{r}\sin\theta}\frac{\partial \widetilde{B}_\theta}{\partial \varphi}+\frac{Pm}{Lu}\left.\widetilde{\Delta}\widetilde{\vec{v}}\right|_\theta\ \ \ ,\\
     \end{split}     
    \end{equation}
    
    \begin{equation}
     \begin{split}
     \label{adim_vphi}
     \vphantom{\left(\frac{B^2}{1}\right)}\frac{\partial \widetilde{v}_\varphi}{\partial \widetilde{t}}+\frac{1}{\widetilde{r}\sin\theta}\left(\widetilde{\vec{v}}_{\vec{m}}\cdot\widetilde{\vec{\nabla}}\right)\left(\widetilde{r}\sin\theta\widetilde{v}_\varphi\right)+ \frac{1}{Lo^2}\frac{\widetilde{v}_\varphi}{\widetilde{r}\sin\theta}\frac{\partial \widetilde{v}_\varphi}{\partial \varphi}=-2\left(\sin\theta\,\widetilde{v}_r+\cos\theta\,\widetilde{v}_\theta\right)-\frac{1}{\widetilde{r}\sin\theta}\frac{\partial\widetilde{p}_1}{\partial\varphi}\\
     +\frac{1}{\widetilde{r}\sin\theta}\left(\widetilde{\vec{B}}_{\vec{p}}\cdot\widetilde{\vec{\nabla}}\right)\left(\widetilde{r}\sin\theta\widetilde{B}_{\varphi}\right)-Lo\frac{\widetilde{B}_\theta}{\widetilde{r}\sin\theta}\frac{\partial \widetilde{B}_\theta}{\partial \varphi}-Lo\frac{\widetilde{B}_r}{\widetilde{r}\sin\theta}\frac{\partial \widetilde{B}_r}{\partial \varphi} +\frac{Pm}{Lu}\left.\widetilde{\Delta}\widetilde{\vec{v}}\right|_{\varphi}\ \ \ ,\\
    \end{split}     
    \end{equation}
    
    The equations for the 3 components of the magnetic field then read:

        \begin{equation}
     \begin{split}
     \label{adim_br}
    \vphantom{\left(\frac{B^2}{1}\right)}\frac{\partial \widetilde{B}_r}{\partial \widetilde{t}}&=\frac{1}{\widetilde{r}\sin\theta}\frac{\partial}{\partial \theta} \left( \sin\theta \widetilde{v}_r\widetilde{B}_\theta- \sin\theta \widetilde{v}_\theta\widetilde{B}_r  \right) \\
     & -\frac{1}{Lo}\frac{1}{\widetilde{r}\sin\theta}\frac{\partial}{\partial \varphi} \left( \widetilde{v}_\varphi \widetilde{B}_r-\widetilde{v}_r\widetilde{B}_\varphi  \right)+\frac{1}{Lu}\left.\widetilde{\Delta}\widetilde{\vec{B}}\right|_{r}
    \end{split}     
    \end{equation}
    
            \begin{equation}
     \begin{split}
     \label{adim_bt}
    \vphantom{\left(\frac{B^2}{1}\right)}\frac{\partial \widetilde{B}_\theta}{\partial \widetilde{t}}&=-\frac{1}{\widetilde{r}}\frac{\partial}{\partial \theta} \left( \widetilde{r} \widetilde{v}_r\widetilde{B}_\theta- \widetilde{r} \widetilde{v}_\theta \widetilde{B}_r  \right) \\
     & +\frac{1}{Lo}\frac{1}{\widetilde{r}\sin\theta}\frac{\partial}{\partial \varphi} \left( \widetilde{v}_\theta \widetilde{B}_\varphi-\widetilde{v}_\varphi\widetilde{B}_\theta  \right)+\frac{1}{Lu}\left.\widetilde{\Delta}\widetilde{\vec{B}}\right|_{\theta}
    \end{split}     
    \end{equation}
    
            \begin{equation}
     \begin{split}
     \label{adim_bp}
    \vphantom{\left(\frac{B^2}{1}\right)}\frac{\partial \widetilde{B}_\varphi}{\partial \widetilde{t}}&=\frac{1}{\widetilde{r}}\frac{\partial}{\partial \widetilde{r}} \left( \widetilde{r} \widetilde{v}_\varphi\widetilde{B}_r- \widetilde{r} \widetilde{v}_r\widetilde{B}_\varphi  \right) \\
     & -\frac{1}{\widetilde{r}}\frac{\partial}{\partial \theta} \left( \widetilde{v}_\theta \widetilde{B}_\varphi-\widetilde{v}_\varphi\widetilde{B}_\theta  \right)+\frac{1}{Lu}\left.\widetilde{\Delta}\widetilde{\vec{B}}\right|_{\varphi}
    \end{split}     
    \end{equation}

And finally the temperature equation reads:

    \begin{equation}
     \label{adim_energie}
     \vphantom{\left(\frac{B^2}{1}\right)}\frac{\partial\widetilde{T}_1}{\partial \widetilde{t}}+\widetilde{\vec{v}}_{\vec{m}}\cdot\widetilde{\vec{\nabla}}\widetilde{T}_1+\frac{1}{Lo}\frac{\widetilde{v}_\varphi}{\widetilde{r}\sin\theta}\frac{\partial \widetilde{T_1}}{\partial \varphi}+\widetilde{v_r}\frac{\partial \overline{T}}{\partial \widetilde{r}}=\frac{Pm}{Lu Pr}\widetilde{\Delta} \widetilde{T}_1\ \ \ ,\\     
    \end{equation}

where $\vec{v_m}=v_r\vec{e_r}+v_\theta\vec{e_\theta}$ is the meridional velocity field and $\vec{B_p}=B_r\vec{e_r}+B_\theta\vec{e_\theta}$ is the poloidal magnetic field.
The tildes indicate the dimensionless quantities. We note that the choice of reference scales in this appendix is slightly different from the one chosen in the next appendix where a scaling analysis of the axisymmetric version of the equations is performed. The variables with a tilde in this appendix are thus different from the tilde-variables of appendix \ref{sec:edd}.

\section{Scaling analysis of the axisymmetric MHD Boussinesq equations}
\label{sec:edd}
    
In the following we present a scale analysis of the axisymmetric MHD equations with the aim of finding a simplified form of these equations that approximates the evolution of our system.  We note that the choice of reference scales to make the axisymmetric equations dimensionless will be slightly different in this appendix than the choice given in Sect. \ref{sec_model} which enabled to produce the full non-axisymmetric set of equations of appendix \ref{sec_eqnonaxi}.

The initial conditions provide the characteristic magnitude of some variables :  the poloidal field $B_0$, the rotation rate $\Omega_0$,  the stable stratification $N=\sqrt{\alpha g_0 \frac{\Delta T}{d}}$, the azimuthal velocity $U^*_{\varphi} = d \Delta \Omega$, the domain size and also the lengthscale of the initial gradients $d=r_o-r_i$. We restrict our analysis to the regime $t_{Ap} = \frac{d \sqrt{4 \pi \rho}}{B_0} \gg t_\Omega = \frac{1}{\Omega_0}$. The toroidal field has no initially prescribed amplitude and there is no physical reason to choose $B_0$. We anticipate instead that a characteristic amplitude is $B^*_\varphi = d \Delta \Omega \sqrt{4 \pi \rho}$,  the magnetic field resulting from the winding-up of the initial poloidal field by the differential rotation $\Delta \Omega$ over an Alfv\'en time $t_{Ap}$. We also need to choose a typical amplitude for the meridional motions $U_m$. Due to the strong stable stratification $N \ge \Omega_0$, we argue that $U_m$ should be small because radial motions are efficiently limited and the mass conservation ensures that latitudinal velocities are of the same order as radial velocities, $v_\theta \sim v_r$. In practice assuming $ U_m \ll d \Delta \Omega \lesssim d \Omega_0$ allows us to simplify the system of equation and to obtain $U_m$ as a result of the scale analysis. The consistency of the assumption $U_m \ll d \Delta \Omega \lesssim d \Omega_0$ is verified
afterwards.
As demonstrated below, such small meridional velocities lead to a thermal-wind balance which in turn determines a typical amplitude
for the temperature fluctuations, $T^* = \frac{\Omega_0^2}{N^2} \frac{\Delta \Omega}{\Omega_0} \Delta T$, and the pressure fluctuations $P^* = \rho d^2\Omega_0 \Delta \Omega$. Finally, as we are interested in the evolution of the angular momentum, the characteristic time scale is chosen from the equation governing this evolution :

\begin{equation}
\label{_AM}
\frac{\partial M}{\partial t}+\left(\vec{v}_m \cdot \vec{\nabla}\right) M =
    \frac{1}{4 \pi \rho}\left(\vec{B}_{\vec{p}} \cdot \vec{\nabla}\right)\left(r\sin\theta B_\varphi\right) + r \sin \theta \nu \left( \Delta - \frac{1}{r^2 \sin^2\theta} \right) v_{\varphi}
    \end{equation}
    
\noindent where  $M = r^2 \sin^2\theta \Omega_0 + r \sin\theta v_\varphi$ is the specific angular momentum. The meridional velocity, $\vec{v}_m=v_r\vec{e_r}+v_\theta\vec{e_\theta}$, advects the angular momentum on a time scale  $t_c = (\Delta \Omega/ \Omega) (d/ U_m)$, where the factor $\Ro = \frac{\Delta \Omega}{\Omega}$  accounts for the effect of the Coriolis force that speed-up the transport when $\Ro < 1$. In our simulations, the initial differential rotation is such that  $\Ro \sim 1$ while in the following  we consider more generally $\Ro \le 1$ regimes.  The other time scale that controls the angular momentum evolution is the poloidal Alfv\'en time $t_{Ap}$ as the time over which the toroidal field produced by the $\Omega$-effect  back reacts onto the rotation. The third time scale is the viscous time $t_\nu$ and it is supposed to be always larger than $t_{Ap}$. Consequently, the relevant time scale to study the angular momentum evolution should be either $t_c$ or $t_{Ap}$. We don't have to choose between these two times yet. But as we already assumed that $t_c \gg t_\Omega$ ( as a consequence of $ U_m \ll d \Ro \Omega_0$) and $t_{Ap} \gg t_\Omega$, we can safely assume that the characteristic time of the angular momentum evolution, denoted $t_*$, verifies $t_* \gg t_\Omega$.   

With these choices, the scaled version of the radial and latitudinal components of the MHD Boussinesq equations read :
 \begin{equation}
     \begin{split}
     \label{_vr}
     \vphantom{\left(\frac{B^2}{1}\right)} 
     \frac{t_{\Omega}^2}{t_* t_c} \frac{\partial \widetilde{v}_r}{\partial \widetilde{t}} & +\Ro \left( \frac{t_{\Omega}}{t_{c}} \right)^2 \left(\widetilde{\vec{v}}_{\vec{m}}\cdot\widetilde{\vec{\nabla}}\widetilde{v}_r-\frac{\widetilde{v}^2_\theta}{\widetilde{r}} \right) - (2 \sin\theta\,\widetilde{v}_\varphi + \Ro \frac{\widetilde{v}^2_\varphi}{\widetilde{r}}) =
     \widetilde{T}_1-\frac{\partial\widetilde{p}_1}{\partial\widetilde{r}}\\
     & + \frac{1}{\Ro} \left( \frac{t_{\Omega}}{t_{Ap}} \right)^2 \left[ \frac{\widetilde{B}_\theta}{\widetilde{r}} \left( \frac{\partial\widetilde{B}_r}{\partial\theta} - \frac{\partial  (\widetilde{r}\widetilde{B}_\theta)}{\partial\widetilde{r}} \right) \right] -
     \Ro \frac{\widetilde{B}_\varphi}{\widetilde{r}} \frac{\partial (\widetilde{r} \widetilde{B}_\varphi)}{\partial\widetilde{r}} +
     \frac{t_{\Omega}^2}{t_\nu t_c}  \left.\widetilde{\Delta}\widetilde{\vec{v}}\right|_r\ \ \ ,\\
     \end{split}
    \end{equation}
    \begin{equation}
     \begin{split}
     \label{_vtheta}
     \vphantom{\left(\frac{B^2}{1}\right)}
\frac{t_{\Omega}^2}{t_* t_c} \frac{\partial \widetilde{v}_\theta}{\partial \widetilde{t}} &+ \Ro \left( \frac{t_{\Omega}}{t_{c}} \right)^2 \left( \widetilde{\vec{v}}_{\vec{m}}\cdot\widetilde{\vec{\nabla}}\widetilde{v}_\theta+\frac{\widetilde{v}_\theta\widetilde{v}_r}{\widetilde{r}} \right) -( 2 \cos\theta\,\widetilde{v}_\varphi + \Ro \cot\theta\frac{\widetilde{v}^2_\varphi}{\widetilde{r}}) =
     - \frac{1}{r}\frac{\partial\widetilde{p}_1}{\partial\theta} \\
     & + \frac{1}{\Ro} \left( \frac{t_{\Omega}}{t_{Ap}} \right)^2 \left[ \frac{\widetilde{B}_r}{\widetilde{r}}\left(\frac{\partial\,(\widetilde{r}\widetilde{B}_\theta)}{\partial\widetilde{r}}-\frac{\partial\widetilde{B}_r}{\partial\theta}\right)
     \right] -
     \Ro \frac{\widetilde{B}_\varphi}{\widetilde{r}\sin\theta}\frac{\partial\,(\sin\theta\widetilde{B}_\varphi)}{\partial\theta} +
     \frac{t_{\Omega}^2}{t_\nu t_c} \left.\widetilde{\Delta}\widetilde{\vec{v}}\right|_\theta\ \ \ ,\\
     \end{split}
    \end{equation}

From these expressions, the inertial terms that do not involve the azimuthal velocity can be neglected because $t_c \gg t_\Omega$ and $t_* \gg t_\Omega$.  Moreover, the viscous terms is negligible if $t_\nu \gg t_\Omega$, and, as long as $\Ro$ is finite and non-zero, the term of the Lorentz force that contains the poloidal field is very small because $t_{Ap} \gg t_\Omega$. We thus simplify Eqs. (\ref{_vr}, \ref{_vtheta}) into :
\begin{equation}
     \begin{split}
     \label{_vr_s}
     \vphantom{\left(\frac{B^2}{1}\right)} 
- 2 \sin\theta\,\widetilde{v}_\varphi - \frac{\Ro}{\widetilde{r}} (\widetilde{v}^2_\varphi - \widetilde{B}^2_\varphi) =
     \widetilde{T}_1-\frac{\partial}{\partial\widetilde{r}}\left( \widetilde{p}_1 + \frac{\Ro}{2} \widetilde{B}^2_\varphi \right)
     \ \ \ ,\\
     \end{split}
    \end{equation}
    \begin{equation}
     \begin{split}
     \label{_vtheta_s}
     \vphantom{\left(\frac{B^2}{1}\right)}
    - 2 \cos\theta\,\widetilde{v}_\varphi - \Ro \frac{\cot\theta}{\widetilde{r}} (\widetilde{v}^2_\varphi - \widetilde{B}^2_\varphi) =  
     - \frac{1}{\widetilde{r}}\frac{\partial}{\partial\theta} \left( \widetilde{p}_1 + \frac{\Ro}{2} \widetilde{B}^2_\varphi \right)
\ \ \ ,\\
     \end{split}
    \end{equation}

The pressure terms, including the magnetic pressure, can be eliminated to get a magneto-thermal wind equation that relates the temperature fluctuations to the differential rotation and the azimuthal field. This relation has been anticipated to determine the characteristic temperature fluctuation  $T^* = \frac{\Omega_0^2}{N^2} \frac{\Delta \Omega}{\Omega_0} \Delta T$ associated with the differential rotation. 
We now turn to the thermal energy equation that relates temperature fluctuations and meridional velocities :
 \begin{equation}
     \label{_energie}
     \vphantom{\left(\frac{B^2}{1}\right)}
     \frac{t_\kappa}{t_*} \frac{\partial \widetilde{T}_1}{\partial \widetilde{t}} +  \Ro \frac{t_\kappa}{t_c} \widetilde{\vec{v}}_{\vec{m}}\cdot\widetilde{\vec{\nabla}}\widetilde{T}_1 + \frac{N^2}{\Omega^2} \frac{t_\kappa}{t_c} \widetilde{v_r}\frac{d \overline{T}}{d \widetilde{r}}=\widetilde{\Delta} \widetilde{T}_1\ \ \ ,\\
    \end{equation}
The advection of the temperature has been split into the advection of temperature fluctuations by meridional motions and the radial advection against the background stratification. This last term is expected to dominate the advection if $ \Ro \frac{\Omega^2}{N^2} \ll 1$. Then, depending on the ratio $t_\kappa/t_*$, it can be balanced either by the time derivative of temperature fluctuations or by the thermal diffusion term. The two cases are now considered separately :

\subsection{Alfv\'en waves and Eddington-Sweet circulation}

If $t_* \gg t_\kappa$ the thermal diffusion term dominates over the temperature time variation in Eq. (\ref{_energie}. Thus the balance between the thermal diffusion transport and the radial advection against the background stratification determines the circulation time  $t_c = t_{\kappa} \frac{N^2}{\Omega^2}$ and the characteristic meridional velocity $U_m = \frac{\kappa}{d}\frac{\Omega^2}{N^2}\frac{\Delta \Omega}{\Omega}$. The scaled thermal energy equation is then : 
\begin{equation}
     \label{_energie_a}
     \vphantom{\left(\frac{B^2}{1}\right)}
     \widetilde{v_r}\frac{d \overline{T}}{d \widetilde{r}}=\widetilde{\Delta} \widetilde{T}_1\ \ \ ,\\
    \end{equation}
where the circulation appears driven by the thermal diffusion of the temperature deviations, that were produced by the differential rotation. It is an Eddington-Sweet type circulation of time scale $t_c =t_{es} = \frac{d^2}{\kappa} \frac{N^2}{\Omega^2}$. We can now verify that the meridional circulation satisfies the condition $t_{c} \gg t_{\Omega}$ necessary to simplify Eqs. (\ref{_vr_s}, \ref{_vtheta_s}) if $t_{es} \gg t_\Omega$. This is satisfied in stars because $t_\kappa = \frac{d^2}{\kappa} \gg t_\Omega$ and $N \ge \Omega$. 
The system of equation is completed by three prognostic equations for $v_\varphi$, $B_\varphi$ and the potential $A$, defined by   $\vec{B}_{\vec{p}} = \vec{\nabla} \times A \vec{e}_\varphi$. 

Their scaled form is : 
\begin{equation}
     \begin{split}
     \label{_vphi_a}
     \vphantom{\left(\frac{B^2}{1}\right)}
     \frac{\partial \widetilde{v}_\varphi}{\partial \widetilde{t}}+\Ro\frac{t_{Ap}}{t_{es}}\frac{1}{\widetilde{r}\sin\theta}\left(\widetilde{\vec{v}}_{\vec{m}}\cdot\widetilde{\vec{\nabla}}\right)\left(\widetilde{r}\sin\theta\widetilde{v}_\varphi\right)
+2 \frac{t_{Ap}}{t_{es}}\left(\sin\theta\,\widetilde{v}_r+\cos\theta\,\widetilde{v}_\theta\right) =
     \frac{1}{\widetilde{r}\sin\theta}\left(\widetilde{\vec{B}}_{\vec{p}}\cdot\widetilde{\vec{\nabla}}\right)\left(\widetilde{r}\sin\theta\widetilde{B}_\varphi\right) +
     \frac{t_{Ap}}{t_\nu} \left( \widetilde{\Delta} - \frac{1}{\widetilde{r}^2 \sin^2\theta} \right) \widetilde{v}_{\varphi}\ \ \ ,\\
     \end{split}
    \end{equation}
    
    \begin{equation}
     \begin{split}
     \label{_toroidal_a}
     \vphantom{\left(\frac{B^2}{1}\right)}
    \frac{\partial \widetilde{B}_\varphi}{\partial\widetilde{t}}+
     \Ro \frac{t_{Ap}}{t_{es}} \widetilde{r}\sin\theta \left(\widetilde{\vec{v}}_{\vec{m}}\cdot\widetilde{\vec{\nabla}}\right)\left(\frac{\widetilde{B}_\varphi}{\widetilde{r}\sin\theta} \right) =
     \widetilde{r}\sin\theta \left(\widetilde{\vec{B}}_{\vec{p}}\cdot\widetilde{\vec{\nabla}}\right)\left(\frac{\widetilde{v}_\varphi}{\widetilde{r}\sin\theta}\right) +
     \frac{t_{Ap}}{t_\eta} \left( \widetilde{\Delta} - \frac{1}{\widetilde{r}^2 \sin^2\theta} \right) \widetilde{B}_{\varphi}\ \ \ ,\\
     \end{split}
    \end{equation}
    
    \begin{equation}
     \begin{split}
     \label{_poloidal_a}
     \vphantom{\left(\frac{B^2}{1}\right)}
     \frac{\partial \widetilde{A}}{\partial\widetilde{t}}+
     \Ro\frac{t_{Ap}}{t_{es}}\frac{1}{\widetilde{r}\sin\theta} \left(\widetilde{\vec{v}}_{\vec{m}}\cdot\widetilde{\vec{\nabla}}\right)\left(\widetilde{r}\sin\theta \widetilde{A}\right) =
     \frac{t_{Ap}}{t_\eta} \left( \widetilde{\Delta} - \frac{1}{\widetilde{r}^2 \sin^2\theta} \right) \widetilde{A}\ \ \ ,\\
     \end{split}
    \end{equation}
where we used $t_* = t_{Ap}$ (but could also have used $t_* = t_{es}$).  

The scale analysis thus led to a simplified system formed by Eqs. (\ref{_vr_s}, \ref{_vtheta_s}, \ref{_energie_a}, \ref{_vphi_a},\ref{_toroidal_a},\ref{_poloidal_a} ), plus the mass conservation equation, $\widetilde{\vec{\nabla}} \cdot \widetilde{\vec{v}}_{\vec{m}} =0$. To be consistent the approximations requires  $t_{Ap} \gg t_\kappa \gg t_\Omega \gg \Ro^{\frac{1}{2}} t_B$  together with $t_\nu \gg t_\Omega$ and $\Ro^{\frac{1}{2}} t_{Ap} \gg t_\Omega$. It intends to describe
axisymmetric motions for time scale of the order of $t_*= t_{Ap}$.
In particular, it should fail when solid body rotation is reached because the Lorentz force term involving the poloidal field component in Eqs. (\ref{_vr_s}, \ref{_vtheta_s}) will no longer be negligible. Also, short time dynamics like gravity waves have been filtered out by the approximation of the scaling analysis. 

In this simplified form, the system is fully determined by the azimuthal velocity and the two components of the magnetic field. The meridional velocity components and the temperature fluctuations are intermediate variables determined by the magneto-thermal wind equilibrium and the thermal equilibrium. Flows, where such equilibrium equation reduces the number of independent variables, are said to have balanced dynamics \citep[e.g.][]{Vallis2006}. Physically, the flow evolves through Alfv\'en wave dynamics and an Eddington Sweet circulation prescribed by the instantaneous angular momentum and azimuthal field distributions.

As compared to the full MHD problem that depends on 5 non-dimensional numbers (plus $Ro = \Delta \Omega/\Omega_0)$, this simplified system has the advantage to depend only on three non-dimensional numbers $\frac{t_{es}}{t_{Ap}} = \frac{Lu}{Pm} \Pra N^2/\Omega_0^2$ ,  $\frac{t_\eta}{t_{Ap}} = Lu$ and $\frac{t_{\nu}}{t_{Ap}} = \frac{Lu}{Pm}$, or equivalently on $\Pra N^2/\Omega_0^2$, $Lu$ and $Pm$. Consequently, for given initial conditions and thus a given $\Ro$, solutions can be expressed in the general form $\widetilde{v}_\varphi = \frac{v_\varphi}{d \Delta \Omega} = f_0(t/t_{Ap}, \vec{r}/d, \Pra N^2/\Omega_0^2,Lu, Pm)$, $\widetilde{B}_\varphi =\frac{B_\varphi}{d \Delta \Omega \sqrt{4 \pi \rho}} = f_1(t/t_{Ap}, \vec{r}/d, \Pra N^2/\Omega_0^2,Lu, Pm)$, $\widetilde{\vec{B}}_{\vec{p}}  =\frac{\vec{B}_{\vec{p}}}{B_0} = f_2(t/t_{Ap}, \vec{r}/d, \Pra N^2/\Omega_0^2,Lu, Pm)$ from which we deduce $B_{\varphi}/B_p = Lo^{-1} f(t/t_{Ap}, \vec{r}/d, \Pra N^2/\Omega_0^2,Lu, Pm)$, that is the expression given in Sect. \ref{submeri}. 

Most of the numerical simulations listed in table \ref{table_cases} meet the requirement of the scaling analysis as they verify $t_\nu \gg t_{Ap} \gg t_\kappa \gg t_\Omega \gg t_B$ together with $Ro = (\Omega_i-\Omega_0)/\Omega_0 \approx 1$. Except for the transient period during which initially excited gravity waves are dissipated, their dependence on $\Pra N^2/\Omega_0^2$ and $Lo$ indicate that they are indeed governed by the simplified equations derived from the present scale analysis.

Below, we consider the case $t_{Ap} \le t_\kappa$. It holds in particular for the run R9 of table \ref{table_cases} for which $\frac{t_\kappa}{t_{Ap}} = \frac{Lu \Pra }{Pm} = 3.125$.
    
\subsection{Alfv\'en waves}
    
If $t_* \ll  t_\kappa$, the balanced thermal energy equation  is : 
\begin{equation}
     \label{_energie_b}
     \vphantom{\left(\frac{B^2}{1}\right)}
 \frac{\partial \widetilde{T}_1}{\partial \widetilde{t}} +   \widetilde{v_r}\frac{d \overline{T}}{d \widetilde{r}}=0\ \ \ ,\\
    \end{equation}with $t_c = t_* \frac{N^2}{\Omega^2}$ or equivalently $U_m = \frac{\Omega^2}{N^2}\frac{\Delta \Omega}{\Omega}  \frac{d}{t_*}$. Then, the conditions $t_* \gg t_\Omega$ and $t_c \gg t_\Omega$ necessary to simplify Eqs. (\ref{_vr_s}, \ref{_vtheta_s}), now read $t_* \gg \frac{t_B^2}{t_\Omega}$. As $N \ge \Omega$, we have $t_c > t_*$ which implies that $t_* = t_{Ap}$ is the more relevant choice for the time scale characterizing the angular momentum evolution. The condition  $t_* = t_{Ap} \gg t_\Omega$ is met because $t_{Ap} \gg t_\Omega$ and $t_\Omega \ge t_B$. The amplitude of the meridional motion is now $U_m = \frac{\Omega^2}{N^2}\frac{\Delta \Omega}{\Omega}  v_{Ap}$. 
    The scaled version of the three 
    prognostic equations for $v_\varphi$, $B_\varphi$ and $\vec{B}_{\vec{p}} = \vec{\nabla} \times A \vec{e}_\varphi$  simplifies into :

\begin{equation}
     \begin{split}
     \label{_vphi_b}
     \vphantom{\left(\frac{B^2}{1}\right)}
     \frac{\partial \widetilde{v}_\varphi}{\partial \widetilde{t}} +2 \frac{\Omega^2}{N^2}\left(\sin\theta\,\widetilde{v}_r+\cos\theta\,\widetilde{v}_\theta\right) =
     \frac{1}{\widetilde{r}\sin\theta}\left(\widetilde{\vec{B}}_{\vec{p}}\cdot\widetilde{\vec{\nabla}}\right)\left(\widetilde{r}\sin\theta\widetilde{B}_\varphi\right) +
     \frac{t_{Ap}}{t_\nu} \left( \widetilde{\Delta} - \frac{1}{\widetilde{r}^2 \sin^2\theta} \right) \widetilde{v}_{\varphi}\ \ \ ,\\
     \end{split}
    \end{equation}
   \begin{equation}
     \begin{split}
     \label{_toroidal_b}
     \vphantom{\left(\frac{B^2}{1}\right)}
    \frac{\partial \widetilde{B}_\varphi}{\partial\widetilde{t}}=
     \widetilde{r}\sin\theta \left(\widetilde{\vec{B}}_{\vec{p}}\cdot\widetilde{\vec{\nabla}}\right)\left(\frac{\widetilde{v}_\varphi}{\widetilde{r}\sin\theta}\right) +
     \frac{t_{Ap}}{t_\eta} \left( \widetilde{\Delta} - \frac{1}{\widetilde{r}^2 \sin^2\theta} \right) \widetilde{B}_{\varphi}\ \ \ ,\\
     \end{split}
    \end{equation}
    \begin{equation}
     \begin{split}
     \label{_poloidal_b}
     \vphantom{\left(\frac{B^2}{1}\right)}
     \frac{\partial \widetilde{A}}{\partial\widetilde{t}} 
     +
     \Ro\frac{\Omega^2}{N^2}\frac{1}{\widetilde{r}\sin\theta} \left(\widetilde{\vec{v}}_{\vec{m}}\cdot\widetilde{\vec{\nabla}}\right)\left(\widetilde{r}\sin\theta \widetilde{A}\right) =
     \frac{t_{Ap}}{t_\eta} \left( \widetilde{\Delta} - \frac{1}{\widetilde{r}^2 \sin^2\theta} \right) \widetilde{A}\ \ \ ,\\
     \end{split}
    \end{equation}
because, as for the thermal energy equation, the advection terms proportional to $\Ro \frac{\Omega^2}{N^2}$ are neglected with respect to the Lorentz force or the $\Omega$-effect term in the $v_\varphi$ and $B_\varphi$ equations, respectively. Although the advection of the poloidal field is also of the order of $\Ro \frac{\Omega^2}{N^2}$, we kept this term in Eq. (\ref{_poloidal_b}) because it may dominate over the magnetic diffusion.

At this stage we can distinguish two sub-regimes depending on the ratio $\Omega/N$. If $\frac{\Omega^2}{N^2} \ll 1$, the Coriolis force term in the angular momentum equation (\ref{_vphi_b}) is negligible. As a consequence, the equations for $v_\varphi$ and $B_\varphi$ are decoupled from the other ones. They describe the evolution of the initial differential rotation through Alfv\'en wave propagation. This regime of the scale analysis requires $t_\kappa \gg t_{Ap}\gg t_\Omega \gg  t_B$ with also $t_\nu \gg t_\Omega$  and $\Ro^{\frac{1}{2}} t_{Ap} \gg t_\Omega$. Under these conditions, the approach of \cite{gaurat2015}, where only the equations for $v_\varphi$ and $B_\varphi$ were solved, appears to be justified.

A second sub-regime corresponding to $\frac{\Omega^2}{N^2} \sim 1$ exists. As $\Ro \frac{\Omega^2}{N^2} \ll 1 $, it implies $\Ro \ll 1$ and the terms $\propto \Ro$ in the thermal wind balance should then be neglected for consistency. Gathering the time scale conditions, this regime holds when $t_\kappa \gg t_{Ap}\gg t_\Omega \sim t_B$  together with $\Ro \ll 1$, $t_\nu \gg t_\Omega$  and $\Ro^{\frac{1}{2}} t_{Ap} \gg t_\Omega$. As shown by a local analysis, this system supports Alfv\'en waves with frequencies (slightly) modified by the stratification and the rotation.

\section{Acheson dispersion relation in the limit of high thermal diffusivity}
\label{sec_acheson}

The procedure used by \citet{Acheson78} to derive his dispersion relation is the following: the MHD equations governing the system with thermal, viscous and magnetic diffusion are linearised
around the background axisymmetric state (which is assumed to be purely toroidal both for the magnetic and the velocity fields). Small amplitude harmonic perturbations in space and time of the following form are then considered:
\begin{equation}
\exp \left[\mi (\ks s+\kz z+m \varphi-\sigma\, t)\right]
\end{equation}

Here $\ks=2\pi/\wls$ ($\kz=2\pi/\wlz$) is the radial (axial) wavenumber
of the instability and $m$ its azimuthal order which is an $O(1)$ integer.
When the imaginary part of $\sigma$ is positive, the applied perturbation
is unstable and grows exponentially at a rate $\gr=\Im(\sigma)$.
 
 In this appendix, we recall the dispersion relation derived by \citet{Acheson78} in the case where all the diffusivities are taken into account (thermal, viscous and magnetic) but when the thermal diffusivity is much higher than the magnetic diffusivity. In this situation, the dispersion relation is reduced to a simpler expression, which corresponds to equation 3.20 in \citet{Acheson78}:

\begin{multline}
\label{a:AchesonDR}
\vA^2 \left[\frac{2\Omega m}{s}+\frac{2(\omega+\iu \nu k^2)}{s}\right]\,\left[m\pdv{\Omega}{h}+(\omega+\iu\md k^2)\pdv{F}{h}-\omega\frac{\eta}{\kappa}\pdv{E}{h}\right]\\+
\left[\frac{k^2}{\kz^2}\left((\omega+ \iu \nu k^2)(\omega+\iu\md k^2) - \frac{m^2 \vA^2}{s^2}\right)-G\frac{\md}{\kappa} \pdv{E}{h}\right]\times
\left[(\omega+\iu\nu k^2)(\omega+\iu\md k^2)-\frac{m^2 \vA^2}{s^2}\right] \\-
\left[(\omega+\iu\md k^2)\pdv{(\Omega s^2)}{h}+m\,\vA^2\pdv{Q}{h}\right] \times
\left[\frac{2\Omega}{s}(\omega+\iu\md k^2) + \frac{2m\vA^2}{s^3}\right] = 0
\end{multline}
where $\omega=\sigma-m\Omega$ is the Doppler-shifted frequency, $\vA=\Bphi/\sqrt{\rho \mpv}$
the \Alfven\ velocity, $k^2=\ks^2+\kz^2$, $E=\ln(P/\rho^\gamma)$, $F=\ln(B/(s\rho))$, $G=g_s-\ks/\kz g_z$ and $Q=\ln(s\Bphi)$. We also defined the meridional
derivative
\beq
\pdv{\,}{h}=\pdv{\,}{s}+\frac{\ks}{\kz}\pdv{\,}{z}
\eeqp

When written as a polynomial equation in the dimensionless frequency $\omegaND=\omega/\Omega_0$,
\refp{a:AchesonDR} reads
\beql{a:AchesonND}
\sum_{i=0}^4 a_i \omegaND^i=0
\eeq
where
\begin{subequations}
\begin{align}
a_4 &= 1+\beta^2\nonumber \\
a_3 &= 2\iu (1+\beta^2)\,(\Rmd^{-1}+\mbox{R}_e^{-1})\nonumber \\
a_2 &= 2q -4 + 2\LophiL^2[b - 1 - (1+\beta^2)m^2] - (1+\beta^2)\,(\Rmd^{-2}+\mbox{R}_e^{-2}+4\Rmd^{-1}\mbox{R}_e^{-1})\nonumber \\
       &\,\,\,\,\,\, - \gamma \frac{\mbox{R}_t}{\Rmd}(\sin\theta-\beta\cos\theta)^2\frac{N^2}{\Omega_0^2} \nonumber \\
a_1 &= -8m\LophiL^2 + 2 \iu \LophiL^2[b-1-(1+\beta^2)m^2](\Rmd^{-1}+\mbox{R}_e^{-1}) \nonumber\\
       &\,\,\,\,\,\,-4 \iu(2-q)\Rmd^{-1} - 2 (1+\beta^2) (\mbox{R}_e^{-2}\Rmd^{-1} + \Rmd^{-2}\mbox{R}_e^{-1})\nonumber \\
       &\,\,\,\,\,\,- \iu\gamma \frac{\mbox{R}_t}{\Rmd} (\Rmd^{-1}+\mbox{R}_e^{-1}) (\sin\theta-\beta\cos\theta)^2\frac{N^2}{\Omega_0^2}\nonumber\\
a_0 &= m^2\LophiL^2\left\{-2q +\LophiL^2[(1+\beta^2)m^2 - 2 (b-1)]\right\} \nonumber \\
       &\,\,\,\,\,\,-2\iu m \LophiL^2(4-q)\Rmd^{-1}+2 (2-q)\Rmd^{-2} \nonumber \\
       &\,\,\,\,\,\,- 2\iu   m \LophiL^2 q \mbox{R}_e^{-1} + 2 \LophiL^2 [m^2(1+\beta^2)-(b-1)] \mbox{R}_e^{-1} \Rmd^{-1} + (1+\beta^2) \mbox{R}_e^{-2} \Rmd^{-2}\nonumber \\
        &\,\,\,\,\,\,+ (m^2 \LophiL^2+ \mbox{R}_e^{-1} \Rmd^{-1}) \gamma \frac{\mbox{R}_t}{\Rmd}(\sin\theta-\beta\cos\theta)^2\frac{N^2}{\Omega_0^2}  \,.
\end{align}
\end{subequations}

We note here that the terms involving the stable stratification are always proportional to $\frac{\mbox{R}_t}{\Rmd}\frac{N^2}{\Omega_0^2}$, which for our cases where $Pm=1$ reduces to our usual parameter $Pr \frac{N^2}{\Omega_0^2}$. So again, we clearly see already that the effect of stable stratification also on the instability will be mainly controlled by this product and not by $N/\Omega_0$ alone. \\

The dispersion relation coefficients depend on six dimensionless parameters: 

The ratio of poloidal wavenumbers (in cylindrical and spherical geometries):
\beq
\beta=\frac{\ks}{\kz}=\frac{\cos\theta \kt +\sin\theta \kr}{\cos\theta \kr - \sin\theta \kt}
\eeqc
Note that when $\kt << kr$ (mostly horizonthal displacement), $\beta=\tan\theta$ and the terms involving the stable stratification in the dispersion relation, all proportional to $(\sin\theta-\beta\cos\theta)$, vanish and the stratification has thus no effect.\\

The shear parameter
\beql{a:q}
q = -\pdv{\ln \Omega}{\ln s} + \beta \frac{s}{z} \pdv{\ln \Omega}{\ln z}
\eeqc
a parameter associated to the field derivatives
\beql{a:b}
b = \frac{1}{2}\left(\pdv{\ln \Bphi^2}{\ln s} - \beta \frac{s}{z} \pdv{\ln \Bphi^2}{\ln z}\right)
\eeqc
the local azimuthal Lorentz number
\beql{a:LophiL}
\LophiL=\frac{\AfphiL}{\Omega}
\eeq
obtained defining the \Alfven\ frequency as $\AfphiL=\Bphi/\sqrt{\mu_0\rho}s$, and finally the
magnetic, kinetic and thermal Reynolds numbers
\beq
\Rmd = \frac{\Omega_0}{\md k^2} \,\,\,\,\,\,\,\, \mbox{R}_e = \frac{\Omega_0}{\nu k^2} \,\,\,\,\,\,\,\, \mbox{R}_t = \frac{\Omega_0}{\kappa k^2}
\eeq
respectively.

\end{document}